\documentclass[12pt,epsfig]{article}
\usepackage{graphicx}
\usepackage{amssymb}
\usepackage{caption2}
\usepackage{longtable}
\usepackage{ltxtable}
\usepackage{afterpage}
\usepackage{lscape}
\def\bfm#1{\mbox{\boldmath$#1$}}
\def\lambdabar{{\lambda\kern -0.5em \raise 0.5 ex \hbox{--}}}

\begin{document}

\newcommand{\beq}{\begin{equation}}
\newcommand{\eps}{\epsilon}
\newcommand{\eeq}{\end{equation}}
\newcommand{\bea}{\begin{eqnarray}}
\newcommand{\eea}{\end{eqnarray}}

\begin{titlepage}

\vspace*{-10mm}
\begin{center}
{\large \bf 
Invited lectures presented at the Joint ICTP-IAEA \\
Advanced Workshop on Model Codes for Spallation Reactions,\\
February 4--8, 2008, ICTP, Trieste, Italy\\
(LANL Report LA-UR-08-2931)
}

\vspace*{30mm}
{\LARGE \bf \sf
CEM03.03 and LAQGSM03.03
Event Generators for the\\

\vspace{1.5mm}
MCNP6, MCNPX, and MARS15
Transport Codes
} 

\vspace{5mm}
{\Large \sf
S. G. Mashnik$^1$, K. K. Gudima$^{2}$,
R. E. Prael$^1$,\\
\vspace{1.5mm}
A. J. Sierk$^1$,
M. I. Baznat$^2$,
and
N. V. Mokhov$^3$
}

\vspace{0.5cm}

$^1${\em  
Los Alamos National Laboratory, Los Alamos, NM 87545, USA}\\

%\vspace{0.5cm}
$^2${\em Institute of Applied Physics,
Academy of Science of Moldova, Chi\c{s}in\u{a}u, %MD-2028, 
Moldova}\\

%\vspace{0.5cm}
$^3${\em Fermi National Accelerator Laboratory, Batavia, IL, USA}\\

\vspace{10mm}
{\large\bf Abstract} \\
\end{center}
\vspace{0.3cm}

A 
%detailed 
description of the IntraNuclear Cascade (INC),
preequilibrium, evaporation, fission, coalescence,
and Fermi breakup models used by the latest versions of 
our CEM03.03 and
LAQGSM03.03 event generators is presented, with a
focus on our most recent developments of these models.
The recently developed ``S'' and ``G'' versions of
our codes, that consider multifragmentation of
nuclei formed after the preequilibrium stage of reactions
when their excitation energy is above $2 A$ MeV using 
the Statistical Multifragmentation Model (SMM)
code by Botvina {\it et al.} (``S'' stands for SMM)
and the fission-like binary-decay model GEMINI by Charity
(``G'' stands for GEMINI), respectively, are briefly
described as well.
Examples of benchmarking our models
against a large variety of experimental data on particle-particle,
particle-nucleus, and nucleus-nucleus reactions are presented.
Open questions on reaction mechanisms and future necessary work 
are outlined.

\vspace*{40mm}
\begin{center} 
{\large April 2008}
\end{center} 

\end{titlepage}

\newpage
\begin{center}
\vspace*{-5mm}
{\large\bf Contents} \\
\end{center}
\vspace*{0.cm}
%\tableofcontents
\begin{tabbing}
0123456789112345678921234567893123456789412345678951234567896123456789712\=1234\kill
 \> \\
{\large\bf 1.  Introduction} \>  \hspace*{14pt}2\\
 \> \\
{\large\bf 2.  A Brief Survey of CEM and LAQGSM Physics} \>  \hspace*{14pt}3\\
 \> \\
{\large\bf 3. The Intranuclear Cascade Mechanism} \>  \hspace*{14pt}4\\
 \> \\
\hspace{9mm} {\large\it 3.1. The INC of CEM03.03} \>  \hspace*{14pt}8\\
 \> \\
\hspace{9mm} {\large\it 3.2. The INC of LAQGSM03.03} \>  \hspace*{7pt}22\\
 \> \\
{\large\bf 4.  The Coalescence Model} \>  \hspace*{7pt}28\\
 \> \\
{\large\bf 5.  Preequilibrium Reactions} \>  \hspace*{7pt}29\\
 \> \\
{\large\bf 6.  Evaporation} \>  \hspace*{7pt}43\\
 \> \\
{\large\bf 7.  Fission} \>  \hspace*{7pt}53\\
 \> \\
\hspace{9mm} {\large\it 7.1. Fission Probability} \>  \hspace*{7pt}53\\
 \> \\
\hspace{9mm} {\large\it 7.2. Mass Distribution} \>  \hspace*{7pt}56\\
 \> \\
\hspace{20mm} {\sf 7.2.a. Asymmetric fission} \>  \hspace*{7pt}56\\
 \> \\
\hspace{20mm} {\sf 7.2.b. Symmetric fission} \>  \hspace*{7pt}56\\
 \> \\
\hspace{9mm}  {\large\it 7.3. Charge Distribution} \>  \hspace*{7pt}57\\
 \> \\
\hspace{9mm} {\large\it 7.4. Kinetic Energy Distribution} \> \hspace*{7pt}57\\
 \> \\
\hspace{9mm} {\large\it 7.5. Modifications to GEM2 in CEM03.01 
and LAQGSM03.01} \> \hspace*{7pt}58\\
 \> \\
{\large\bf 8. The Fermi Breakup Model} \>  \hspace*{7pt}62\\ 
 \> \\
{\large\bf 9. Reactions Involving Pions and Photons; Excitation} \>  \\
\hspace{6mm} {\large\bf Functions}
 \>  \hspace*{7pt}67\\
 \> \\
{\large\bf 10. CEM03.S1, CEM03.G1, LAQGSM03.S1, and} \> \\
\hspace{9mm} {\large\bf  LAQGSM03.G1} \> \hspace*{7pt}70\\
 \> \\
{\large\bf
Acknowledgments} \>  \hspace*{7pt}83\\
 \> \\
{\large\bf
References} \>  \hspace*{7pt}84\\
\end{tabbing}

\newpage

\begin{center}
{\large\bf 1.  Introduction} \\
\end{center}
\vspace{0.3cm}

Following an increased interest in intermediate-
and high-energy nuclear data 
in relation to such projects as the Accelerator Transmutation of nuclear 
Wastes (ATW),  
%the 
Accelerator Production of Tritium (APT), 
%the 
Spallation Neutron Source (SNS), 
%the 
Rare Isotope Accelerator (RIA), 
Proton Radiography (PRAD) as a radiographic probe for the Advanced 
Hydro-test Facility, NASA needs, and others,  the US Department 
of Energy has supported during the last decade our work on the 
development of improved versions of the Cascade-Exciton Model (CEM)
and of the Los Alamos version of the Quark Gluon String Model (LAQGSM)
which has led to our intermediate-
and high-energy event generators CEM03.03 and LAQGSM03.03,
respectively, and their modifications described below.

The main focus of our workshop is nucleon-induced reactions
for Accelerator Driven Systems (ADS) and spallation neutron sources
(up to 2--3 GeV), usually described well enough by our intermediate-energy
code CEM03.03 without a need to use our high-energy code LAQGSM03.03. 
This is why we focus below mostly on CEM03.03 and its modifications.
However, as discussed below in Section 3.1, 
CEM does not consider the so-called ``trawling'' effect (depletion
of target nucleons during a cascade),
therefore does not describe well reactions on very light nuclei like
C at incident energies above about 1 GeV. 
Therefore, in transport codes that use both CEM and our high-energy 
code LAQGSM as event generators, we recommend simulating nuclear 
reactions with CEM at incident energies up to about 1 GeV for 
light nuclei like C and up to about 5 GeV for actinide nuclei, 
and to switch to simulations by LAQGSM, which does consider the ``trawling" 
effect, at higher energies of transported particles. 
This is the reason we have included in the present lectures a
brief description of LAQGSM as well. Even at energies of ADS applications
below about 3 GeV, we recommend using LAQGSM instead of CEM in 
the case of light target nuclei. 

The Cascade-Exciton Model (CEM) of nuclear reactions 
was proposed almost 30 years ago at the Bogoliubov
Laboratory of Theoretical Physics, Joint Institute for
Nuclear Research, Dubna, USSR by Gudima, Mashnik,
and Toneev \cite{CEMP,CEM}. It is based on the standard
(non time-dependent)
Dubna IntraNuclear Cascade (INC) \cite{Book,UPN73}
and the Modified Exciton Model (MEM) \cite{MEM,MODEX}.
CEM was extended later to consider photonuclear reactions \cite{cemphoto}
and to describe fission cross sections
using different options for nuclear masses, fission barriers, 
and level densities \cite{pit95} and its 1995 version, CEM95, 
was released to the public via NEA/OECD, Paris as the code
IAEA1247, and via the Radiation Safety Information Computational 
Center (RSICC) at Oak Ridge, USA, as the RSICC code package 
PSR-357 \cite{CEM95}.

The {\it International Code Comparison for Intermediate
Energy Nuclear Data}~\cite{nea94a,nagel95}
organized during 1993--1994 at NEA/OECD in Paris to address
the subject of codes and models used to calculate nuclear reactions
from 20 to 1600 MeV showed that CEM95 (more exactly, CEM92m, 
which is almost identical in its physics content 
to the publicly released CEM95 version)
had one of the best predictive powers
to describe nucleon-induced reactions at energies above about 150 
MeV when compared to other models and codes available at that time. 

The code LAQGSM03.03 described briefly below
is the latest modification \cite{LAQGSM03.03}
of LAQGSM~\cite{LAQGSM}, which in its turn is an improvement 
of the  Quark-Gluon String Model (QGSM) \cite{QGSM}.
It describes reactions induced by both particles and 
nuclei at incident energies up to about 1 TeV/nucleon, 
generally, as a three-stage process: IntraNuclear Cascade (INC), 
followed
by preequilibrium emission of particles during the equilibration of the
excited residual nuclei formed after the INC, followed by evaporation 
of particles from and/or fission of the compound nuclei. 

CEM95 and/or its predecessors and successors like
CEM97~\cite{CEM97,CEM97f},
CEM2k~\cite{CEM2k}, \\
CEM2k+GEM2~\cite{SATIF6}--\cite{fitaf},
CEM03~\cite{ND2004,JNRS05}, including the latest version available
to the public from RSICC and NEA OECD,
CEM03.01~\cite{CEM03.01},
as well as different versions of LAQGSM and its predecessor QGSM 
are used as stand-alone codes to study different
nuclear reactions for applications and fundamental nuclear physics
(see, {\it e.g.}, \cite{YF88}--\cite{Pavia06} and references therein). 
Parts of different versions of the CEM, LAQGSM, or QGSM codes are
used in many other stand-alone codes, like
{\bf PICA95}~\cite{PICA95},
{\bf PICA3}~\cite{PICA3},
{\bf CASCADO}~\cite{CASCADO}, 
{\bf CAMO}~\cite{CAMO},
{\bf MCFX}~\cite{MCFX},
{\bf ECM}~\cite{ECM}, and
{\bf NUCLEUS}~\cite{NUCLEUS}.
Different versions of CEM and LAQGSM, or of the older QGSM,
are incorporated wholly, or in part
in many transport codes used in a variety of applications,
like
{\bf CASCADE}~\cite{CASCADE},
{\bf MARS}~\cite{MARS},
{\bf MCNPX}~\cite{MCNPX}
{\bf GEANT4}~\cite{GEANT4,GEANT4Preq},
{\bf SHIELD}~\cite{SHIELD},
{\bf RTS\&T}~\cite{RTSandT},
{\bf SONET}~\cite{SONET},
{\bf CALOR}~\cite{CALOR}, 
{\bf HETC-3STEP}~\cite{HETC-3STEP}, 
{\bf CASCADE/INPE}~\cite{CASCADE/INPE},
{\bf HADRON} \cite{HADRON}, 
{\bf CASCADE-2004}~\cite{CASCADE-2004},
and others. %e.g., FRITIOF, QMD modified by Uzhinsky.

In these lectures, we present a brief description of all models,
approximations, and systematics used in the latest versions of our
CEM and LAQGSM event generators.
 
\begin{center}
{\large\bf 2.  A Brief Survey of CEM and LAQGSM Physics} \\
\end{center}
%\vspace{0.3cm}

The basic version of both our CEM and LAQGSM
event generators is the so-called ``03.01'' version, namely
CEM03.01 \cite{CEM03.01} and LAQGSM03.01 \cite{LAQGSM03.01}.
The CEM03.01 code calculates nuclear reactions induced by nucleons, 
pions, and photons. It assumes
that the reactions occur generally in three stages. 
The first stage is the IntraNuclear Cascade (INC),
in which primary particles can be re-scattered and produce secondary
particles several times prior to absorption by, or escape from the nucleus.
When the cascade stage of a reaction is completed, CEM03.01 uses the
coalescence model
to ``create" high-energy d, t, $^3$He, and $^4$He by
final-state interactions among emitted cascade nucleons, already 
outside of the target.
The emission of the cascade particles determines the particle-hole 
configuration, Z, A, and the excitation energy that is
the starting point for the second, preequilibrium stage of the
reaction.  
The subsequent relaxation of the nuclear excitation is
treated in terms of an improved version of the modified exciton 
model of preequilibrium decay 
followed by the equilibrium evaporation/fission stage of the reaction.
Generally, all four components may contribute to experimentally measured 
particle spectra and other distributions. 
But if the residual nuclei after the INC have atomic numbers 
with  $A \le 12$,  CEM03.01 uses the Fermi breakup model 
to calculate their further disintegration 
instead of using the preequilibrium and evaporation models. 
Fermi breakup is much faster to calculate and gives results very similar 
to the continuation of the more detailed models to
much lighter nuclei. 
   As already mentioned in the Introduction, LAQGSM03.01 also 
describes nuclear reactions, generally, as a three-stage process: 
IntraNuclear Cascade (INC), followed
by preequilibrium emission of particles during the equilibration of the
excited residual nuclei formed after the INC, followed by evaporation 
of particles from or fission of the compound nuclei. 
LAQGSM was developed with a primary focus on describing reactions
induced by nuclei, as well as induced by most elementary particles,
at high energies, up to about 1 TeV/nucleon. The INC of LAQGSM 
is completely different from the one in CEM. LAQGSM03.01 also
considers Fermi breakup of nuclei with $A \le 12$
produced after the cascade, and the coalescence model
to ``produce" high-energy d, t, $^3$He, and $^4$He from
nucleons emitted during the INC.

The main difference of the following, so-called ``03.02'' versions
of CEM and LAQGSM from the basic ``03.01'' versions is that 
the latter use the Fermi breakup model 
to calculate the disintegration of light nuclei
instead of using the preequilibrium and evaporation models
only after the INC, when the excited nuclei after the INC have
a mass number  $A \le 12$, but do not use the Fermi breakup model
at the preequilibrium, evaporation, and fission stages, when, due
to emission of preequilibrium particles or due to evaporation or to
a very asymmetric fission, we get an excited nucleus or
a fission fragment with $A \le 12$.
This problem was solved in the 03.02 versions of CEM and LAQGSM,
where the Fermi breakup model is used at any stage of a reaction,
when we get an excited nucleus with $A \le 12$.

In addition, the routines that describe the Fermi breakup model 
in the basic 03.01 version of our codes
were written some twenty years ago in the group
of Prof. Barashenkov at JINR, Dubna, Russia, and are far
from being perfect, though they are quite reliable and are still used 
currently without any changes in some
 transport codes.
First, these routines allow in rare cases production of some light unstable
fragments like $^5$He, $^5$Li, $^8$Be, $^9$B, etc., as a result
of a breakup of some light excited nuclei. Second, these
routines allowed in some very rare cases even production of
``neutron stars'' (or ``proton stars''), {\ i.e.}, residual ``nuclei''
produced via Fermi breakup that consist of only neutrons (or only protons).
Lastly, 
%as was discovered by Dick Prael, 
in some very rare cases,
these routines could even crash the code, due to cases of 0/0.
All these problems of the Fermi breakup model routines are
addressed and solved 
%by Dick Prael 
in the 03.02 version
of our codes \cite{CEM03.02}. Several bugs are also fixed in 03.02
in comparision with its predecessor. On the whole, the 03.02 versions
describe nuclear reactions on intermediate and light nuclei,
and production of fragments heavier than $^4$He from heavy targets
much better than their predecessors, almost do not produce
any unstable unphysical final products, and are free of the fixed bugs.

However, even after solving these problems and after 
implementing the improved Fermi breakup
model into CEM03.02 and LAQGSM03.02 \cite{CEM03.02}, in some very 
rare cases, our event generators still could produce some unstable 
products via very asymmetric fission, when the
excitation energy of such fragments was below 3 MeV and they were
not checked and not disintegrated with the Fermi breakup model
(see details in \cite{LAQGSM03.03}). This problem was addressed
in the 03.03 versions of our codes, where we  
force such unstable products to disintegrate 
via Fermi breakup independently of their excitation energy. 
Several more bugs were fixed on the 03.03 version as well.
A schematic outline of a nuclear reaction calculation by
CEM03.03 or LAQGSM03.03 is shown in Fig.\ 1.  We emphasize that
the occurrence of these problems even in the 03.01 versions is
quite rare, allowing stand-alone calculations of many nuclear reactions
to proceed without problems, but are unacceptable when the event 
generators are used inside transport codes doing large-scale 
simulations.

%\clearpage            % Use to start references on new page.

\begin{figure}[ht]                                                 %Fig.\ 1
\centering
\includegraphics[height=140mm,angle=-0]{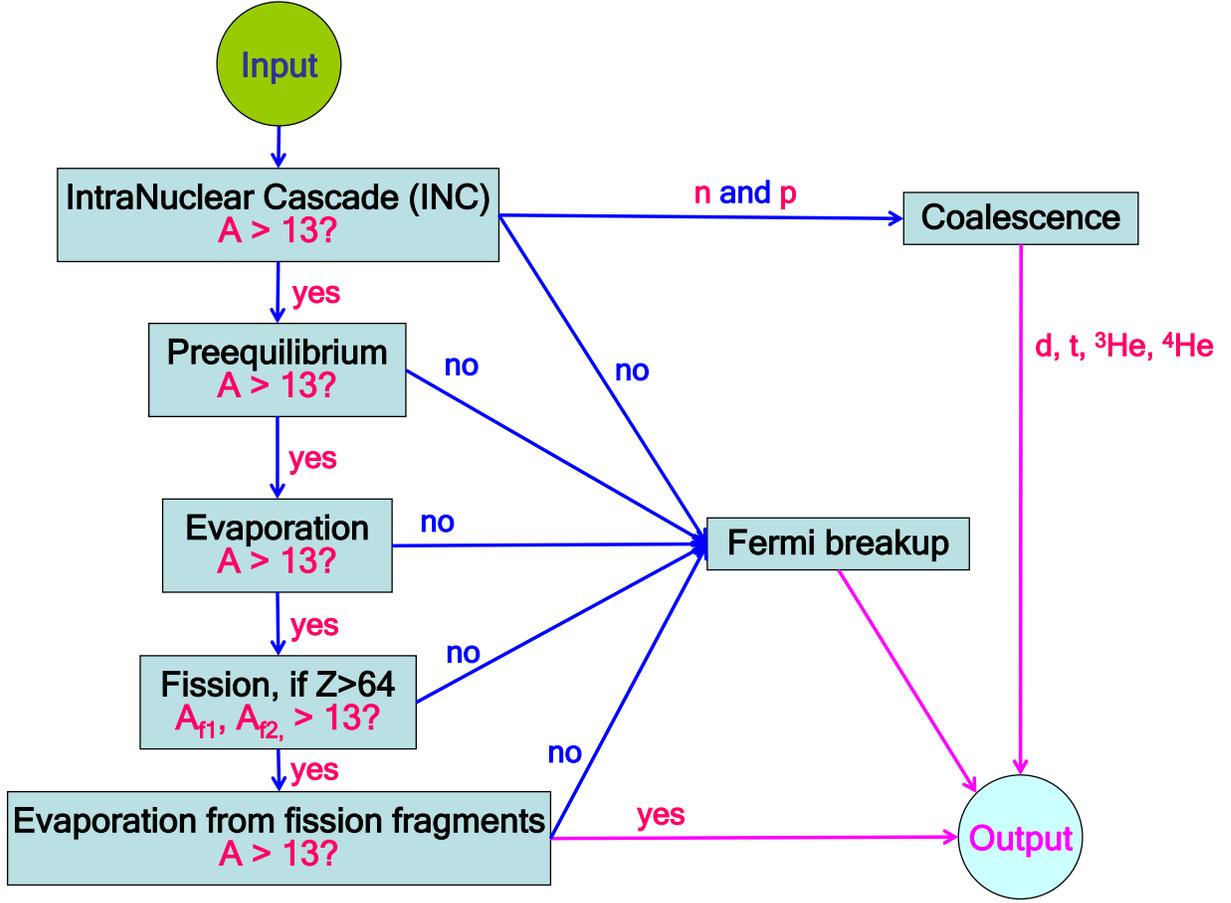}
\caption{Flow chart of nuclear-reaction calculations by
CEM03.03 and LAQGSM03.03.}
\end{figure}

In the following Sections, we highlight the main assumptions 
of the models contained in CEM and LAQGSM.\\

{\large\bf 3. The Intranuclear Cascade Mechanism} \\

The inelastic interaction of a high-energy particle with a nucleus,
and even more the collisions of two nuclei, is a very complex and 
multi-faceted phenomenon whose analytical description encounters 
considerable difficulties \cite{Book,UPN73}. In recent years
calculations of such interactions have been carried out
by statistical simulations using the Monte-Carlo method.

The INC approach was apparently first developed by Goldberger 
\cite{Goldberger48}, who in turn based his work on the
ideas of Heisenberg and Serber \cite{Serber47}, who regarded intranuclear 
cascades as a series of successive quasi-free collisions of the
fast primary particle with the individual nucleons of the nucleus.

Let us recall here the main basic assumptions of the INC, 
following \cite{Book,UPN73}.
The main condition for the applicability of the intranuclear-cascade
model is that the DeBroglie wavelength $\lambdabar$
of the particles participating in the interaction be sufficiently small:
It is necessary that for most of these particles  $\lambdabar$
be less than the average distance between the intranuclear
nucleons $\Delta \sim 10^{-13}$ cm.
Only in this case does the particle acquire quasi-classical features and 
can we speak approximately of particle trajectory and two-particle 
collisions inside the nucleus. It is clear that for this to be the case the
primary particle kinetic energy $T$ must be greater than several tens of MeV.

Another important condition for applicability of the INC is the requirement
that the time in which an individual two-particle intranuclear
collision occurs on the average, $\tau \sim 10^{-23}$ sec, be less than 
the time interval between two such consecutive interactions
$$ \Delta t = l/c \gtrsim 4 \pi R^3 /3 A \sigma c \gtrsim 3 \cdot 10^{-22} /
\sigma \mbox{ (mb) sec,}$$
where $l$ is the mean range of the cascade particle before the
interaction, $c$ is the velocity of light, $R = r_0 A^{1/3}$ is the
mean radius of the nucleus, and $\sigma$ is
the cross section for interaction with an intranuclear nucleon. 
This permits the interaction of the incident particle with the nucleus to be
reduced to a set of individual statistically independent intranuclear
collisions.

The requirement $\tau < \Delta t$ is equivalent to the
requirement that the intranuclear interaction cross section be
sufficient small: $\sigma \lesssim 100 \xi$ mb, where the coefficient 
$\xi \sim 1$.

Since the energy of the particles participating in the cascade is rather 
large---as a rule significantly greater than the binding energy of the 
intranuclear nucleons---the same characteristics can be used for interaction
of cascade particles inside the nucleus as for the interaction
of free particles. The effect of other intranuclear nucleons
is taken into account by introduction of some average potential $V$,
and also by the action of the Pauli principle.\footnote{The 
nucleus is considered to be a degenerate Fermi gas of nucleons
enclosed in the nuclear volume. According to the Pauli principle
the nucleons, after an intranuclear collision, must have energy
above the Fermi energy; otherwise such an interaction is forbidden. The 
action of the Pauli principle leads in effect to an increase of the mean free
path of fast particles inside the nucleus.}

We can say that a high-energy particle which has entered the nucleus 
passes through a gas of free nucleons, producing a cascade (avalanche)
of secondary particles. A fraction of these secondary particles leaves 
the nucleus, and the remaining fraction is absorbed, exciting the
nucleus to some energy $E^*$.

Following  \cite{Book,UPN73},
after the choice of a nuclear model and an algorithm for determination
of the elementary particles involved in the INC with the intranuclear nucleons
(for this purpose it is necessary to store in the computer memory
also the values of the integrated cross sections for elastic and inelastic
interactions $\sigma_{el}(T)$ and $\sigma_{in}(T)$), 
calculation of the intranuclear cascade is carried out according
to the scheme shown in Fig.\ 2.
The turquoise boxes 1, 2, 4, 5, 8--10, 12, and 14
in the diagram denote operations which are definite 
logically closed parts of the INC program. The yellow boxes 
3, 6, 7, 11, and 13 denote 
logical operations which control the various branchings of the
program (transfer conditions).
 
Box 1 takes into account the change in primary-particle momentum due
to the effect of the intranuclear potential and to refraction and 
reflection of the DeBroglie wave of the particle at the nuclear boundary.

\begin{figure}[ht]                                                 %Fig.\ 2
\centering
\hspace*{-10mm}
\includegraphics[height=190mm,angle=90]{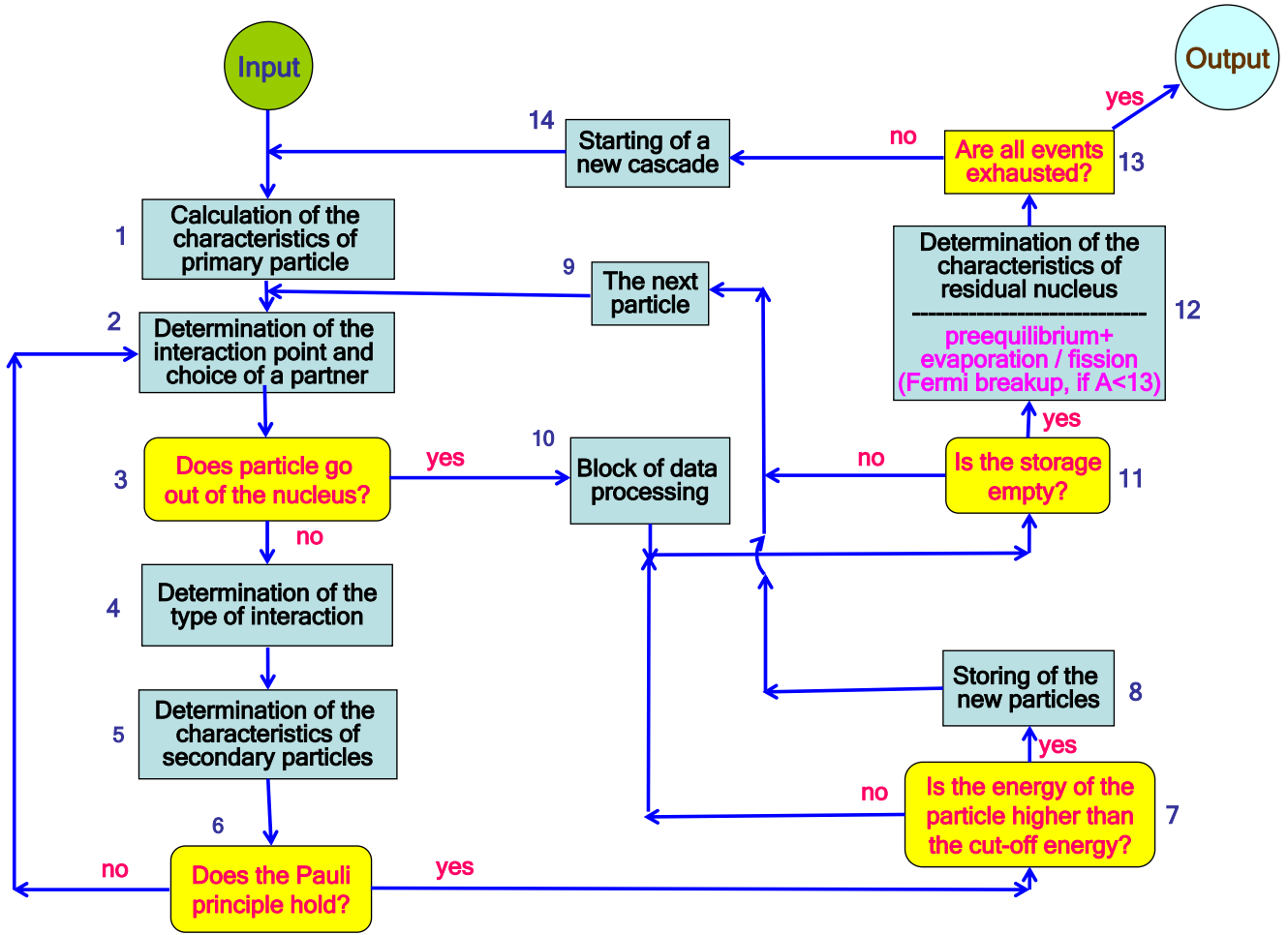}

\vspace*{-5mm}
\caption{Flow chart for the intranuclear cascade calculation.}
\end{figure}

In box 2 is chosen the momentum and isospin (proton or neutron)
of the intranuclear nucleon with which the interaction occurs
(for brevity we will call this nucleon the partner), and from the given 
elementary cross section
$\sigma_{tot} (t) = \sigma_{el} (t) + \sigma_{in} (t)$
(where $t$ is the relative energy of the primary particle and the
partner taking part in the intranuclear motion)
the mean free path of the particle in nuclear matter
$L = L(\sigma_{tot})$ is calculated and the point of
interaction is determined.

Box 3 tests whether this point of interaction is inside
the nucleus. If it is not, then the particle is assumed to have passed 
through the nucleus without interaction.  The ratio of the number of
such particles to the total number of interactions considered with 
the nucleus $N_{tot}$ obviously characterizes the reaction
cross section $\sigma_{in}(t)$.

If the point of interaction is inside the nucleus, then the type
of interaction: elastic or inelastic, is determined from the known cross
sections $\sigma_{el}(t)$ and  $\sigma_{in}(t)$ in box 4.

In box 5 the secondary-particle characteristics are determined in 
accordance with the type of interaction selected
(the nature, number, energy, and the emission angle).

Box 6 is a test of whether the Pauli principle is satisfied.
Interactions which do not satisfy this principle are considered
forbidden and the particle trajectory is followed beyond the point of the
forbidden interaction.

In box 7 the particle energy $T$ is compared with some initially 
specified cutoff energy $T_{cut}$ which determinates whether this
particle is sufficiently
energetic ($T > T_{cut}$) to take further part in development of the
intranuclear cascade or whether its energy is so small
($T \leq T_{cut}$ that the particle is simply absorbed by the nucleus.
In the first case the particle is followed further as was described above.
(For this the parameters of all cascade particles with energy $T > T_{cut}$
are stored in the memory in box 8 and later the cascade calculation is 
repeated for each of them in turn by going to boxes 9 and 2.)
In the second case the INC treatment of this particle is terminated;
if this particle is a nucleon,
in box 10 it contributes to the energy of the residual nucleus
and become an exciton to be further treated by the Modified-Exciton
Model in box 12.

The calculation is carried out until all particles are absorbed
or leave the nucleus. The operations in boxes 8, 9, and 11 are
responsible for this. If the history of one particle which entered the 
nucleus had been completed ({\it i.e.}, if the computer memory is empty;
see box 11), a possible preequilibrium, followed by evaporation/fission,
or/and Fermi breakup stage of this event is simulated in box 12
until the excitation energy of the residual nucleus is below the
binding energy of a neutron or other particles that could be
emitted from this nucleus,
then, the history of the next particle ({\it i.e.}, next ``event'')
is simulated (boxes 13 and 14), and so forth, until all events
are simulated and we get the needed statistics.

Any cascade calculation at not very high energies
where it is still possible to neglect many-particle
interactions and the change in density of the intranuclear
nucleons can be fitted into the general scheme shown in Fig.\ 2.
The specific form of the box operations and their complexity
are determined by the choice of the nuclear model and by
the number and variety of elementary processes which it is 
considered necessary to take into account in a given calculation. The 
individual boxes can be studied in more detail in
Refs.~\cite{Barashenkov68,Barashenkov68b}, as well as in the
quite old, but still one of the best monographs
on the INC and other nuclear reaction models we highly recommend
to readers interested in details of 
high-energy nuclear reactions \cite{Book}.
Some specific details on the INC of CEM and LAQGSM are
provided in the following Sections 3.1 and 3.2.\\

%\vspace{4mm}
{\large\it 3.1. The INC of CEM03.03} \\

The intranuclear cascade model
in CEM03.03 is based on the standard (non-time-dependent)
version of the Dubna cascade model 
\cite{Book,UPN73,Barashenkov68,Barashenkov68b}.
All the cascade 
calculations are carried out in a three-dimensional geometry.
The nuclear matter density $\rho(r)$
is described by a Fermi distribution with two parameters 
taken from the analysis of electron-nucleus scattering, namely
\begin{equation}
\rho(r) = \rho_p(r) + \rho_n(r) = \rho_0 \{ 1 + exp [(r-c) / a] \} \mbox{ ,}
%\label{a2}
\end{equation}
where $c = 1.07 A^{1/3}$ fm, $A$ is the mass number of the target, and
$a = 0.545$ fm.  For simplicity,
the target nucleus is divided by concentric spheres into 
seven zones in which the nuclear density is considered
to be constant.
The energy spectrum of the target nucleons is estimated in 
the perfect Fermi-gas approximation with the local Fermi energy
$T_F(r) = \hbar^2 [3\pi^2 \rho(r)]^{2/3}/(2m_N)$, where $m_N$ is the nucleon
mass.  An example of the nucleon density and the Fermi energy used by
CEM03.01 to calculate nuclear reactions on $^{208}$Pb is shown in Fig.\ 3.

The influence of intranuclear nucleons on the incoming projectile is
taken into account by adding to its laboratory kinetic
energy an effective real potential $V$, as well as by considering
the Pauli principle which forbids a number of intranuclear collisions
and effectively increases the mean free path of cascade particles inside
the target.
For incident nucleons $V \equiv V_N (r) = T_F(r) + \epsilon$,
where $T_F(r)$ is the corresponding Fermi
energy and $\epsilon$ is the binding energy of the nucleons. 
For pions, CEM03.01 uses a square-well nuclear potential 
with the depth $V_{\pi} \simeq 25$ MeV, independently of the nucleus and
pion energy, as was done in the initial Dubna INC \cite{Book,UPN73}.

\begin{figure}[ht]                                                 %Fig.\ 3
\centering
\hspace*{-1mm}
\includegraphics[height=160mm,angle=90]{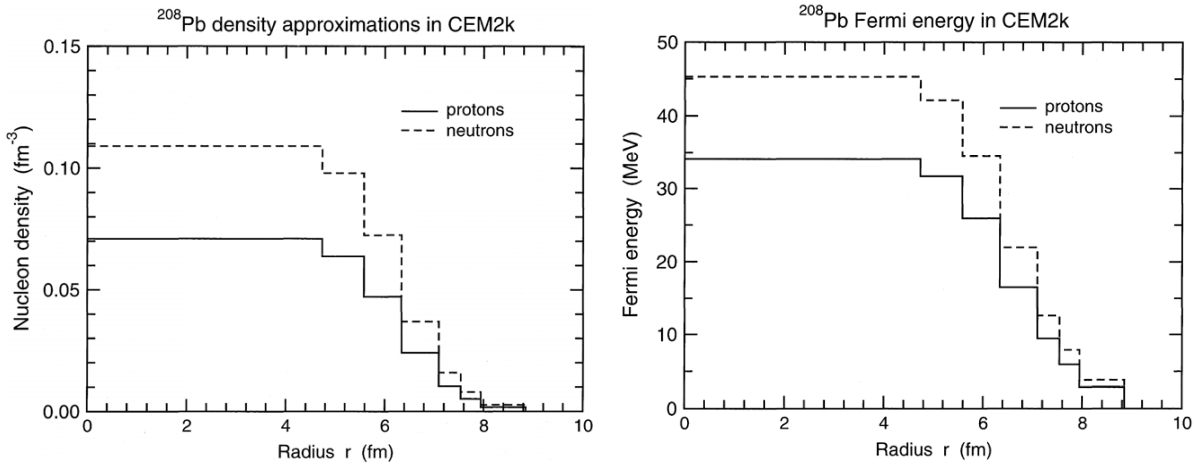}

\vspace*{-1mm}
\caption{Examples of the nucleon density and the Fermi energy used by
CEM03.01 to calculate nuclear reactions on $^{208}$Pb.}
\end{figure}

The interaction of the incident particle with the nucleus is approximated as
a series of successive quasi-free collisions of the fast cascade
particles ($N$,  $\pi$, or $\gamma$) with intranuclear nucleons:
\begin{eqnarray}
NN \to NN , \qquad NN \to \pi NN , \qquad  NN \to \pi _1,\cdots,\pi _i NN 
\mbox{ ,} \\
\pi N \to \pi N, \qquad  \pi N \to \pi _1,\cdots,\pi _i N  \qquad  (i \geq 2) 
\ .
\label{a3}
\end{eqnarray}

In the case of pions, besides the elementary processes (3), CEM03.01 
also takes into account pion absorption on nucleon pairs 
\begin{equation}
\pi NN \to NN .
\label{a4}
\end{equation}
The momenta of the two nucleons participating in the absorption are chosen
randomly from the Fermi distribution, and the pion energy is distributed 
equally between these nucleons in the center-of-mass system
of the three particles participating in the absorption. The direction 
of motion of the resultant nucleons in this system is taken as
isotropically distributed in space. The effective cross section for
absorption is related (but not equal) to the experimental cross sections 
for pion absorption by deuterons.

In the case of photonuclear reactions \cite{JNRS05}, CEM03.01 follows
the ideas of the photonuclear version of the Dubna INC
proposed initially 35 years ago by
one of us (KKG) in collaboration with Iljinov and Toneev \cite{Dubna69}
to describe photonuclear reactions at energies
above the Giant Dipole Resonance (GDR) region \cite{Dubna74}.
[At photon energies $T_{\gamma} = 10$--$40$ MeV, the DeBroglie 
wavelength $\lambdabar$ is of the order of $20$--$5$ fm,  
greater than the average inter-nucleonic distance in the nucleus; 
the photons interact with the nuclear
dipole resonance as a whole, thus the INC is not applicable.]
Below the pion-production threshold, the Dubna INC considers
absorption of photons on only ``quasi-deuteron" pairs according
to the Levinger model \cite{Levinger}:
\begin{equation}
\sigma_{\gamma A} = L \frac{Z(A-Z)}{A} \sigma_{\gamma d} \mbox{ ,} 
\end{equation}
where $A$ and $Z$ are the mass and charge numbers of the nucleus,
$L \approx 10$, and $ \sigma_{\gamma d}$ is 
the total photo-absorption cross section on deuterons as
defined from experimental data.

At photon energies above the pion-production threshold, the Dubna INC
considers production of one or two pions; the specific
mode of the reaction is chosen by the Monte-Carlo method according to the
partial cross sections (defined from available experimental data):
\begin{eqnarray}
\gamma + p & \to & p + \pi^0  \mbox{ ,} \\ 
           & \to & n + \pi^+ \mbox{ ,}  \\
           & \to & p + \pi^+ + \pi^-  \mbox{ ,}  \\
           & \to & p + \pi^0 + \pi^0  \mbox{ ,}  \\
           & \to & n + \pi^+ + \pi^0  \mbox{ .} 
\end{eqnarray}
The cross sections of $\gamma + n$ interactions are derived from 
consideration of isotopic invariance, {\it i.e.} it is assumed that
$\sigma (\gamma + n) = \sigma (\gamma + p)$. The Compton effect on 
intranuclear nucleons is neglected, as its cross section is less
than $\approx 2$\% of other reaction modes (see, {\it e.g.}
Fig.\ 6.13 in Ref.\ \cite{LockMeasday}). 
The Dubna INC does not consider processes
involving production of three and more pions; this limits the model's
applicability to photon energies $T_{\gamma} \lesssim 1.5$ GeV
[for $T_\gamma$ higher than the threshold for three-pion production,
the sum of the cross sections (8)--(10) is assumed to be equal to 
the difference
between the total inelastic $\gamma + p$ cross section 
and the sum of the cross sections of the two-body reactions (6)--(7)].

The integral cross sections for the free $NN$, $\pi N$, and
$\gamma N$ interactions (2)--(10) are approximated 
in the Dubna INC model~\cite{Book} used in CEM95 \cite{CEM95}
and its predecessors
using a special algorithm of interpolation/extrapolation
through a number of picked points, mapping as well as possible the
experimental data.
This was done very accurately by the group of Prof.~Barashenkov using all
experimental data available at that time, about 35 years ago.
Currently the experimental data on cross sections is much 
more complete than at that time; therefore we have 
\vspace*{-5mm}

\newpage
\begin{figure}[h!]                                                 %Fig.\ 4

\centering
\includegraphics[height=160mm,angle=-90]{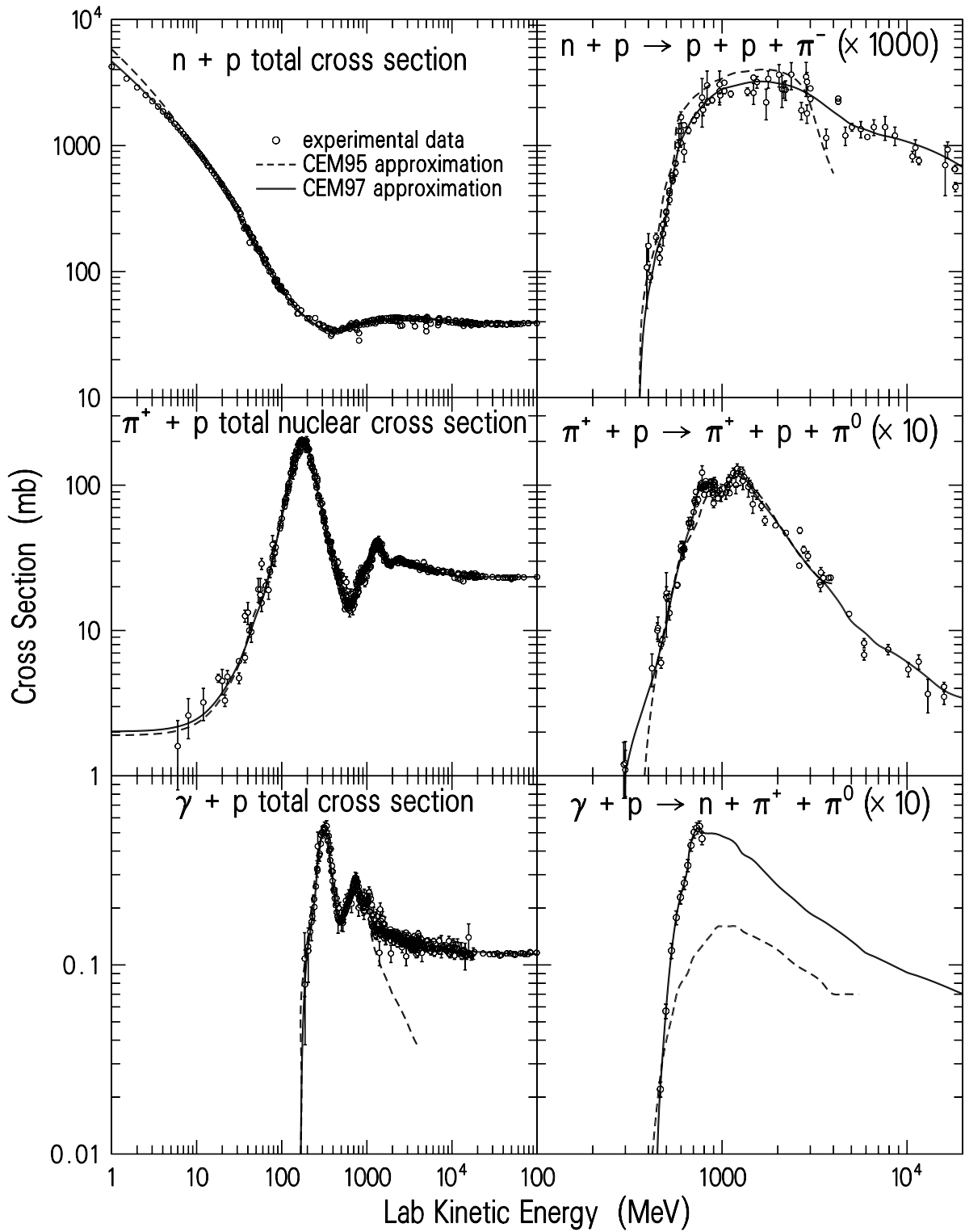}

%\vspace*{+0.8cm}
\caption{
Energy dependence of the  $n p$, $\pi^+ p$, and $\gamma p$ total cross
 sections and of the $n p \to p p \pi^-$, $\pi^+ p \to \pi^+ p \pi^0$, and
$\gamma p \to n \pi^+ \pi^0$ ones.
Experimental points are from our compilation \cite{CEM97}.
Solid lines are results using our new approximations; dashed lines
show the standard Dubna INC approximations \cite{Book} used 
in CEM95 \cite{CEM95}. 
}
\end{figure}

%\newpage

\newpage

\vspace{-5mm}
%%%%%%%%%%%%%%%%%%%%%%%%%%%%%%%%%%%%%%%%%%%%%%%%%%%%%%%%%%%%%% Fig. 5
\begin{figure}[!ht]
\begin{center}
\includegraphics[width=14.0cm]{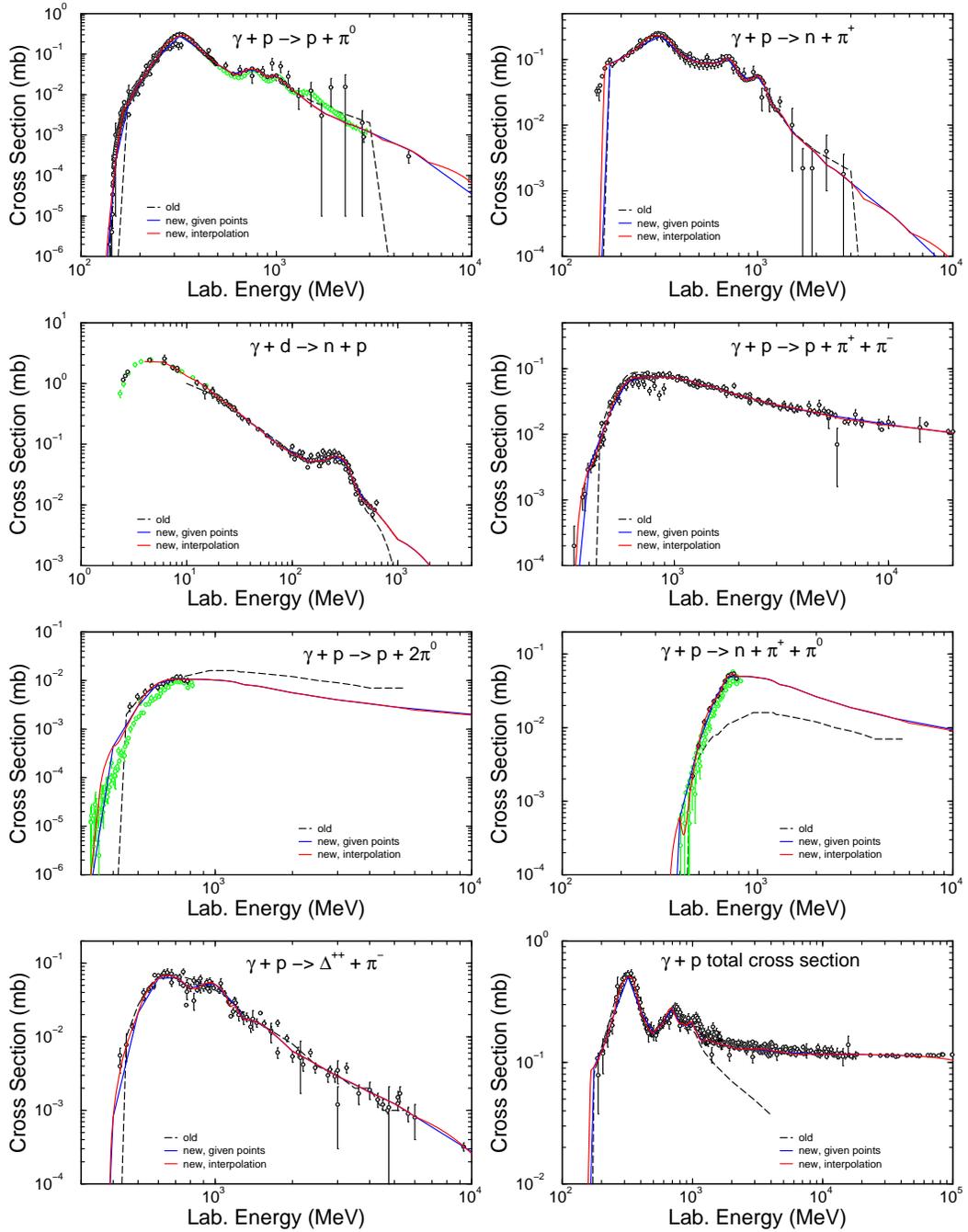}
\vspace{-2mm}
\caption{
Comparison of eight experimental total $\gamma + p(d)$ cross sections
with the old approximations used in the Dubna INC \cite{Book}
and with
our approximations incorporated into the 
CEM03.01 and LAQGSM03.01 codes. The red curve gives the code results
using parabolic interpolation, while the blue solid curve uses linear 
interpolation between our tabulated points.  Where no blue curve is
visible, it is coincident with the red curve.
References to experimental data  shown by black and green circles
may be found in our recent paper \cite{JNRS05}.
The green circles show recent
experimental data that became available to us after
we completed our fit; Although these recent
data agree reasonably well with our
approximations, a refitting would slightly improve the agreement.
}
\end{center}
\end{figure}

\newpage
{\noindent
revised the 
approximations of all the integral elementary
cross sections used in CEM95 \cite{CEM95} and its predecessors. 
We started by collecting all published 
experimental data from all available sources.
}
Then we developed an improved, as compared with the standard 
Dubna INC \cite{Book},
algorithm for approximation of cross sections and developed simple
and fast approximations for elementary cross sections which fit very well
presently available experimental data not only to 5 GeV, the upper
recommended energy for the present version of the CEM, but up  to 50--100
GeV and higher, depending on availability of data
(see details in \cite{CEM97,JNRS05}).
So far, we have in CEM03.01 new
approximations for 34 different types of elementary cross sections
induced by nucleons, pions, and gammas. Integral cross sections for other
types of interactions taken into account in CEM03.01 are calculated from
isospin considerations using the former as input.

Examples of several
compiled experimental cross sections together with our new approximations
and the old approximations from CEM95 \cite{CEM95}
are shown in Figs.~4 and 5. We see that
our new approximations describe indeed very well all data.
Although presently we have much more data than 35 years ago when 
Barashenkov's group produced their approximations used in CEM95, for 
a number of interaction modes like the total cross sections shown 
in the left panel of Fig.\ 4,
 the original approximations also agree very well with
presently available data, in the energy region where the Dubna INC was 
developed to work.  This is a partial explanation of
why the old
Dubna INC \cite{Book} and the younger CEM95 \cite{CEM95} work so well
for the majority of characteristics of nuclear reactions. 
On the other hand, for some modes of elementary interactions 
like the ones shown in the right panel of Fig.\ 4, the old approximations 
differ significantly from the present data, demonstrating the need for our 
recent improvements for a better description of all modes of nuclear 
reactions.

The kinematics of two-body elementary interactions
and absorption of photons and pions
by a pair of nucleons is completely defined by a given direction of
emission of one of the secondary particles.
The cosine of the angle of emission of secondary particles 
in the c.m.\ system is calculated by the Dubna INC \cite{Book}
as a function of a random number $\xi$, distributed uniformly in the 
interval [0,1] as
\begin{equation}
\cos \Theta = 2 \xi ^{1 / 2} \left[ \sum_{n=0}^{N} a_n \xi^n
+ (1- \sum_{n=0}^N a_n ) \xi^{N+1} \right] -1 \mbox{ ,}
\end{equation}
where $N = M = 3$,
\begin{equation}
a_n = \sum_{k=0}^M a_{nk} T_{i}^k \mbox{ .}
\end{equation}
The coefficients $a_{nk}$ were fitted to the then available 
experimental data at a number of
incident kinetic energies $T_i$ , then interpolated and extrapolated
to other energies (see details in \cite{Book,Dubna69,Dubna74} and 
references therein).
The distribution of secondary particles over the azimuthal angle
$\varphi$ is assumed isotropic. For elementary interactions 
with more than two particles in the final state, 
the Dubna INC uses the statistical model to simulate the angles 
and energies of products (see details in \cite{Book}).

For the improved version of the INC in CEM03.01,
we use currently available experimental data and recently published
systematics proposed by other authors and have developed new
approximations for angular and energy distributions of
particles produced in nucleon-nucleon and photon-proton interactions.
So, for $pp$, $np$, and $nn$ interactions at energies up to
2 GeV, we did not have to develop our own approximations
analogous to the ones described by Eqs.\ (11) and (12),
since reliable systematics have been developed recently
by Cugnon {\it et al.} for the Liege INC \cite{INCL}, then 
improved still further by Duarte for the BRIC code \cite{BRIC1.4};
we simply incorporate into CEM03.01 the systematics by Duarte \cite{BRIC1.4}. 
%Similarly, for
%$\gamma N$ interactions, 
%we take advantage of the event generators for $\gamma p$ and
%$\gamma n$ reactions from the Moscow INC \cite{Iljinov97}
%kindly sent us by Dr.\ Igor Pshenichnov. 
%In CEM03.01, we use part of a data file with smooth approximations 
%through presently available experimental data, developed
%for the Moscow INC \cite{Iljinov97} and have ourselves developed 
%a simple and fast algorithm to simulate unambiguously
%$d \sigma / d \Omega$ and to
%choose the corresponding value of $\Theta$ for any $E_\gamma$,
%using a single random number $\xi$ uniformly
%distributed in the interval [0,1] (see details in \cite{JNRS05}). 

Examples of angular distributions of secondary particles from
$np$  reactions at several energies are shown
in Fig.\ 6. The new approximations reproduce the experimental data
much better than the old Dubna INC used in our previous code 
versions (and in several other codes developed from the 
Dubna INC).

%\newpage
\begin{figure}[h!]                                                 %Fig.\ 6

\centering
\includegraphics[height=160mm,angle=-0]{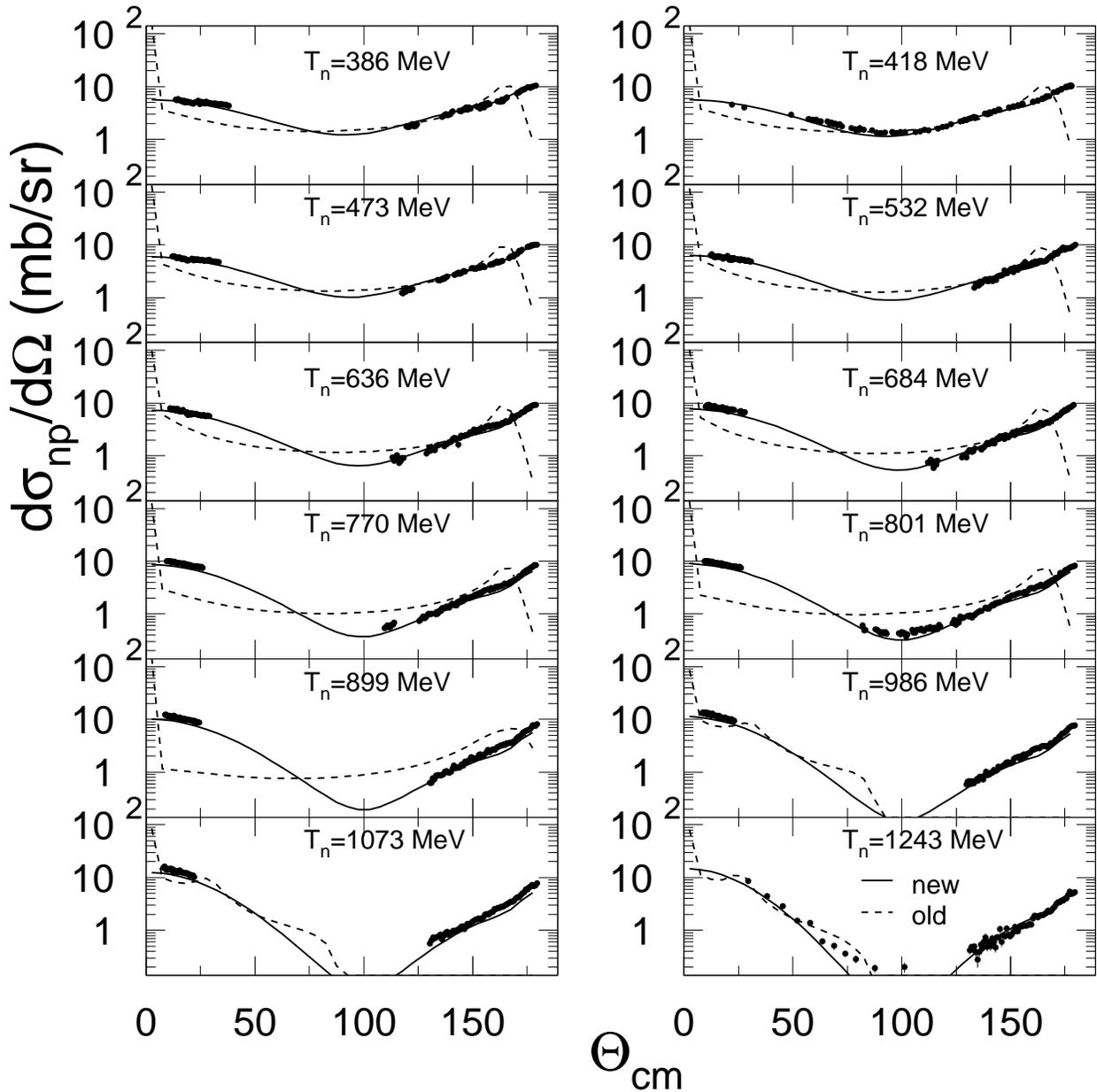}

\vspace*{5mm}
\caption{Example of twelve
angular distributions of $n$ from $np$ elastic interactions
as functions
of $\Theta^{n}_{c.m.}$ at $T_n$ from 386 to 1243 MeV.
The dashed lines show the old approximations from the Dubna INC 
while the solid lines are the new approximations incorporated into 
CEM03 and LAQGSM03.
References to 
experimental data shown
here by black circles 
may be found in our paper \cite{ND2004}.
}

\end{figure}

In the case of $\gamma p$ reactions (6) and (7), 
we chose another way:
Instead of fitting
the parameters $a_n$ from Eq.\ (11) at different 
$E_\gamma$ we found data
(see, {\it e.g.}, Fig.\ 7) and finding the energy dependence
of parameters $a_{nk}$ in Eq.\ (12) using the values obtained for $a_n$,
we took advantage of the event generator for $\gamma p$ and
$\gamma n$ reactions from the Moscow INC \cite{Iljinov97}
kindly sent us by Dr.\ Igor Pshenichnov. 
That event generator
includes a data file with smooth approximations through presently
available experimental data 
at 50 different gamma energies from 117.65 to 6054 MeV (in the
system where the $p$ or $n$ interacting with $\gamma$ is at rest)
for the c.m.\ angular distributions 
$d \sigma / d \Omega$ of secondary particles as functions
of $\Theta$ tabulated for values of  $\Theta$ from 0 to 180 deg.,
with the step $\Delta \Theta = 10$ deg.,
for 60 different channels of $\gamma p$ and $\gamma n$ reactions 
considered by the Moscow INC (see details in \cite{Iljinov97}).
We use part of that data file with data for reactions (6) and (7),
and have written an algorithm to simulate unambiguously
$d \sigma / d \Omega$ and to
choose the corresponding value of $\Theta$ for any $E_\gamma$,
using a single random number $\xi$ uniformly
distributed in the interval [0,1]. This is straightforward due to the 
fact that the function $\xi (\cos \Theta)$
$$
\xi(\cos \Theta) =
\int\limits_{-1}^{\cos \Theta}  d \sigma / d \Omega \: d \cos \Theta \bigg/\!\!
\int\limits_{-1}^{1}  d \sigma / d \Omega \: d \cos \Theta $$
is a smooth monotonic function increasing from 0 to 1  as
$\cos \Theta$ varies from -1 to 1. Naturally, when $E_\gamma$
differs from the values tabulated in the data file, we perform
first the needed interpolation in energy.
We use this procedure
to describe in CEM03.01 angular distributions of secondary particles from
reactions (6) and (7), as well as for isotopically symmetric
reactions $\gamma + n \to n + \pi^0$ and $\gamma + n \to p + \pi^-$.

Examples of eight angular distributions 
%of $\pi^{0}$ from $\gamma p \to \pi^{0} p$  and 
of $\pi^{+}$ from $\gamma p \to \pi^{+} n$ 
as functions of $\Theta^{\pi}_{c.m.s}$ are
presented in Fig.\ 7. We see that the approximations
developed in CEM03.01 (solid histograms)
agree much better with the available experimental data
than the old Dubna INC approximations (11)--(12)
used in all predecessors of CEM03 (dashed histograms).

%\newpage
%%%%%%%%%%%%%%%%%%%%%%%%%%%%%%%%%%%%%%%%%%%%%%%%%%%%%%%%%%%%%%%%%% Fig. 7
\begin{figure}[h!]

%\centering
\begin{center}
\includegraphics[width=170mm]{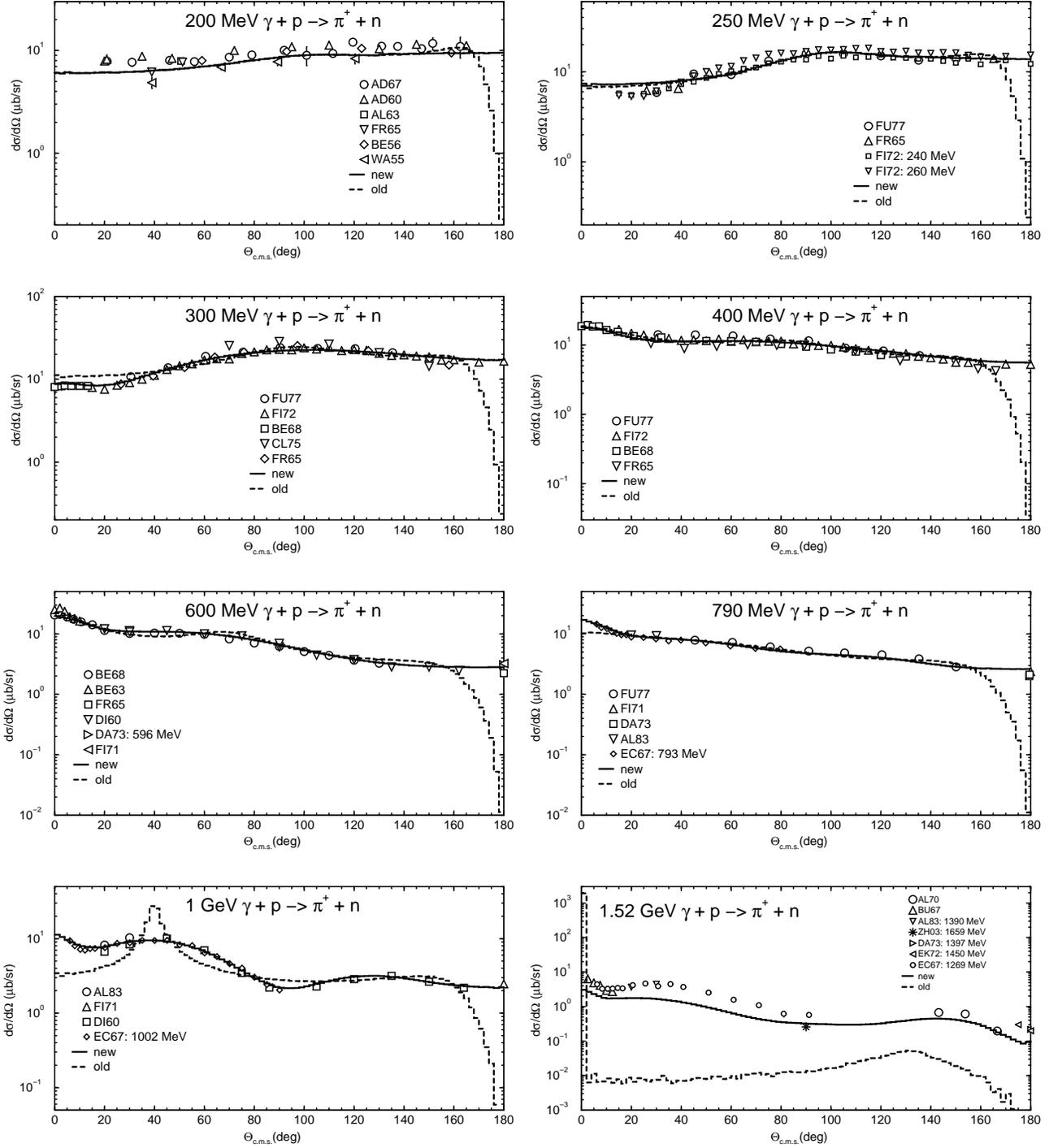}
\caption{
Example of eight
angular distributions of $\pi^{+}$ from
$\gamma p \to \pi^{+} n$ 
as functions
of $\Theta^{\pi}_{c.m.s}$ at photon energies from 200 MeV to 1.52 GeV.
The dashed lines show the old approximations used in the Dubna INC PRM 
while the solid lines are our new approximations incorporated into the 
CEM03 and LAQGSM03 codes. References to 
experimental data shown by symbols
may be found in our recent paper \cite{JNRS05}.
}
\end{center}
\end{figure}

The analysis of experimental data has shown that the channel (8)
of two-pion photo-production proceeds mainly through the decay of the 
$\Delta^{++}$ isobar listed in the last Review of Particle Physics by the
Particle Data Group as having the mass $M = 1232$ MeV
\begin{eqnarray}
\gamma + p  & \to & \Delta^{++} + \pi^-  \mbox{ ,} \nonumber \\
\Delta^{++} & \to & p + \pi^+  \mbox{ ,} 
\end{eqnarray}
whereas the production cross section of other isobar components
$\left({3\over 2}, {3\over 2} \right)$ are small and can be neglected.
The Dubna INC uses the Lindenbaum-Sternheimer resonance model \cite{Delta}
to simulate the reaction (13).
In this model, the mass of the isobar $M$ is determined from
the distribution
\begin{equation}
{{\mathrm{d} W}\over {\mathrm{d} M}} \sim F(E,M) \sigma(M) \mbox{ ,}
\end{equation}
where $E$ is the total energy of the system, $F$ is the two-body phase
space of the isobar and $\pi^-$ meson, and $\sigma$ is the 
isobar production cross section which is assumed to be equal to the
cross section for elastic $\pi^+ p$ scattering.

The c.m.\ emission angle of the isobar is approximated
using Eqs.\ (11) and (12) with the coefficients $a_{nk}$ listed in Tab.\ 3 of 
Ref.\ \cite{Dubna74}; isotropy of the decay of the isobar in its c.m.\ system 
is assumed.

In order to calculate the kinematics of the non-resonant
part of the reaction (8) and the
two remaining three-body channels (9) and (10),
the Dubna INC uses the statistical model. The total energies of the two 
particles 
(pions) in the c.m.\ system are determined from the distribution
\begin{equation}
{ {\mathrm{d} W}\over {\mathrm{d} E_{\pi_1} \mathrm{d} E_{\pi_2}} }
\sim (E-E_{\pi_1}-E_{\pi_2}) E_{\pi_1}E_{\pi_2} / E \mbox{ ,}
\end{equation}
and that of the third particle (nucleon, $N$) from 
conservation of energy.
%The law of conservation of momentum is taken into account in the same
%manner as in the model of inelastic $\pi + N$ and $N + N$ collisions
%(see Sec. 3 in Ref.\ \cite{INC68b} and \S 49 of the monograph \cite{Book}).
The actual simulation of such reactions is done as follows:
Using a random number $\xi$,
we simulate in the beginning the energy of the first pion using
$$E_{\pi_1} = m_{\pi_1} + \xi (E^{max}_{\pi_1} - m_{\pi_1}) ,$$
where 
$$E^{max}_{\pi_1} = [ E^2 + m^2_{\pi_1} - (m_{\pi_2} + m_N )^2 ] / 2 E . $$
Then, we simulate the energy of the second pion $E_{\pi_2}$ 
according to Eq.\ (15) using the Monte-Carlo rejection method.
The energy of the nucleon is calculated as 
$E_N = E - E_{\pi_1} - E_{\pi_2}$, following which we check that the 
``triangle law" for momenta
$$| p_{\pi_1} - p_{\pi_2} | \leq p_N \leq |p_{\pi_1} + p_{\pi_2}|$$
is fulfilled, otherwise this sampling is rejected and the procedure
is repeated. The angles $\Theta$ and $\varphi$ of the pions
are sampled assuming an isotropic distribution of particles in the
c.m.\ system, 
$$ \cos \Theta_{\pi_1} = 2 \xi_1 -1, \qquad
 \cos \Theta_{\pi_2} = 2 \xi_2 -1, \qquad 
\varphi_{\pi_1} = 2 \pi \xi_3,\qquad 
\varphi_{\pi_2} = 2 \pi \xi_4,$$
and the angles of the nucleon are defined from momentum
conservation, $\vec{p}_N = - (\vec{p}_{\pi_1} + \vec{p}_{\pi_2} )$.
More details on our new approximations for differential 
elementary cross sections may be found in \cite{ND2004,JNRS05}.

The Pauli exclusion principle at the cascade stage of the reaction is
handled by assuming that nucleons of the target
occupy all the energy levels up to the Fermi energy. Each simulated
elastic or inelastic interaction of the projectile (or of a cascade
particle) with a nucleon of the target is considered forbidden if the 
``secondary" nucleons have energies smaller than the Fermi energy. If they do,
the trajectory of the particle is traced further from the forbidden point and
a new interaction point, a new partner and a new interaction mode are
simulated for the traced particle, {\it etc.}, until the Pauli principle
is satisfied or the particle leaves the nucleus.

In this version of the INC, 
the kinetic energy of the cascade particles is increased or decreased 
as they move from one of the seven potential regions (zones) to another, 
but their directions remain unchanged. That is, in our calculations, 
refraction 
or reflection of cascade nucleons at potential boundaries is neglected.
CEM03.01 allows us to take into account refractions and 
reflections of cascade nucleons at potential boundaries; for this,
one needs to change the value of the parameter {\bf irefrac} from
0 to 1 in the subroutine {\bf initial}. But this option provides
somewhat worse overall agreement of calculations with some experimental 
data, therefore the option of no refractions/reflections was chosen 
as the default in CEM03.01.

The INC in CEM does not take into account the
so-called ``trawling" effect~\cite{Book}. That is, 
in the beginning of the simulation of each event, the nuclear density
distributions for the protons and neutrons of the target are calculated 
according
to Eq.\ (1) and a subsequent decrease of the nuclear density with the 
emission of cascade particles is not taken into account.
Our detailed analysis 
of different characteristics of nucleon- and pion-induced reactions 
for targets from C to Am has shown that this
effect may be neglected at incident energies below about 5 GeV
in the case of heavy targets like actinides and below about
1 GeV for light targets like carbon.
At higher incident energies the progressive decrease of nuclear 
density with the development of the intranuclear 
cascade has a strong influence on the calculated characteristics and this 
effect has to be taken into account~\cite{Book}. 
Therefore, in transport codes that use as event generators both CEM03.01
\cite{CEM03.01}
and our high-energy code LAQGSM03.01 \cite{LAQGSM03.01}, we recommend
simulating nuclear reactions with CEM03.01 at incident
energies up to about 1 GeV for light nuclei like C and up to about 5 GeV
for actinide nuclei, and to switch to simulations
using LAQGSM03.01, which considers the ``trawling" effect,
at higher energies of transported particles.

An important ingredient of the CEM is the criterion for transition 
from the intranuclear cascade to the preequilibrium model. In 
conventional cascade-evaporation models
(like the Bertini INC used in MCNPX \cite{MCNPX}),
fast particles are traced down to some minimal energy, the cutoff energy
$T_{cut}$ (or one compares the duration of the cascade stage of a reaction 
with a cutoff time, in ``time-like" INC models, such as the 
Liege INC \cite{INCL}).
This cutoff is usually less than $\simeq 10$ MeV above the Fermi energy,
below which particles are considered to be absorbed by the
nucleus. The CEM uses a different criterion to decide when a primary
particle is considered to have left the cascade.

An effective local optical absorptive potential $W_{opt.\: mod.}(r)$ is 
defined from the local interaction cross section of the particle,
including Pauli-blocking effects. This imaginary potential is compared
to one defined by a phenomenological global optical model
$W_{opt.\: exp.}(r)$. We characterize the degree of similarity or difference
of these imaginary potentials by the parameter 
\begin{equation}
{\cal P} =\mid (W_{opt.\: mod.}-W_{opt.\: exp.}) / W_{opt.\: exp.} \mid . 
\end{equation}

When $\cal P$ increases above an empirically chosen value, the particle
leaves the cascade, and is then considered to be an exciton.
From a physical point of view, such a smooth transition from the cascade
stage of the reaction seems to be more attractive than the
``sharp cut-off" method. In addition, as was shown  in Ref.~\cite{CEM},
this improves the agreement between the calculated and experimental spectra
of secondary nucleons, especially at low incident energies and backward
angles of the detected nucleons (see {\it e.g.}, Figs.\ 3 and 11 
of Ref.~\cite{CEM}).
More details about this can be found in \cite{CEM,CEM2k,Mashnik97}.

CEM03.01 uses a fixed value $\cal P$ = 0.3 
(at incident energies below 100 MeV), just as all its predecessors did.
With this value, we find that the cascade stage of the CEM is generally 
shorter than that in other cascade models. This fact leads to an 
overestimation of preequilibrium particle emission at incident energies 
above about 150 MeV, and correspondingly to an underestimation of 
neutron production from such reactions, as was established in 
Ref.\ \cite{CEM2k}. In Ref.\ \cite{CEM2k}, this problem was
solved temporarily in a very rough way by using the transition
from the INC to the preequilibrium stage
according to Eq.\ (16) when the incident energy of the projectile is
below 150 MeV, and by using the ``sharp cut-off" method with
a cutoff energy $T_{cut} = 1$ MeV for higher incident energies.
This ``ad hoc" rough criterion solved the problem of underestimating
neutron production at high energies, providing meanwhile a reasonably
good description of reactions below 150 MeV. But it provides
an unphysical discontinuity in some observables calculated by
MCNPX using CEM2k \cite{CEM2k} as an event generator, observed
but not understood by Broeders and Konobeev \cite{Broeders05}.
In CEM03.01, this problem is solved by using a smooth
transition from the first criterion to the second one in the energy
interval from 75 to 225 MeV, so that no discontinuities are 
produced in results from CEM03.01.

Beside the changes to the Dubna INC mentioned above, we also made 
in the INC a number of other improvements and refinements, 
such as imposing momentum-energy conservation for each simulated event
(the Monte-Carlo algorithm previously used in the CEM 
provided momentum-energy conservation only 
statistically, on the average, but not exactly for each simulated
event) and using real binding energies for nucleons in the cascade 
instead of the approximation of a constant
separation energy of 7 MeV used in previous versions of the CEM.
We have also
improved many algorithms used in the Monte-Carlo simulations
in many subroutines, decreasing the computing time by up to a
factor of 6 for heavy targets, which is very important when performing
practical simulations with transport codes like MCNPX or MARS.

Let us mention that in the CEM the initial configuration for the 
preequilibrium decay (number of excited particles and holes, {\it i.e.} 
excitons $n_0 = p_0 + h_0$, excitation energy $E^*_0$, 
linear momentum ${\bfm P}_0$, and angular momentum ${\bfm L}_0$
of the nucleus) differs significantly from that 
usually postulated in exciton models. Our calculations 
\cite{CEM,cemphys,smolenice} have shown 
that the distributions of residual nuclei 
remaining  after the cascade stage of the reaction, {\it i.e.} before the 
preequilibrium emission, with respect to $n_0$, $p_0$, $h_0$, $E^*_0$, 
${\bfm P}_0$, and ${\bfm L}_0$ are rather 
broad.\footnote{Unfortunately, this fact was misunderstood by the authors
of the code HETC-3STEP~\cite{HETC-3STEP}.
In spite of the fact that it has been stressed explicitly, and
figures with distributions of excited nuclei after the cascade stage
of a reaction with respect to the number of excitons and other
characteristics were  shown
in a number of publications (see, {\it e.g.}, Fig.\ 5 in Ref.~\cite{CEM},
Fig.\ 1 in Ref.~\cite{smolenice}, p.\ 109 in Ref.~\cite{cemphys}, and
p.\ 706 in Ref.~\cite{NP94}),
the authors of Ref.~\cite{HETC-3STEP} misstated this fact as
{\em ``Gudima {\it et al.}\ assumed the state of two particles and one
hole at the beginning $\cdots$ Hence, their assumption is not valid for
the wide range of incident energy"}, claiming this as a 
weakness of the CEM and a
priority of the code HETC-3STEP, where smooth distributions of excited
nuclei after the cascade stage of reactions  with respect to $n_0$ 
are used. This had already been done in the CEM~\cite{CEMP,CEM}.}\\

CEM03.01 (just like LAQGSM03.01 and
many other INC-based models) calculates the 
total reaction cross section, $\sigma_{in}$, by the Monte-Carlo method
using the geometrical cross section, $\sigma_{geom}$, and the
number of inelastic, $N_{in}$, and elastic, $N_{el}$, simulated
events, namely: 
$\sigma_{in} = \sigma_{geom} N_{in} / (N_{in} + N_{el})$.
The value of the total reaction cross section
calculated this way is printed in the beginning
of the CEM03.01 output labeled as
{\it Monte Carlo inelastic cross section}.
This approach provides a good agreement with available data 
for reactions induced by nucleons, pions, and photons at incident
energies above about 100 MeV, but is not reliable enough at 
energies below 100 MeV
(see, {\it e.g.}, Figs.\ 8 and 9 below). 

To address this problem, we have incorporated \cite{SantaFe02} into
CEM03.01 the NASA systematics by Tripathi {\it et al.}\ \cite{Tripathi}
%$^{33}$ 
for all incident protons and for neutrons with energies above the 
maximum in the NASA
reaction cross sections, and the Kalbach systematics \cite{Kalbach98}
%$^{34}$ 
for neutrons of lower energy. For reactions induced by monochromatic and
bremsstrahlung photons, we incorporate \cite{JNRS05} into CEM03.01
the recent systematics by Kossov \cite{Kossov}. Details on these
systematics together with 
examples of several total inelastic cross sections calculated
with them compared with available experimental
data may be found in \cite{SantaFe02,JNRS05}. 
Our analysis of many different reactions has shown that at incident energies
below about 100 MeV these systematics generally describe the total
inelastic cross sections better that the Monte-Carlo method 
does, and no worse than the Monte-Carlo method 
at higher energies. Therefore we choose these systematics as the default
for normalization of all CEM03.01 results. The total reaction cross
sections calculated by these systematics are printed in the CEM03.01 
output labeled as {\it Inelastic cross section used here}. 
(Of course, users may re-normalize all the CEM03.01 results 
to the Monte-Carlo total reaction cross sections by making a small
change to the code in the subroutine {\bf typeout}).

Let us note, however, that in applications, when
CEM03.01 and LAQGSM03.01, or any other codes, are
used as event generators in transport codes,
it does not matter how they calculate the total reaction 
cross sections (normalization):
All transport codes use their own routines or systematics 
to calculate the total elastic and inelastic cross sections
of the projectiles, before   
starting to simulate with an event generator
an inelastic interaction of the traced projectile with a 
nucleus of the thick target.

To summarize this Section, 
in comparison with the initial version of the Dubna
INC \cite{Book,UPN73}

\newpage

\begin{figure}[ht]                                                 %Fig.\ 8.
\centering
\vspace*{-20mm}
\includegraphics[angle=-0,width=14.0cm]{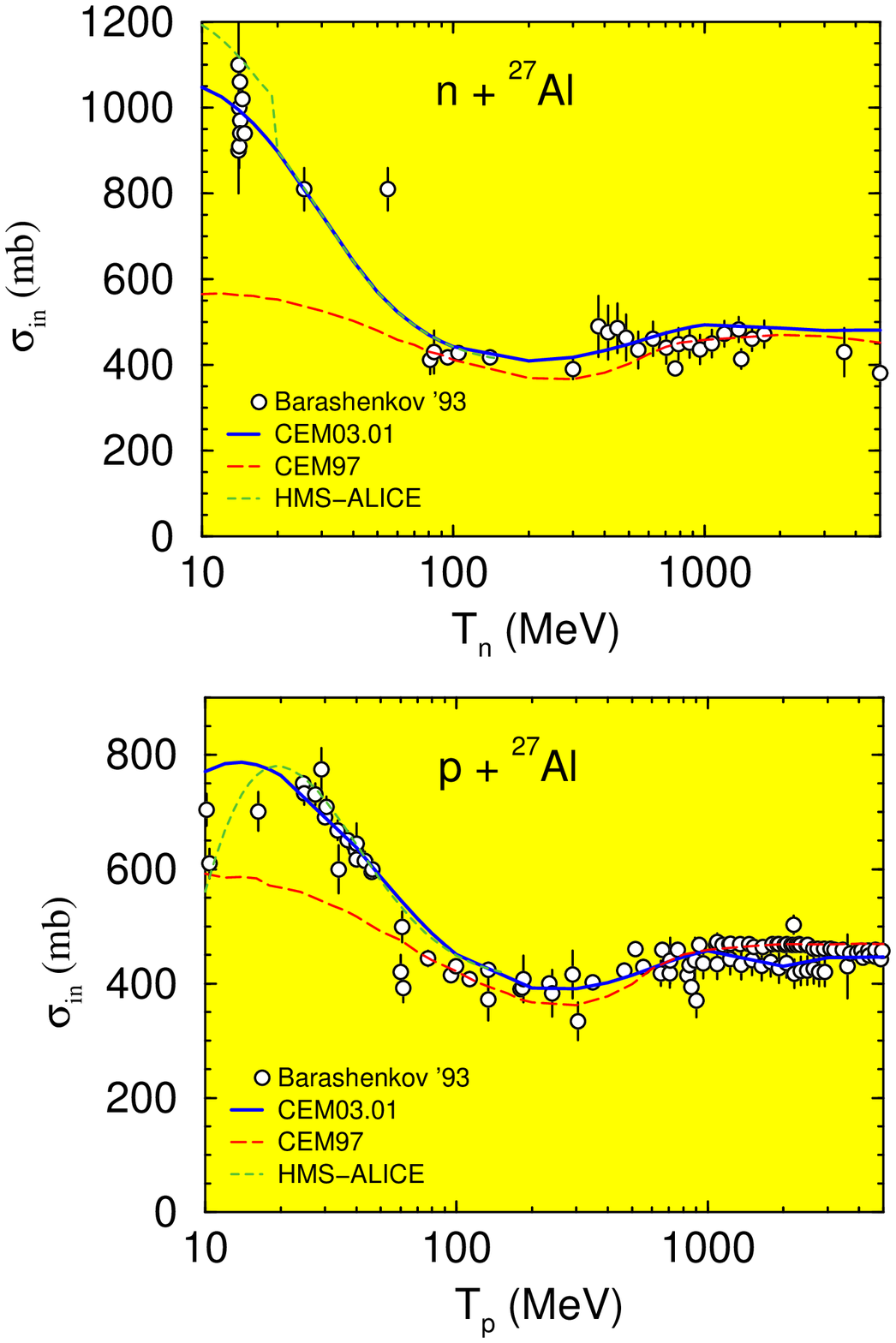}

%\vspace*{-29mm}
\caption{
Total reaction cross sections for $n$- and $p$-induced 
reactions on Al calculated by CEM03.01 and its predecessor CEM97
with experimental 
data compiled by Barashenkov \cite{Barashenkov93}
 and calculations
from the HMS-ALICE code \cite{HMS-ALICE}.
}

\end{figure}
%\clearpage

%%%%%%%%%%%%%%%%%%%%%%%%%%%%%%%%%%%%%%%%%%%%%%%%%%%%%%%%%%%%%%%%%%% Fig. 9
\begin{figure}[!h]
\begin{center}
\vspace*{-5mm}
\includegraphics[width=15.0cm]{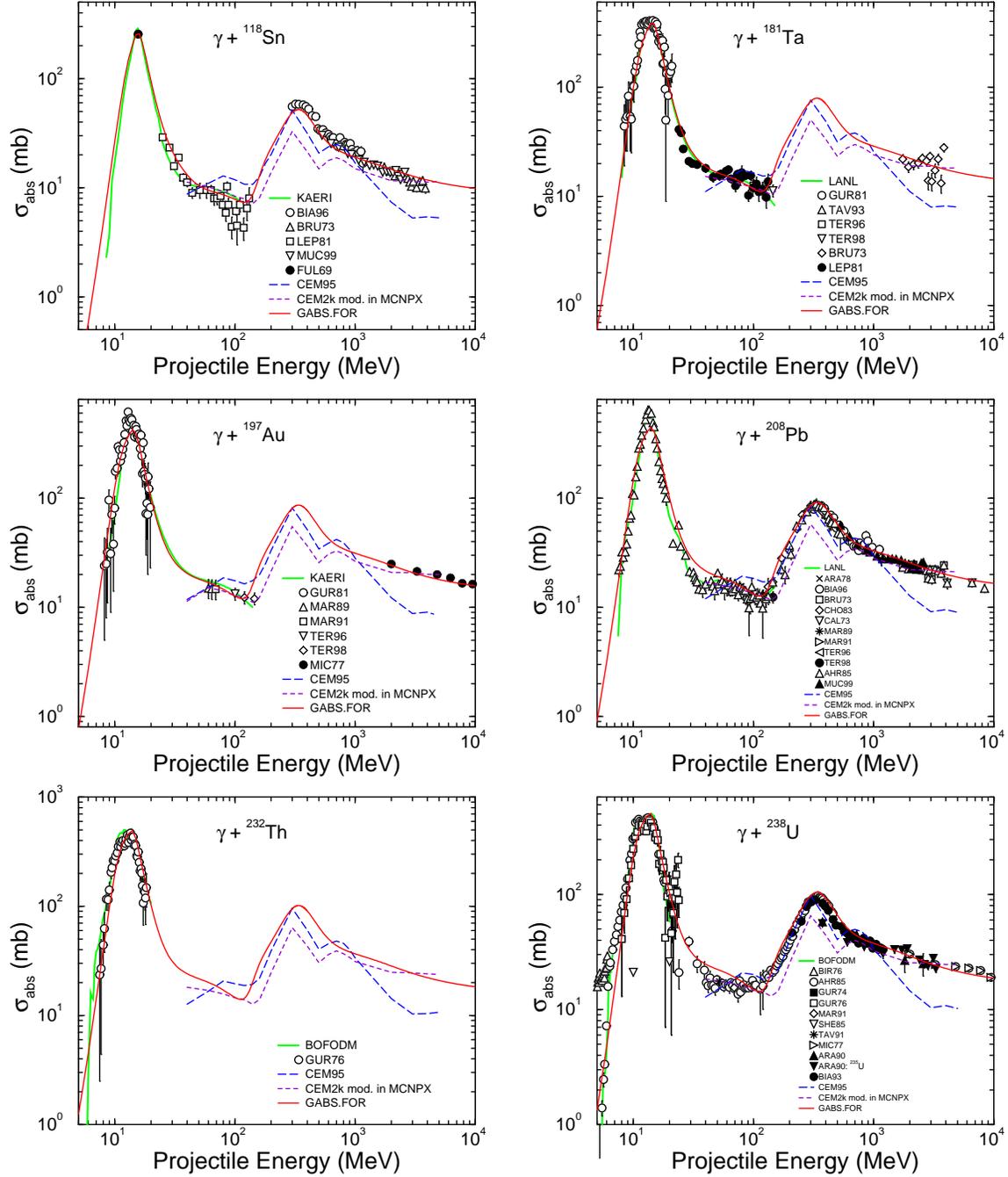}
\caption{
Examples of total photoabsorption cross sections for 
$^{118}$Sn, $^{181}$Ta, $^{197}$Au, $^{208}$Pb, $^{232}$Th, and $^{238}$U
as functions of photon energy. The red lines marked as ``GABS.FOR" are
results by a CEM03.01
subroutine written to reproduce Kossov's \cite{Kossov}
systematics, as described in \cite{JNRS05}. 
The green line marked as ``LANL", ``KAERI", or ``BOFODM" 
show the
evaluations by LANL, KAERI, or by a collaboration between IPPE/Obninsk
and CDFE/Moscow (the BOFOD(MOD) Library) from the IAEA Photonuclear Data 
Library \cite{IAEA}. Results from the 
photonuclear version of CEM95 \cite{cemphoto} and
from CEM2k as modified for MCNPX by Gallmeier \cite{Franz04}
are shown by the blue and brown dashed lines, respectively.
References to experimental data shown by different symbols
may be found in our recent paper \cite{JNRS05}. 
}
\end{center}
\end{figure}

\clearpage

used in CEM95 \cite{CEM95}, 
for CEM03.03 we have:

1) developed better approximations for the total elementary cross sections;

2) developed new approximations to describe more accurately experimental
    elementary energy and angular distributions of secondary particles from
    hadron-hadron and photon-hadron interactions;

3) normalized the photonuclear  reactions to detailed systematics developed by
    M. Kossov and the 
    nucleon-induced reactions, to NASA and Kalbach systematics;

4) changed the condition for transition from the INC stage  of a reaction 
    to preequilibrium; on the whole, the INC stage in CEM03.01 is 
    longer while the preequilibrium stage is shorter in comparison 
    with previous versions;

5) incorporated real binding energies for nucleons in the cascade 
    instead of the approximation of a constant separation energy of 7 MeV 
    used in the initial versions of the CEM; 

6) imposed  momentum-energy conservation for each 
    simulated even (provided only ``on the average'' by the initial versions);

7) changed and improved the algorithms of many INC routines 
    and almost all INC routines were rewritten, which speeded up the 
    code significantly;

8) fixed some preexisting bugs.

%{\noindent
On the whole, the INC of CEM03.01 describes nuclear reactions better
and much faster than the the initial version of the Dubna
INC \cite{Book,UPN73} used in CEM95 \cite{CEM95}.
One example of results by the INC from our CEM03.01 is shown in Fig.\ 10, 
namely,  $\pi^0$ spectra from 500 MeV 
$\pi^-$ + Cu.\\
%}

%\newpage
\begin{figure}[h]                                                 %Fig. 10
\centering
\includegraphics[width=95mm,angle=-90]{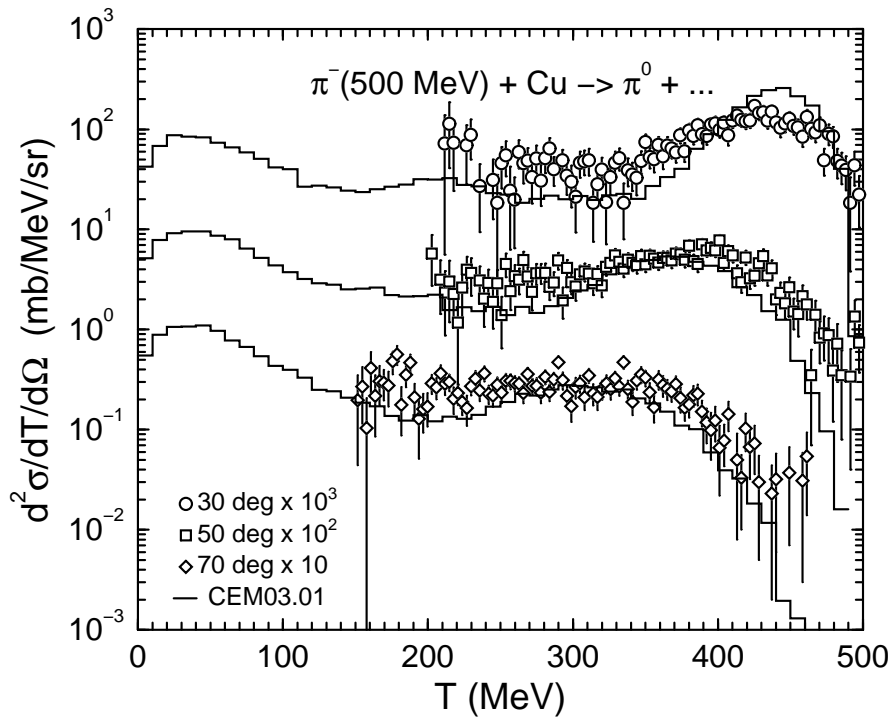}
\caption{
Experimental $\pi^0$ spectra from 500 MeV 
$\pi^-$ + Cu \cite{Mashnik00,ouyang} compared with CEM03.01 results.
Let us recall here that as
pions are produced by CEM03.01 only at the INC stage of reactions,
calculated pion spectra do not depend at all on how
other reaction mechanisms like coalescence,
evaporation, fission, or Fermi breakup are calculated.
}
\end{figure}

\clearpage

%\vspace{4mm}
{\large\it 3.2. The INC of LAQGSM03.03} \\

The INC of LAQGSM03.03 is described with a recently improved version 
\cite{LAQGSM03.03,LAQGSM03.01,Varenna06}
of the time-dependent intranuclear cascade model developed 
initially at JINR in Dubna, often referred to in the literature as 
the Dubna intranuclear Cascade Model, DCM (see 
\cite{Toneev:83}
and references 
therein). The DCM models interactions of fast cascade particles 
(``participants") with nucleon spectators of both the target and 
projectile nuclei and includes as well interactions of two 
participants (cascade particles). It uses experimental cross 
sections at energies below 4.5 GeV/nucleon,
or those calculated by the Quark-Gluon String Model 
%\cite{Amelin:90,Amelin86,Kaidalov87,Amelin84,Amelin89,Toneev89,Toneev90}
\cite{Amelin:90}--\cite{Toneev90}
at higher energies 
to simulate angular and energy distributions of cascade particles, 
and also considers the Pauli exclusion principle.

In contrast to the CEM03.01 version of the INC described above, DCM uses a 
continuous nuclear density distribution (instead of the approximation 
of several concentric zones, where inside each the nuclear density is 
considered to be constant); therefore, it does not need to consider 
refraction and reflection of cascade particles inside or on the 
border of a nucleus. It also keeps track of the time of an 
intranuclear collision and of the depletion of the nuclear density 
during the development of the cascade (the so-called ``trawling effect" 
mentioned above) and takes into account the hadron formation time
(see Fig.\ 11).

%\clearpage
\begin{figure}[ht]                                                 %Fig.\ 11.
\centering
%\vspace*{-10mm}
\includegraphics[angle=-0,width=13.5cm]{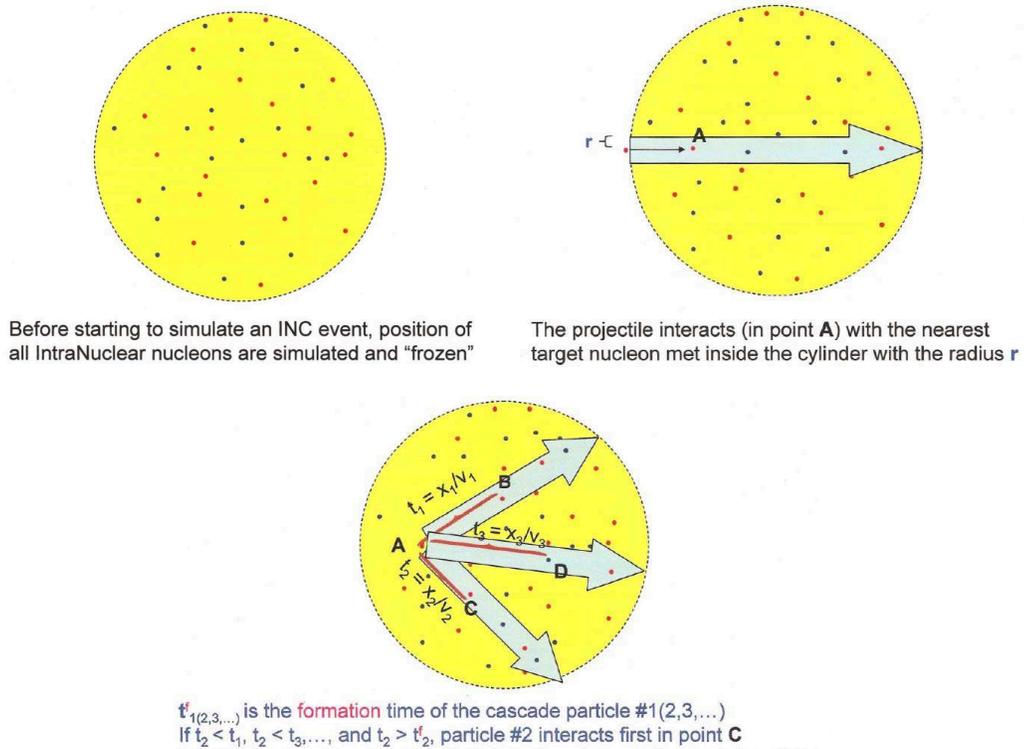}

%\vspace*{-29mm}
\caption{
An illustrative scheme of a target nucleus, of interaction points
of cascade particles (participants) with intranuclear nucleons
(spectators), and of selection of the corresponding time of such 
interactions, as performed in the INC used in LAQGSM.
}
\end{figure}

In the INC used in LAQGSM, all the new approximations 
developed recently for the INC of CEM03.01 to describe 
total cross sections and 
%experimental 
elementary energy and angular distributions of secondary particles 
from hadron-hadron interactions have been incorporated 
\cite{LAQGSM03.01}. 
Then, a new high-energy photonuclear reaction model based on the
of the event generators for $\gamma p$ and
$\gamma n$ reactions from the Moscow INC \cite{Iljinov97}
kindly provided us by Dr.\ Igor Pshenichnov,
and on the latest photonuclear version of CEM \cite{JNRS05}
was developed and incorporated \cite{Varenna06}
into the INC of LAQGSM, which allows us to calculate reactions 
induced by photons with energies of up to tens of GeV.
Finally,
the algorithms of many LAQGSM INC routines were changed and some INC routines 
were rewritten, which speeded up the code significantly; some preexisting 
bugs in the DCM were fixed; many useful comments were added \cite{LAQGSM03.03}.

In the latest version of LAQGSM, LAQGSM03.03 \cite{LAQGSM03.03}, the INC was 
modified for a better description of nuclear reactions at very high energies 
(above 20 GeV/nucleon), namely:

1) The latest fits to currently available evaluated experimental databases
for the total and elastic $\pi^+ p$, $\pi^- p$, 
$pp$, and $pn$ cross sections (see Chapter 
40 in the last Review of Particle Physics \cite{pdg06} and references therein) 
have been incorporated into LAQGSM.
LAQGSM03.03 uses now these approximations at energies above 20--30 GeV, 
and its own approximations developed for CEM03.01 \cite{CEM03.01}
at lower energies.

2) Previously, LAQGSM was used only at energies below 800 GeV. 
In  \cite{LAQGSM03.03},  
the possibility of using LAQGSM03.03 at ultra-relativistic energies, above 
1 TeV was studied. It was found that to describe ultra-high energy 
reactions, the value of the  parameter  $\sigma_{\perp} = 0.51$ GeV/c 
in the transverse 
momentum distribution of the constituent quarks of QGSM (see Eq.\ (12) 
in \cite{LAQGSM} or Eq.\ (10) in Ref.\ \cite{Amelin84})
has to be increased from 0.51 GeV/c at $T_p \leq  200$ GeV \cite{LAQGSM}
to $\approx 2$ GeV/c at $T_p \simeq 21$ TeV.

More details on the INC and other nuclear reaction models considered 
by different versions of LAQGSM may be found in Refs.
\cite{LAQGSM03.03,LAQGSM,LAQGSM03.01,Varenna06,01s1g1}.
Several examples of recent results by LAQGSM are shown in Figs.\ 12--16.

Fig.\ 12 shows a test of LAQGSM03.03 on inclusive pion
production spectra in proton-beryllium
collisions at 6.4, 12.3, and 17.5 GeV/c obtained from 
data taken by the already quite old E910
measurements at  Brookhaven National Laboratory, but analyzed and published
only several months ago \cite{Chemakin08}. 
Let us recall again that 
pions are produced only at the INC stage of reactions, without any
contributions from other reaction mechanisms,
so that such results test just the INC part of any model.
We see that LAQGSM03.03 describes
these pion spectra quite well, just as we obtained and published with 
previous versions of LAQGSM for other
spectra of different particles measured by the E910 experiment.

Fig.\ 13 presents part of the recent extensive experimental data
on fragmentation cross
sections of $^{28}$Si on H, C, Al, Cu, Sn, and Pb at energies from 290 to 1200
MeV/nucleon \cite{Zeitlin07}. Such measurements are of interest for 
NASA to plan long-duration space-flights and to test the models 
used to 
evaluate radiation exposure in flight,
and were performed at many incident energies in this energy range
at the Heavy Ion Medical
Accelerator in Chiba (HIMAC) and at Brookhaven National Laboratory
(see details in \cite{Zeitlin07} and references therein).
We calculate in our model practically all these data,
but here limit ourselves to examples of results for only three energies,
for each measured target. For comparison, we present in Fig.\ 13 results from
both LAQGSM03.03 (solid lines) and its predecessor LAQGSM03.02 (dashed lines). 
In general, LAQGSM03.03 describes these new data slightly
better than LAQGSM03.02 \cite{CEM03.02}, although this is not obvious
on the scale of the figure.
The agreement of our calculations with these data is excellent, especially 
considering that the results presented in this figure, just as all our other 
results shown in these lectures, 
are obtained without fitting any parameters in our codes;

\clearpage            % Use to start references on new page.

\begin{figure}[ht]                                                 %Fig.\ 12

\vspace*{-20mm}
\centering

\includegraphics[height=200mm,angle=-0]{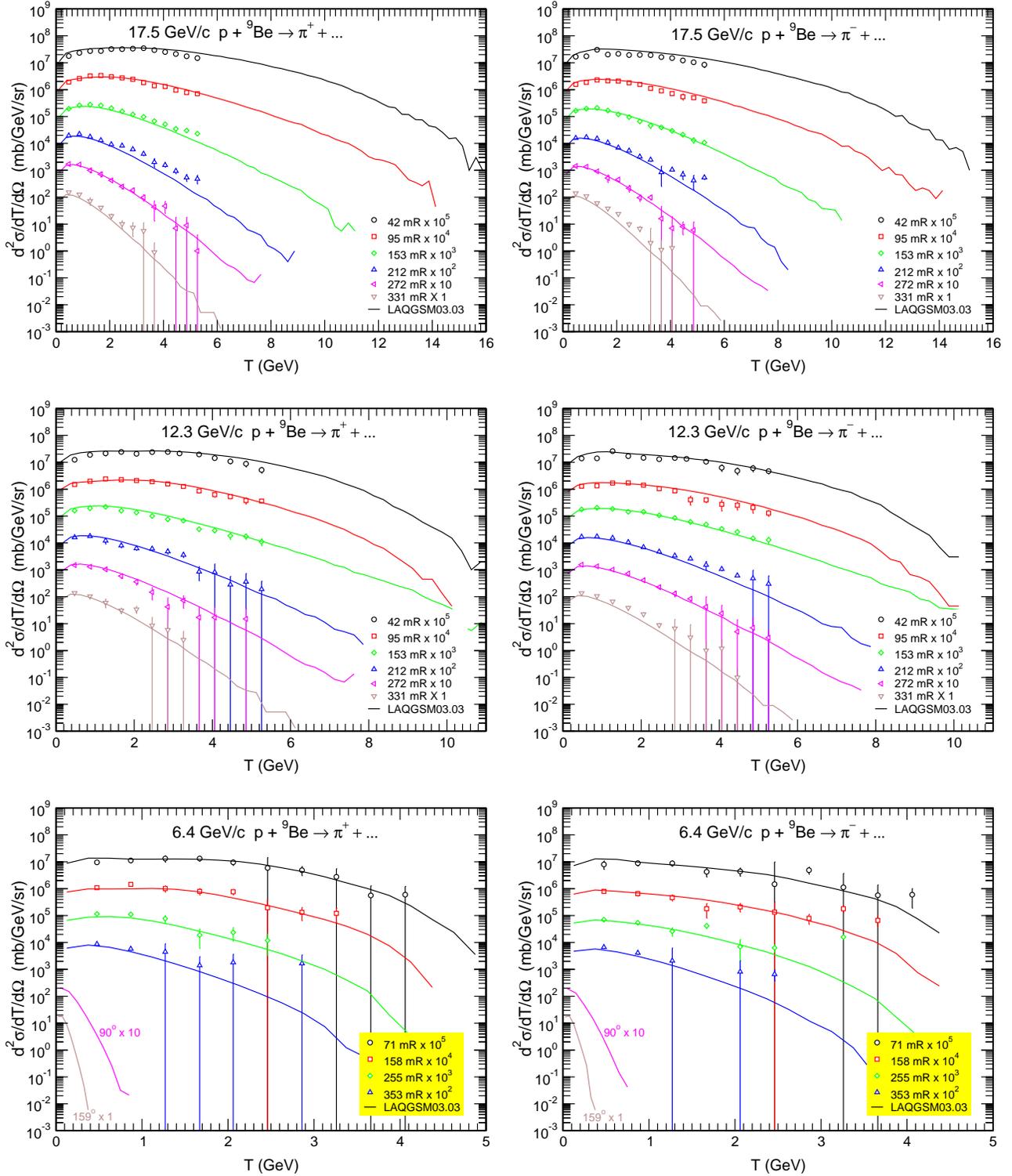} 

\caption{Measured inclusive forward $\pi^+$ and $\pi^-$ spectra 
from 6.4, 12.3, and 17.5 GeV/c p + $^9$Be %[47] 
\cite{Chemakin08}
compared with LAQGSM03.03 results at angles 
of detection as indicated in the plots. For reactions induced by 6.4 GeV/c
protons,
we also show LAQGSM03.03 predictions \cite{LAQGSM03.03}
for unmeasured spectra at 90 and 159 degrees. 
}
\end{figure}

\clearpage

\begin{figure}[ht]                                                   %Fig.\ 13

\vspace*{-30mm}
\centering

\hspace*{-10mm} 
\includegraphics[height=265mm,angle=-0]{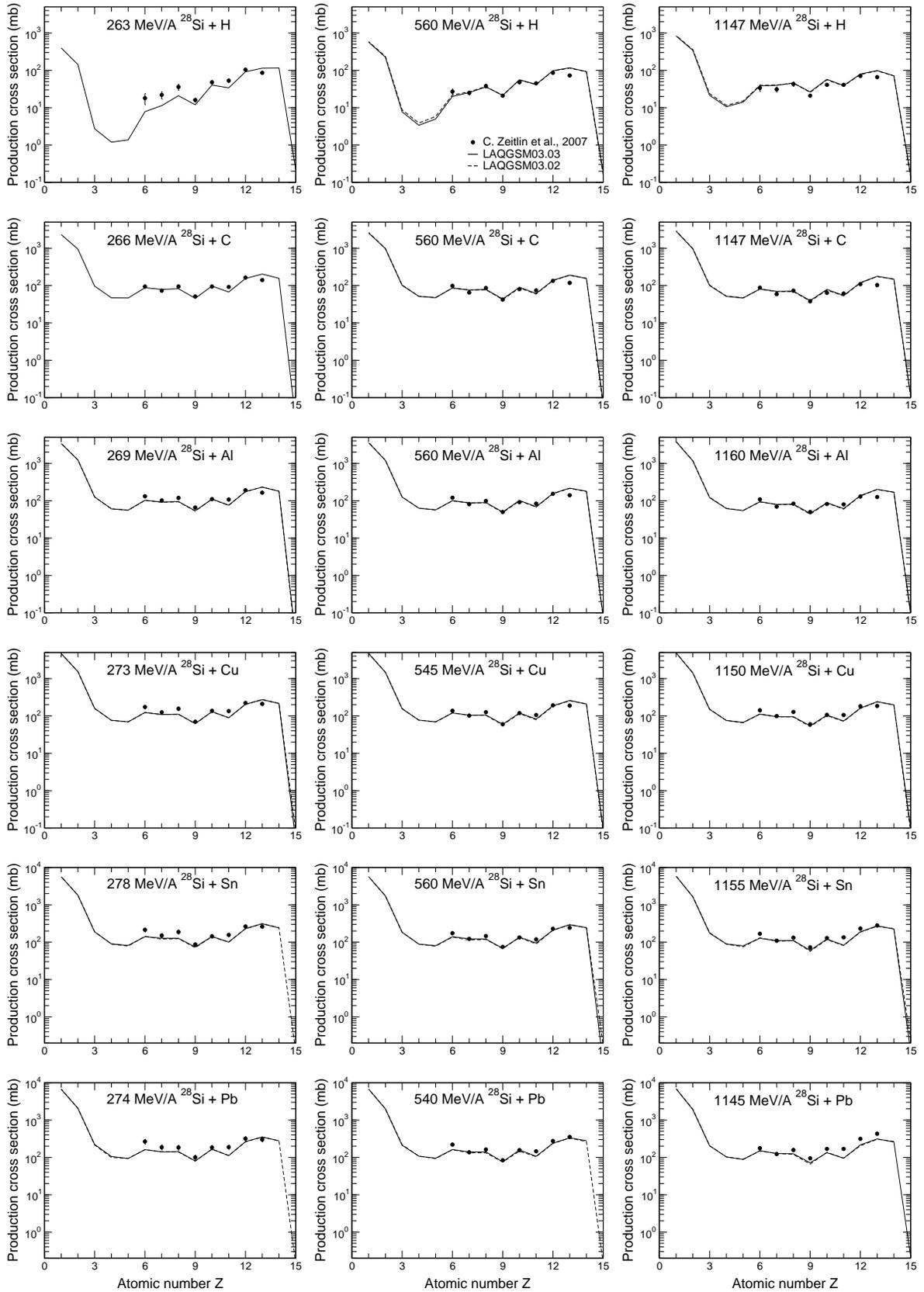}
\vspace*{-30mm}
\caption{Atomic-number dependence of the fragment-production cross sections 
from the interactions of $^{28}$Si of about 270, 560, and 1150 MeV/nucleon
with H, C, Al, Cu, Sn, and Pb, as indicated.
Filled circles are measurements by Zeitlin {\it et al.}
\cite{Zeitlin07}; 
solid lines are results from LAQGSM03.03 \cite{LAQGSM03.03}, 
while dashed lines are results from LAQGSM03.02 
\cite{CEM03.02}.
}
\end{figure}

\clearpage            % Use to start references on new page.

{\noindent
 we 
simply input $A$ and $Z$ of the projectile and target and the energy of
the projectile, then calculate without changing or fitting anything.
}

Fig.\ 14. shows proton spectra  spectra at 30, 60, 90, 120 and 150 
degrees from
interaction of bremsstrahlung $\gamma$ quanta of maximum energy 
4.5 GeV with $^{12}$C,  $^{27}$Al,  $^{63}$Cu, and $^{208}$Pb.
Experimental data shown by symbols in the figures are 
quite old, measured about 30 years ago by Alanakyan {\it et al}.
\cite{Alanakyan77,Alanakyan81},
however, to the best of our knowledge, we have described with LAQGSM03.01
these data for the first time: We do not know of
any publications or oral presentations where these measurements
were reproduced by a theoretical model, event generator,
or transport code.

\begin{figure}[ht]                                                   %Fig.\ 14

%\vspace*{-30mm}
\centering

\hspace*{-10mm} 
\includegraphics[height=167mm,angle=-0]{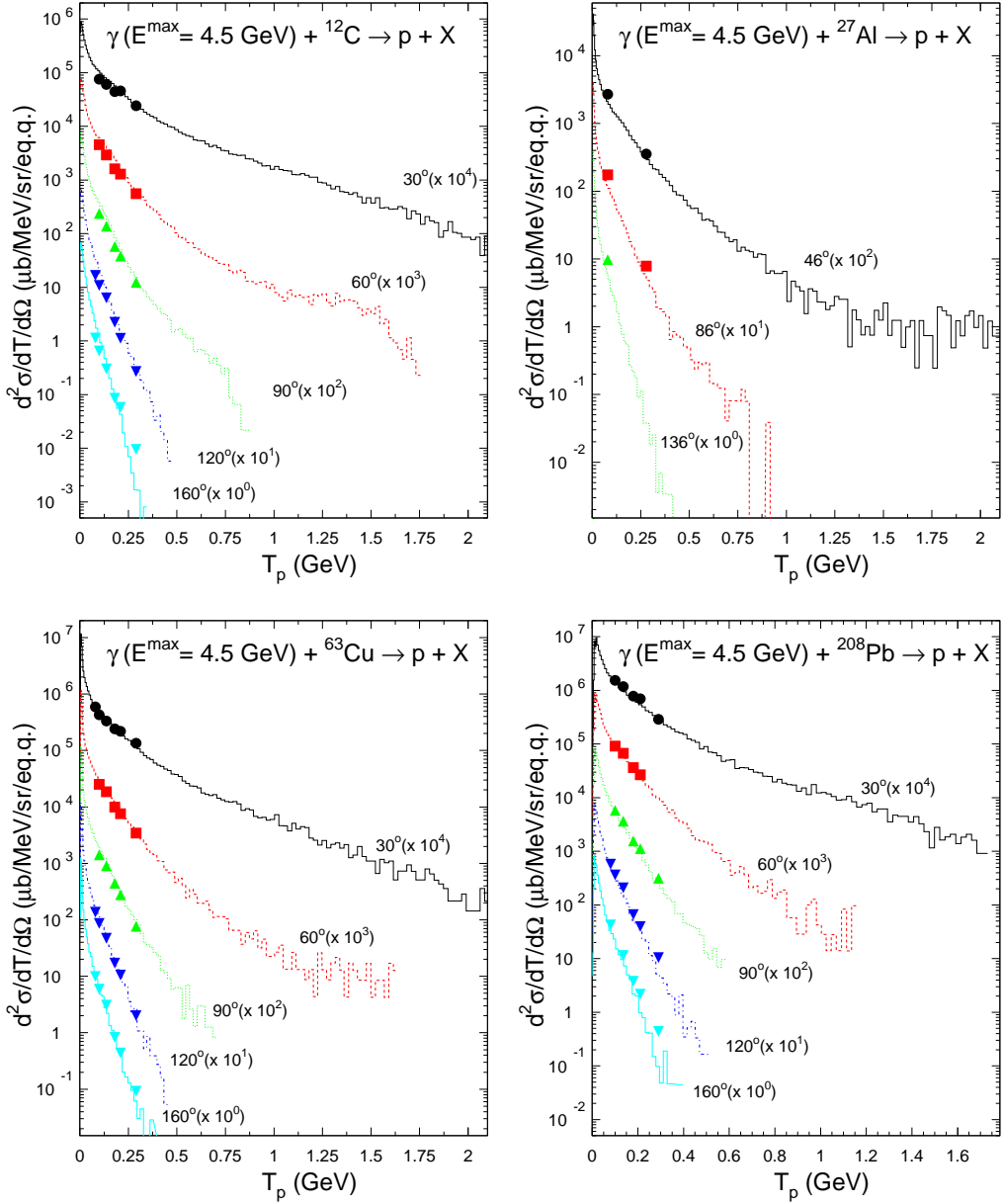}
\caption{Proton spectra at 30, 60, 90, 120 and 150 degrees from
interaction of bremsstrahlung $\gamma$ quanta of maximum energy 
4.5 GeV with $^{12}$C,  $^{27}$Al,  $^{63}$Cu, and $^{208}$Pb.
Experimental values shown by symbols are from %[20,21]
\cite{Alanakyan77,Alanakyan81}
while histograms show results from LAQGSM03.01.
}
\end{figure}

Fig.\ 15. shows an example of neutron spectra from intermediate-energy 
nucleus-nucleus interactions studied lately by many authors because
of a great interest
in such reactions for radiotherapy and because
such processes contribute substantially to the dose and dose
equivalent in space-flight. Namely, in Fig.\ 15, we compare results
from LAQGSM03.01 for neutron spectra at 5, 10, 20, 30, 40, 60, and
80 degrees from interactions of 600 MeV/nucleon  $^{20}$Ne,  
on $^{12}$C, $^{27}$Al, $^{64}$Cu, and  $^{208}$Pb
with measurements by Iwata {\it et al.} \cite{Iwata01}
and calculations by JQMD \cite{JQMD} and HIC \cite{HIC}.
We see that LAQGSM describes these data reasonably well,
generally as well as or better than do JQMD or HIC.

\begin{figure}[ht]                                                   %Fig.\ 15

%\vspace*{-30mm}
\centering

\hspace*{-10mm} 
\includegraphics[height=140mm,angle=-0]{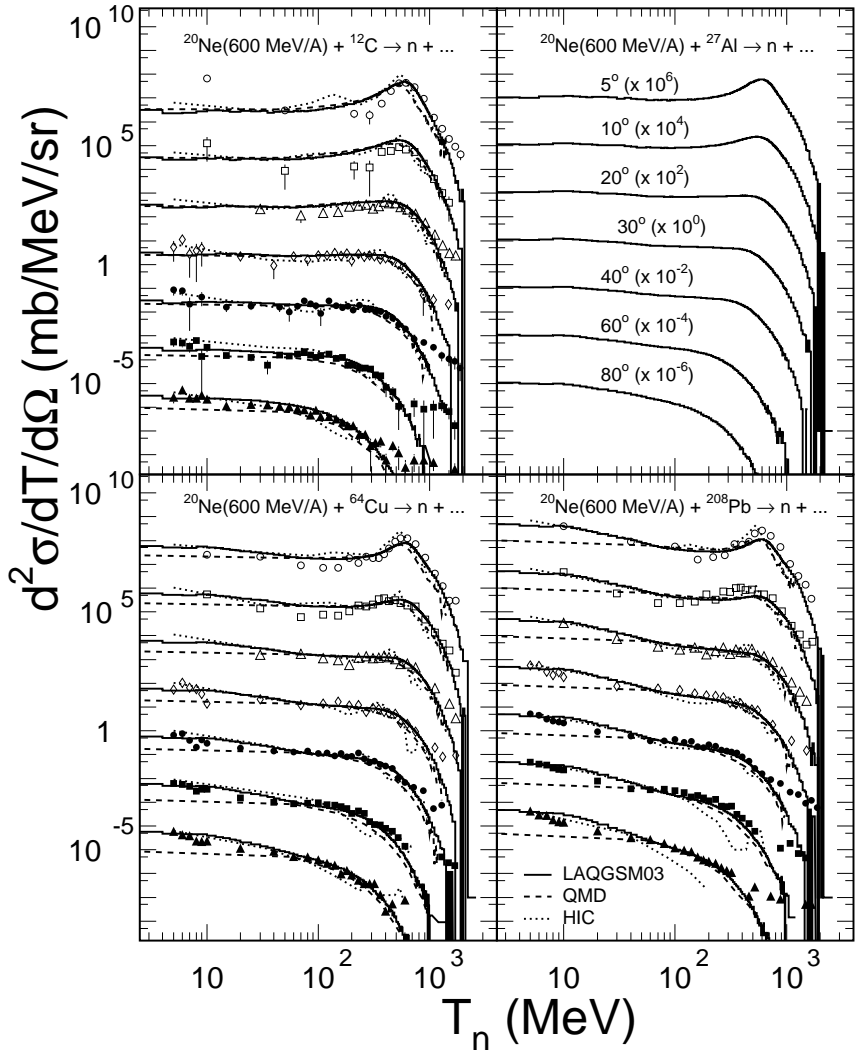}
\caption{
Comparison of measured \cite{Iwata01} double differential cross
sections of neutrons from 600 MeV/nucleon $^{20}$Ne,  
on $^{12}$C, $^{27}$Al, $^{64}$Cu, and  $^{208}$Pb with our
LAQGSM03.01 results and calculations by JQMD \cite{JQMD}
and HIC \cite{HIC}. Experimental data for these reactions on
Al are not yet available so we present here \cite{ND2004_AA_n}
only our predictions from LAQGSM03.01.
}
\end{figure}

Finally, Fig.\ 16 shows an example of product mass yields measured
recently at GSI in inverse kinematics \cite{Taieb03,Bernas03},
namely, product yields of 9
isotopes from Zn to Hg produced from interactions of a $^{238}$U beam
with a liquid-hydrogen target as calculated by LAQGSM03.01
as a stand-alone code and by the transport code MARS15 \cite{MARS}
using LAQGSM03.01 as its event generator \cite{MARS15_Madeira}.
The results from MARS15 using LAQGSM03.01
agree very well with the results from LAQGSM03.01 as a stand-alone code
and with experimental data.

\begin{figure}[ht]                                                    %Fig.\ 16

%\vspace*{-30mm}
\centering

\hspace*{-10mm} 
\includegraphics[height=100mm,angle=-0]{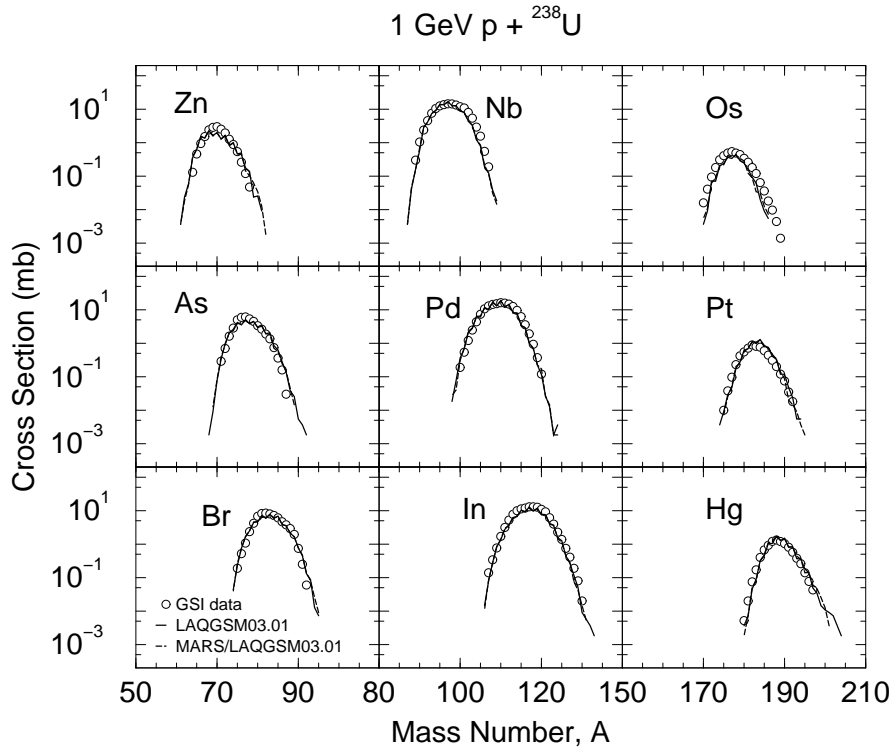}
\caption{
Experimental mass yield in 1 GeV p + $^{238}$U measured in inverse kinematics
at GSI 
\cite{Taieb03,Bernas03}
compared with results by MARS15 using LAQGSM03.01 as its event generator
(dashed lines) and by LAQGSM03.01 as a stand-alone code (solid lines).
}
\end{figure}

\vspace{4mm}
{\large\bf 4.  The Coalescence Model} \\

When the cascade stage of a reaction is completed, CEM03.0x
and LAQGSM03.0x use the
coalescence model described in Refs.~\cite{Toneev:83,Gudima:83a}
to ``create" high-energy $d$, $t$, $^3$He, and $^4$He by
final-state interactions among emitted cascade nucleons, already outside 
of the target nucleus. In contrast to most other
coalescence models for heavy-ion-induced reactions,
where complex-particle spectra are estimated simply by
convolving the measured or calculated inclusive spectra of nucleons
with corresponding fitted coefficients (see, {\it e.g.}, \cite{Kapusta:80}
and references therein), CEM03.0x and LAQGSM03.0x use in their simulations of
particle coalescence real information about all emitted cascade nucleons
and do not use integrated spectra. We assume that
all the cascade nucleons having differences in their momenta 
smaller than $p_c$ and the correct isotopic content form an appropriate
composite particle. This means that the formation probability for,
{\it e.g.} a deuteron is
\begin{equation}
W_d(\vec p,b) = \int \int d \vec p_p  d \vec p_n
\rho^C(\vec p_p,b) \rho^C(\vec p_n,b) \delta(\vec p_p + \vec p_n - \vec p)
\Theta(p_c - |\vec p_p - \vec p_n|) ,
\end{equation}
where the particle density in momentum space is related to the
one-particle distribution function $f$ by
\begin{equation}
\rho^C(\vec p,b) = \int d \vec r f^C (\vec r, \vec p,b) .
\end{equation}
Here, $b$ is the impact parameter for the projectile interacting
with the target nucleus and the
superscript index $C$ shows that only cascade nucleons are taken into
account for the coalescence process. The coalescence radii $p_c$
were fitted for each composite particle in Ref.~\cite{Toneev:83}
to describe available data for the reaction Ne+U at 1.04 GeV/nucleon,
but the fitted values turned out to be quite universal and were
subsequently found to satisfactorily describe high-energy  
complex-particle production for a variety of reactions induced 
both by particles and nuclei at incident energies up to 
about 200 GeV/nucleon,
when describing nuclear reactions with different versions of
LAQGSM \cite{LAQGSM03.03,CEM03.02,LAQGSM}
or with its predecessor, the Quark-Gluon String
Model (QGSM) \cite{Amelin:90}.
These parameters are:
\begin{equation}
p_c(d) = 90 \mbox{ MeV/c; \hspace{0.5cm} }
p_c(t) =  p_c(^3{\mbox He}) = 108 \mbox{ MeV/c; \hspace{0.5cm} }
p_c(^4{\mbox He}) = 115 \mbox{ MeV/c .}
\end{equation}
As the INC of CEM03.0x is different from those of LAQGSM or QGSM, 
it is natural to expect different best values for $p_c$ as well.
Our recent studies show that the values of parameters $p_c$ 
defined by Eq.\ (19) are also good for CEM03.01 for projectile particles 
with kinetic energies $T_0$ lower than 300 MeV and equal to or above 
1 GeV. For incident energies in the interval
300 MeV $< T_0 \le 1$ GeV, a better overall agreement with the
available experimental data is obtained by using values of $p_c$
equal to 150, 175, and 175 MeV/c for $d$, $t$($^3$He), and $^4$He,
respectively. These values of $p_c$ are fixed as defaults in CEM03.01.
If several cascade nucleons are chosen to coalesce into composite
particles, they are removed from the distributions of nucleons and
do not contribute further to such nucleon characteristics as spectra,
multiplicities, {\it etc.} 

In comparison with the initial version \cite{Toneev:83,Gudima:83a},
in CEM03.0x and LAQGSM03.0x, several coalescence routines have been 
changed/deleted and have been tested against a large variety of 
measured data on nucleon- and nucleus-induced reactions at different 
incident energies.

Two examples of results from the coalescence model are shown in Figs.\ 
17 and 18. We see that for a reaction between an intermediate-energy
neutron and a medium-mass nucleus (Fig.\ 17; 96 MeV n + Fe),
where the mean multiplicity of the secondary nucleons is small, the
contribution from coalescence to the total angle-integrated
energy spectra of complex particles
is very low, less than a few percent. 
On the other hand, in the case
of interaction of a high-energy proton with a heavy nucleus
(Fig.\ 18, 70 GeV p + Pb), very energetic secondary complex
particles produced at forward angles: deuterons with 14 GeV/c
and tritons with 19 GeV/c were measured \cite{Abramov8587} at 160 mrad.
Probably all such extremely energetic $d$ and $t$ at
forward angles are produced only via coalescence of complex particles
particles from energetic nucleons emitted during the INC stage of the reaction:
We do not know any other interaction mechanisms that would produce
$d$ and $t$ of such high energies from this reaction. 
The coalescence model in LAQGSM03.0x
reproduces these experimental spectra quite well. It is clear that 
the coalescence mechanism is more important for high-energy
heavy-ion reactions, where the multiplicity of secondary INC nucleons
is much higher than in the case of nucleon-induced reactions.\\

%\vspace{4mm}
{\large\bf 5.  Preequilibrium Reactions} \\

The subsequent  preequilibrium interaction stage of nuclear reactions
is considered by our current CEM and LAQGSM in the framework of 
the latest version of the 
Modified Exciton Model (MEM)~\cite{MEM,MODEX}
as implemented in CEM03.01 \cite{CEM03.01}.
At the preequilibrium stage of a reaction we take into account all
possible nuclear transitions changing the number of excitons $n$
with 
$\Delta n = +2, -2$, and 0, as well as all possible multiple subsequent
emissions of $n$, $p$, $d$, $t$, $^3$He, and  $^4$He. The corresponding
system of master equations describing the behavior of a nucleus at
the preequilibrium stage is solved by the Monte-Carlo 
technique~\cite{CEMP,CEM}.

\clearpage
\begin{figure}[ht]                                                    %Fig.\ 17

\vspace*{-10mm}
\centering

\hspace*{-5mm} 
\includegraphics[height=190mm,angle=-0]{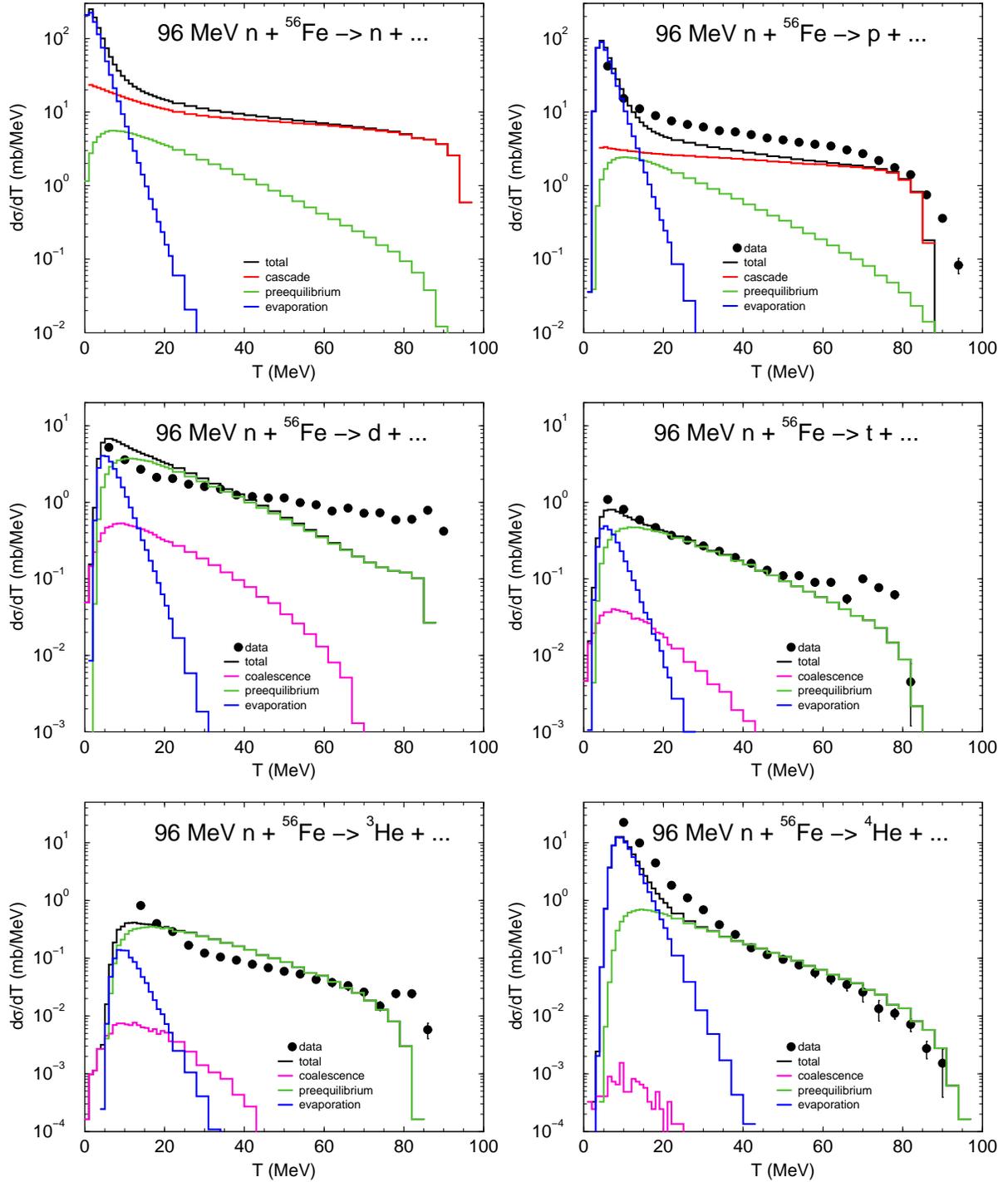}
\caption{
Angle-integrated energy spectra of $n$, $p$, $d$, $t$, $^3$He,
and $^4$He from 96 MeV $n$ + $^{56}$Fe
calculated by CEM03.01 compared
with experimental data by Blideanu {\it et al.}
\cite{Blideanu04}. The black histograms show the total calculated
spectra (all reaction mechanisms considered by CEM03.0x),
while the color histograms show separately contributions to the total spectra
from INC, coalescence, preequilibrium, and evaporation, respectively,
as indicated in the corresponding legends of plots.
}
\end{figure}

\clearpage

\begin{figure}[ht]                                                    %Fig.\ 18

\vspace*{-5mm}
\centering

\hspace*{-5mm} 
\includegraphics[height=160mm,angle=-0]{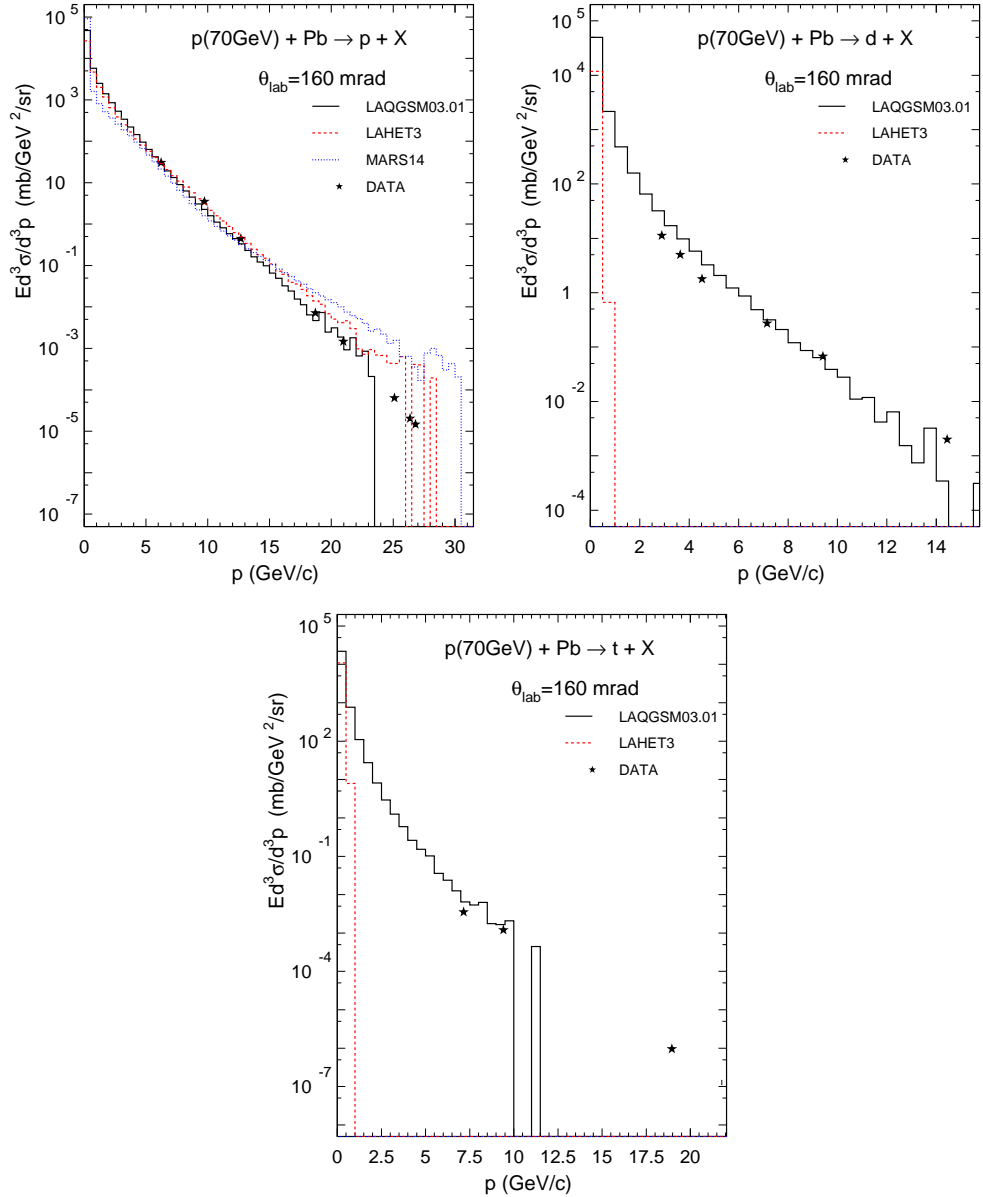}
\caption{
Invariant cross sections $E d^3 \sigma /d^3 p$ for forward production
of p, d, and t at 160 mrad (9.17 deg) as functions of
particle momentum $p$ from 70 GeV protons on $^{208}$Pb.
Experimental data shown by stars  are from 
Ref.\ \cite{Abramov8587},
while calculations by LAQGSM03.01, LAHET3 \cite{LAHET3}, and MARS14 \cite{MARS}
are shown by histograms, as indicated in the legends.
}
\end{figure}

For a preequilibrium nucleus with excitation energy $E$ and number of
excitons $n=p+h$, the partial transition probabilities changing the 
exciton number by $\Delta n$ are
\begin{equation}
\lambda_{\Delta n} (p,h,E) =
\frac{2\pi}{\hbar}|M_{\Delta n}|^2 \omega_{\Delta n} (p,h,E) \mbox{ .}
\label{a10}
\end{equation}
The emission rate of a nucleon of type $j$ into the continuum is
estimated according to the detailed balance principle
\begin{eqnarray}
\Gamma_{j}(p,h,E) & = & \int\limits_{V_{j}^{c}} ^{E-B_j}
\lambda_c^j(p,h,E,T)dT \mbox{ ,} \nonumber \\
\lambda_c^j(p,h,E,T) & = & \frac{2s_j + 1}{\pi^2 \hbar^3}
\mu_j \Re_j(p,h) \frac{\omega(p-1,h,E-B_j-T)}
{\omega(p,h,E)} T \sigma_{inv}(T) \mbox{ ,}
\label{a11}
\end{eqnarray}
where $s_j$, $B_j$, $V_j^c$, and $\mu_j$ are the spin, binding energy,
Coulomb barrier, and reduced mass of the emitted particle, respectively.
The factor $\Re_j(p,h)$ ensures the condition for
the exciton chosen to be the particle of type $j$ and can easily be 
calculated by the Monte-Carlo technique.

Assuming an equidistant level scheme with the single-particle
density $g$, we have the level density of the $n$-exciton
state as~\cite{ericson}
\begin{equation}
\omega(p,h,E) = \frac{g (gE)^{p+h-1}}{p! h! (p+h-1)!} \mbox{ .}
\label{a12}
\end{equation}
This expression should be substituted into Eq.\ (21). For the transition 
rates (20), one needs the number of states taking into account
the selection rules for intranuclear exciton-exciton scattering.
The appropriate formulae have been derived by Williams~\cite{williams}
and later corrected for the exclusion principle and indistinguishability 
of identical excitons in Refs.\ \cite{williams2,ribansky}:
\begin{eqnarray}
\omega _+ (p,h,E) & = &
\frac{1}{2} g \frac{[gE-{\cal A}(p+1,h+1)]^2}
{n+1}
\biggl[ \frac{gE - {\cal A}(p+1,h+1)}{gE - {\cal A}(p,h)} \biggr] ^{n-1}
\mbox{ ,} \nonumber \\
\omega _0 (p,h,E) & = &
\frac{1}{2} g \frac{[gE-{\cal A}(p,h)]}{n} [p(p-1)+4ph+h(h-1)]
\mbox{ ,} \nonumber \\
\omega _- (p,h,E) & = &
\frac{1}{2} gph(n-2) \mbox{ ,}
\label{a13}
\end{eqnarray}
where ${\cal A}(p,h) = (p^2 +h^2 +p-h)/4 - h/2$.
By neglecting the difference of matrix elements with different $\Delta n$,
$M_+ = M_- = M_0 = M$, we estimate the value of $M$ for a given 
nuclear state by associating the $\lambda_+ (p,h,E)$ transition with the
probability for quasi-free scattering of a nucleon
above the Fermi level on a nucleon of the target nucleus.
Therefore, we have
\begin{equation}
\frac{ < \sigma (v_{rel}) v_{rel} >}{V_{int}} =
\frac{\pi}{\hbar} |M|^2 
\frac{g [ gE-{\cal A}(p+1,h+1)]}{n+1}
\biggl[ \frac{gE - {\cal A}(p+1,h+1)}{gE - {\cal A}(p,h)} \biggr] ^{n-1}
\mbox{ .}
\label{a14}
\end{equation}
Here, $V_{int}$ is the interaction volume estimated as 
$V_{int} = {4 \over 3} \pi (2 r_c + \lambda / 2 \pi)^3$, 
with the de Broglie wave 
length $\lambda / 2 \pi$ corresponding to the relative velocity 
$v_{rel} = \sqrt{2 T_{rel} /m_N}$. A value of the order of the nucleon radius
is used for $r_c$ in the CEM: $r_c = 0.6$ fm.

The averaging in the left-hand
side of Eq.\ (24) is carried out over all excited states taking into account
the Pauli principle in the approximation
\begin{equation}
< \sigma (v_{rel}) v_{rel} > \simeq < \sigma (v_{rel}) > < v_{rel} > 
\mbox{ .}
\label{a15}
\end{equation}
The averaged cross section $< \sigma (v_{rel}) >$ is calculated 
by the Monte-Carlo simulation method
and by introducing a factor $\eta$ effectively taking into
account the Pauli principle exactly as is done in the Fermi-gas model
(see, {\it e.g.}, \cite{kikuchi})\footnote{Unfortunately, 
formula (27) as presented in Ref.~\cite{CEM} had some misprints;
in the prior publication~\cite{CEMP}, it was correct.}
\begin{equation}
\sigma (v_{rel}) = {1\over2} [ \sigma_{pp} (v_{rel} )
+ \sigma_{pn} (v_{rel}) ] \eta(T_F/T) \mbox{ , where}
\label{a6}
\end{equation}
\begin{equation}
\eta(x) =
\cases{1 - {7 \over 5} x,&if $x \le 0.5$ ,\cr
 1 - {7 \over 5} x + {2 \over 5} x (2-{1 \over x})^{5/2},&if $x > 0.5$ .\cr}
\label{a7}
\end{equation}
Here, $v_{rel}$ is the relative velocity of the excited nucleon (exciton) and
the target nucleon in units of the speed of light and $T$
is the kinetic energy of the exciton. 
The free-particle interaction cross sections 
$\sigma_{pp} (v_{rel} )$ and $\sigma_{pn} (v_{rel})$ in Eq.\ (26) 
are estimated using the
relations suggested by Metropolis {\it et al.}\ \cite{metropolis1}
\begin{eqnarray}
\sigma_{pp}(v_{rel})={10.63 \over v_{rel}^2}-{29.92 \over v_{rel} }+42.9 
\mbox{ ,}
\nonumber \\
\sigma_{pn}(v_{rel})={34.10 \over v_{rel}^2}-{82.2 \over v_{rel} }+82.2 
\mbox{ ,}
\label{a8}
\end{eqnarray}
where the cross sections are given in mb.

The relative kinetic energy of colliding particles necessary to
calculate $< v_{rel} >$ and the factor $\eta$ in Eqs.\ (26,27) are estimated 
in the so-called ``right-angle-collision" approximation~\cite{MEM},
{\it i.e.} as a sum of the mean kinetic energy of an excited particle (exciton)
measured from the bottom of the potential well 
$T_p = T_F + E/n$ plus the mean kinetic energy of an intranuclear
nucleon partner
$T_N = 3T_F/5$, that is
$T_{rel} = T_p + T_N = 8T_F/5 + E/n$.

Combining (20), (22) and (24), we get finally for the transition rates:
\begin{eqnarray}
\lambda _+ (p,h,E) & = &
\frac{ < \sigma (v_{rel}) v_{rel} >}{V_{int}} \mbox{ ,} \nonumber \\
\lambda _0 (p,h,E) & = &
\frac{ < \sigma (v_{rel}) v_{rel} >}{V_{int}}
\frac{n+1}{n}
\biggl[ \frac{gE - {\cal A}(p,h)}{gE - {\cal A}(p+1,h+1)} \biggr] ^{n+1}
\frac{p(p-1)+4ph+h(h-1)}{gE-{\cal A}(p,h)} \mbox{ ,} \nonumber \\
\lambda _- (p,h,E) & = &
\frac{ < \sigma (v_{rel}) v_{rel} >}{V_{int}}
\biggl[ \frac{gE - {\cal A}(p,h)}{gE - {\cal A}(p+1,h+1)} \biggr] ^{n+1}
\frac{ph(n+1)(n-2)}{[gE-{\cal A}(p,h)]^2} \mbox{ .}
\label{a16}
\end{eqnarray}

CEM considers the possibility of 
fast $d$, $t$, $^3$He, and $^4$He emission at the 
preequilibrium stage of a reaction in addition to the
emission of nucleons. We assume that in
the course of a reaction $p_j$ excited nucleons (excitons) are able to
condense with probability $\gamma_j$ forming a complex particle
which can be emitted during the preequilibrium state. 
A modification of Eq.\ (21) for the complex-particle emission rates is 
described in detail in Refs.~\cite{CEMP,CEM}.
The ``condensation" probability $\gamma_j$ is estimated in those
references as the overlap
integral of the wave function of independent nucleons with that
of the complex particle (cluster)
\begin{equation}
\gamma_j \simeq p^3_j (V_j / V)^{p_j - 1} =  p^3_j (p_j / A)^{p_j - 1}
\mbox{ .}
\label{a17}
\end{equation}

This is a rather crude estimate.  
In the usual way the values $\gamma_j$ are taken from fitting the 
theoretical preequilibrium spectra to the experimental ones,
which gives rise to an additional, as compared to (30), dependence of the 
factor $\gamma_j$ on $p_j$ and excitation energy
(see, {\it e.g.}, Refs.~\cite{betak76,wuchang}), for each considered reaction. 

The single-particle density $g_j$
for complex-particle states is found in the CEM by assuming the
complex particles move freely in a uniform potential well whose 
depth is equal to the binding energy of this particle in a 
nucleus~\cite{CEM}
\begin{equation}
g_j(T)  = { V(2s_j+1)(2\mu_j)^{3/2}  \over
4 \pi^2 \hbar^3 } (T+B_j)^{1/2}
\mbox{ .}
\label{a18}
\end{equation}

As we stated previously, this is a crude approximation
and it does not provide a good prediction of emission of 
preequilibrium $\alpha$ particles
(see, {\it e.g.}, \cite{Mashnik97} and references therein).  
In CEM03.0x, to improve the description of preequilibrium 
complex-particle emission, we estimate $\gamma_j$ by multiplying the
estimate provided by Eq.\ (30) by an empirical coefficient $M_j(A,Z,T_0)$
whose values are fitted to available nucleon-induced
experimental complex-particle spectra. We fix the fitted values of 
 $M_j(A,Z,T_0)$ in data commons of CEM03.0x and complement them with
routines {\bf gambetn} and {\bf gambetp} for their interpolation 
outside the region covered by our fitting. As shown already above in Fig.\ 17
and proved in several more figures below,
after fitting 
$M_j(A,Z,T_0)$, CEM and LAQGSM describe quite well the measured
spectra of all complex particles, providing a much better
agreement with experimental data than all their predecessors did.

CEM and LAQGSM predict forward-peaked (in the laboratory system) angular
distributions for preequilibrium particles.
For instance, CEM03.0x assumes that a 
nuclear state with a given excitation energy $E^*$ should
be specified not only by the exciton number $n$ but also by the
momentum direction $\Omega$. Following
Ref.~\cite{mantzouranis}, the master equation (11) from
Ref.~\cite{CEM} can be generalized for this case
provided that the angular dependence for the transition rates
$\lambda _+$, $\lambda _0$, and $\lambda _-$ (Eq.\ (29)
is factorized. In accordance with
Eqs.\ (24) and (25), in the CEM it is assumed that
\begin{equation}
<\sigma> \to <\sigma> F(\Omega) \mbox{ ,}
\end{equation}
where
\begin{equation}
F(\Omega) = {d \sigma^{free}/ d \Omega \over
\int d \Omega '  d \sigma^{free} / d \Omega '} \mbox{ .} 
\end{equation}
The scattering cross section $ d \sigma^{free}/ d \Omega$ is
assumed to be isotropic in the reference frame of the interacting excitons,
thus resulting in an asymmetry in both the nucleus
center-of-mass and laboratory frames.
The angular distributions of preequilibrium complex particles are assumed
\cite{CEM} to be similar to those for the nucleons in each nuclear state.

This calculation scheme is easily realized by the Monte-Carlo technique.
It provides a good description of double differential spectra
of preequilibrium nucleons and a not-so-good but still satisfactory
description of complex-particle spectra from different types
of nuclear reactions at incident energies from tens of MeV to several GeV.
For incident energies below about 200 MeV, 
Kalbach \cite{Kalbach88} has developed a phenomenological systematics
for preequilibrium-particle angular distributions by fitting
available measured spectra of nucleons and complex particles.
As the Kalbach systematics are based on measured spectra, they
describe very well the double-differential spectra of preequilibrium
particles and generally provide a better agreement of 
calculated preequilibrium
complex-particle spectra with data than does the CEM approach based on 
Eqs.\ (32,33). This is why we have incorporated into CEM03.0x and LAQGSM03.0x
the Kalbach systematics \cite{Kalbach88} to describe angular distributions
of both preequilibrium nucleons and complex particles at incident
energies up to 210 MeV. At higher energies, we use
the CEM approach based on Eqs.\ (32,33).

By  ``preequilibrium particles" we mean particles
which are emitted after the cascade stage of a reaction but
before achieving statistical equilibrium at a
time $t_{eq}$, which is fixed by
the condition $\lambda_+(n_{eq},E) = \lambda_-(n_{eq},E)$ from
which we get 
\begin{equation}
n_{eq} \simeq \sqrt{2gE} \mbox{ .}
\end{equation}
At $t \ge t_{eq}$
(or  $n \ge n_{eq}$), the behavior of the remaining excited compound
nucleus is described in the framework of both the Weisskopf-Ewing 
statistical theory of particle evaporation~\cite{weisewing} 
and fission competition according to Bohr-Wheeler theory~\cite{borwil}.

The parameter $g$ entering into Eqs.\ (29) and (34) is related to the
level-density parameter of single-particle states
$a = \pi^2 g /6$. At the preequilibrium stage, we
calculate the level-density parameter $a$
with our own approximation \cite{CEM97f}
in the form proposed initially by Ignatyuk {\it et al.} \cite{Ignatyuk}, 
following the method by Iljinov {\it et al.}\ \cite{iljinov92}:
\begin{equation}
a(Z,N,E^*) = \tilde a(A) \left\{ 1 + \delta W_{gs}(Z,N)
{{f(E^* - \Delta)} \over {E^* - \Delta}} \right\},
\end{equation}
where
\begin{equation}
\tilde a(A) = \alpha A + \beta A^{2/3} B_s 
\end{equation}
is the asymptotic Fermi-gas value of the level-density parameter at
high excitation energies.
Here, $B_s$ is the ratio of the surface area of the nucleus to the
surface area of a sphere of the same volume (for the ground state of a
nucleus, $B_s \approx 1$), and
\begin{equation}
f(E) = 1 - exp (-\gamma E) \mbox{ .}
\end{equation}
$E^*$ is the total excitation energy of the nucleus, related to the
``thermal" energy $U$ by: 
$U = E^* - E_R - \Delta$, where
$E_R$ and $\Delta$ are the rotational and pairing energies, respectively.

We use the shell correction $\delta W_{gs}(Z,N)$
by M\"oller {\it et al.}\ \cite{moller95}
and the pairing energy shifts from 
M\"oller, Nix, and Kratz \cite{moller97}.
The values of the parameters
$\alpha$, $\beta$, and $\gamma$ were derived in Ref.\ \cite{CEM97f}
by fitting the the same data analyzed by Iljinov {\it et al}.\ \cite{iljinov92} 
(we discovered that Iljinov {\it et al}.\ used $11 / \sqrt{A}$ for the 
pairing energies $\Delta$ 
in deriving their level-density systematics instead of the value of  
$12 / \sqrt{A}$ stated in Ref.\ \cite{iljinov92} and we also found 
several misprints 
in the nuclear level-density data shown in their Tables.~1 and 2 used 
in the fit). 
We find:
$$\alpha = 0.1463 \mbox{, } \beta = -0.0716 \mbox{, and } 
\gamma = 0.0542 \mbox{ .}$$ 

As mentioned in Section 3.1, the standard
version of the CEM \cite{CEM} provides an overestimation of
preequilibrium particle emission from different 
reactions we have analyzed (see more details in
\cite{CEM2k,SATIF6}). One way to solve this problem 
suggested in Ref.\ \cite{CEM2k} is to change the
criterion for the transition from the cascade stage
to the preequilibrium one, as described in Section 3.1. Another easy
way suggested in Ref.\ \cite{CEM2k} to shorten the preequilibrium 
stage of a reaction is to arbitrarily allow only transitions that 
increase the number of excitons, $\Delta  n = + 2$, {\it i.e.},
only allow the evolution of a nucleus toward the compound nucleus.
In this case, the time of the equilibration will be shorter and fewer
preequilibrium particles will be emitted, leaving more excitation
energy for the evaporation. Such a ``never-come-back" 
approach is used by some other exciton
models, for instance, by the Multistage Preequilibrium Model (MPM)
used in LAHET \cite{LAHET} and by FLUKA \cite{FLUKA}. This 
approach was used in the CEM2k \cite{CEM2k} version of the CEM
and it allowed us to describe much better the p+A reactions
measured at GSI in inverse kinematics at energies around 1 GeV/nucleon.
Nevertheless, the ``never-come-back" approach seems unphysical,
therefore we no longer use it. 
We now address the problem of emitting
fewer preequilibrium particles in the CEM by following 
Veselsk\'{y} \cite{sigpre}. We assume that the ratio
of the number of quasi-particles (excitons) $n$ at each preequilibrium
reaction stage to the number of excitons in the equilibrium configuration
$n_{eq}$, corresponding to the same excitation energy, to be a
crucial parameter for determining the probability of preequilibrium
emission $P_{pre}$. This probability for a given preequilibrium
reaction stage is evaluated using the formula
\begin{equation}
P_{pre}(n/n_{eq}) = 1 - \exp \Bigl( - { {(n/n_{eq} -1)} 
\over {2\sigma_{pre} ^2} }\Bigr)
\end{equation}
for $n \leq n_{eq}$ and equal to zero for $n > n_{eq}$. 
The basic assumption leading to Eq.\ (38) is that $P_{pre}$ depends
exclusively on the ratio $n/n_{eq}$ as can be deduced from the
results of B\"{o}hning \cite{Bohning} where the density of particle-hole
states is approximately described using 
a Gaussian centered at $n_{eq}$. The parameter $\sigma_{pre}$ 
is a free parameter and we assume no dependence on
excitation energy \cite{sigpre}. Our calculations of
several reactions using different values of $\sigma_{pre}$ show that
an overall reasonable agreement with available data can be obtained
using $\sigma_{pre} = 0.4$--$0.5$ (see Fig.\ 11 in Ref.\ \cite{SATIF6}). 
In CEM03.0x, we choose the fixed value $\sigma_{pre} = 0.4$ 
and use Eqs.\ (34,38) as criteria for the transition from the preequilibrium 
stage of reactions to evaporation, instead of using  the ``never-come-back"
approach along with Eq.\ (34), as was done in CEM2k.

Algorithms of many preequilibrium routines are changed and 
almost all these routines are rewritten, which has speeded up the 
code significantly. Finally, some bugs were fixed as previously mentioned.

Several examples with results from the preequilibrium model used in our 
current event generators are shown in Figs.\ 19--24.

The energy spectra
of secondary particles emitted from 96 MeV $n$ + $^{56}$Fe interactions
presented in Fig.\ 17 show that the main contribution
to the total spectra of complex particles
as calculated by CEM03.01 comes
from preequilibrium emission: The coalescence mechanism
contributes only a few percent to the total spectra
from these reactions, while
evaporation is important only in the production of low energy particles,
below $\sim 30$ MeV. 
Fig.\ 19 shows several
more detailed spectra for the same reactions, namely,
CEM03.01 results for 
double-differential spectra of  $d$ and $t$ at eight different angles
compared with the measured data \cite{Blideanu04}.
Let us recall that CEM03.0x produces
almost all these deuterons and tritons via preequilibrium emission.
We see that CEM03.0x describes quite well these 
double-differential spectra, except the very high-energy tails of
spectra at the most forward angles, where we should expect a 
contribution from direct processes like pick-up and knock-out, not
considered so far in our models.

Fig.\ 20 shows examples of angle-integrated energy spectra,
energy-integrated angular distributions,
and double-differential spectra of nucleons and complex particles
from a reaction induced %already 
by intermediate-energy protons,
namely from 62.9 MeV $p$ + Pb, as 
calculated by CEM03.01 and compared with the recent measurements
by Guertin {\it et al.} \cite{Guertin05}.               
CEM03.01 produces complex particles from this reaction also
mainly via preequilibrium emission, and we see that it describes
these experimental spectra quite well too.

Fig.\ 21 shows another example of 
double-differential spectra of complex particles
from proton-induced reactions, at higher energies, namely
CEM03.01 calculated $^4$He spectra at 20, 40, 60, 100, 120, and 140
degrees from 160 MeV $p$ + $^{27}$Al, $^{59}$Co, and $^{197}$Au
compared with the Cowley {\it et al.} data \cite{Cowley96}.
We see again a good agreement between our calculations and the
measurements; the main contribution in the production of 
these $^4$He by CEM03.01 is again from preequilibrium emission.

One more example of 
double-differential spectra of complex particles
from proton-induced reactions, at higher energies, namely
CEM03.01 calculated $^3$He and $^4$He spectra at 20, 90, and 160
degrees from 210, 300, and 480 MeV proton-silver
interactions
compared with the Green and Korteling data \cite{Green78}
is presented in Fig.\ 22.
Again a fairly good agreement between the calculations and the data
can be seen, and again 
the main contribution in the production of 
these particles by CEM03.01 is from preequilibrium emission.

Finally, Fig.\ 23 show an example of 
double-differential spectra of complex particles
from neutron-induced reactions, at even higher energies, namely
CEM03.01 calculated 
double-differential spectra of 
$p$, $d$, and $t$ 
at 54, 68, 90, 121, and 164 
degrees from interactions of
542 MeV neutrons with copper and bismuth
compared with the measurements by Franz {\it et al.} \cite{Franz90}.
We see again a  good agreement between the calculations and the data
and again the main contribution in the production of $d$ and $t$
from these reactions by CEM03.01 is from preequilibrium emission.
Let us note that we analyzed these data
in 1992 \cite{NP94}, with the version of CEM we had at that time,
CEM92m \cite{cemphys}. If we compare the agreement with
the experimental data \cite{Franz90}
of the results we got in 1992 \cite{NP94}
with our current CEM03.01 results, there is a tremendous improvement
in the description of these data.

To conclude this Section, Fig.\ 24 shows several examples of 
gas production cross-section calculations by CEM03.01
from $n + ^{56}$Fe, $p + ^{208}$Pb, and $n + ^{238}$U
reactions compared with experimental data
\cite{Haight04}--\cite{Leya05}
and results by TALYS from \cite{Raeymackers03},
McGNASH \cite{Talou06}, and IPPE-99 \cite{Ignatyuk00}.
The overall agreement of the results by CEM03.01 with these data is 
similar to that achieved by TALYS, McGNASH, and  IPPE-99.

For the production of $^4$He from $n + ^{238}$U (bottom plot in Fig.\ 24),
we show not only
the total gas production cross section,
but also contributions to the total yield
from preequilibrium emission, from events with evaporation
that are not followed by fission, from evaporation before fission 
and from
evaporation of fission fragments in events with fission,
the total evaporation component (both with and without fission),
and from coalescence, as calculated by CEM03.01. 
We see that the preequilibrium contribution to the total
yield of $^4$He from this reaction is about one order of magnitude
higher than contributions from other reaction mechanisms considered by 
our model.

\begin{figure}[ht]                                                   %Fig.\ 19

%\vspace*{-10mm}
\centering

%\hspace*{-5mm} 
\includegraphics[height=100mm,angle=-0]{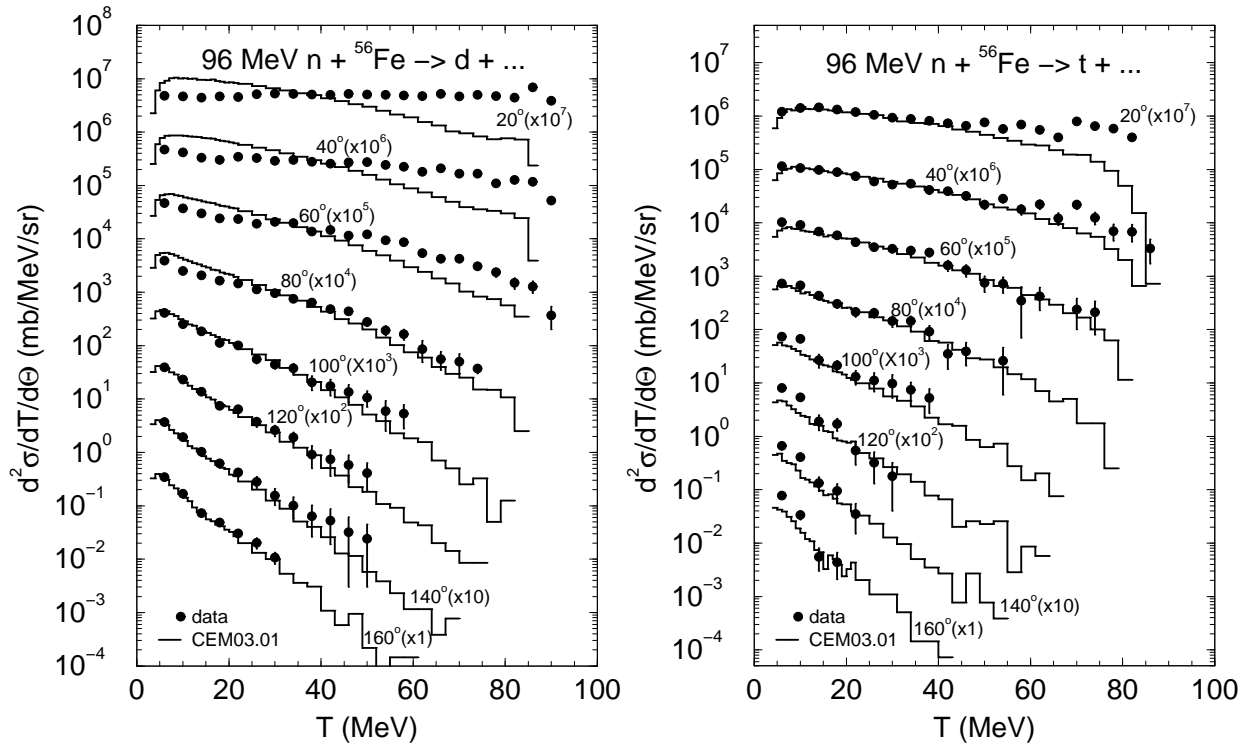}
\caption{
Double-differential 
spectra of  $d$ and $t$ at 20, 40, 60, 80, 100, 120, 140, and 160 degrees
from 96 MeV $n$ + $^{56}$Fe
calculated by CEM03.01 compared
with experimental data by Blideanu {\it et al.} \cite{Blideanu04}.
}
\end{figure}

\newpage

%\vspace*{-15mm}
\begin{figure}[ht]                                                 %Fig. 20
\centering
\includegraphics[width=170mm,angle=-0]{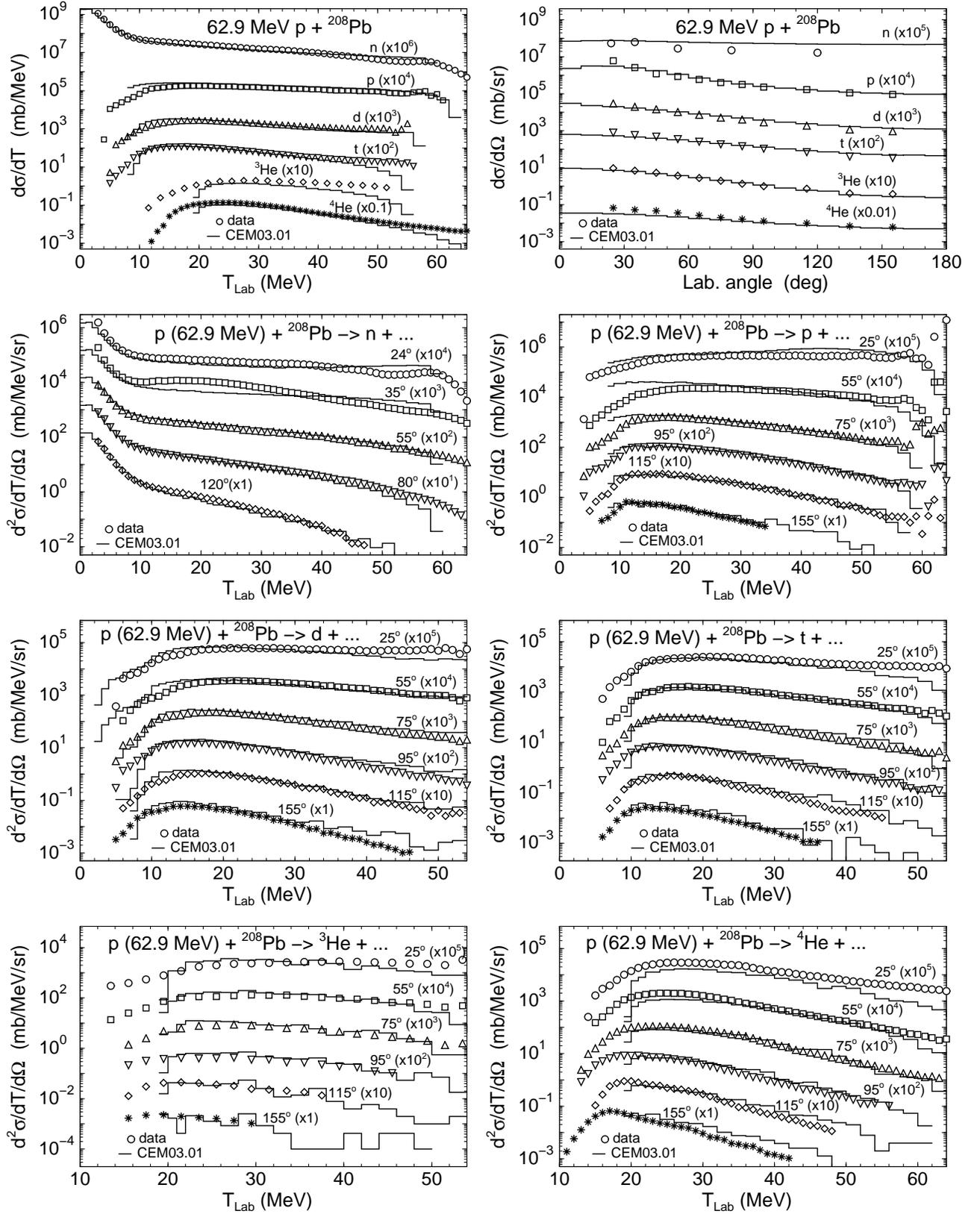}
\caption{
Experimental angle-integrated energy spectra (upper left plot),
energy-integrated angular distributions (upper right plot),
and double-differential spectra of nucleons and complex particles
from 62.9 MeV $p$ + Pb
\cite{Guertin05}               
compared with CEM03.01 results.
}

%\vspace*{-40mm}
\end{figure}
\clearpage

\begin{figure}[ht]                                                   %Fig.\ 21

\vspace*{-10mm}
\centering

\hspace*{-5mm} 
\includegraphics[height=200mm,angle=-0]{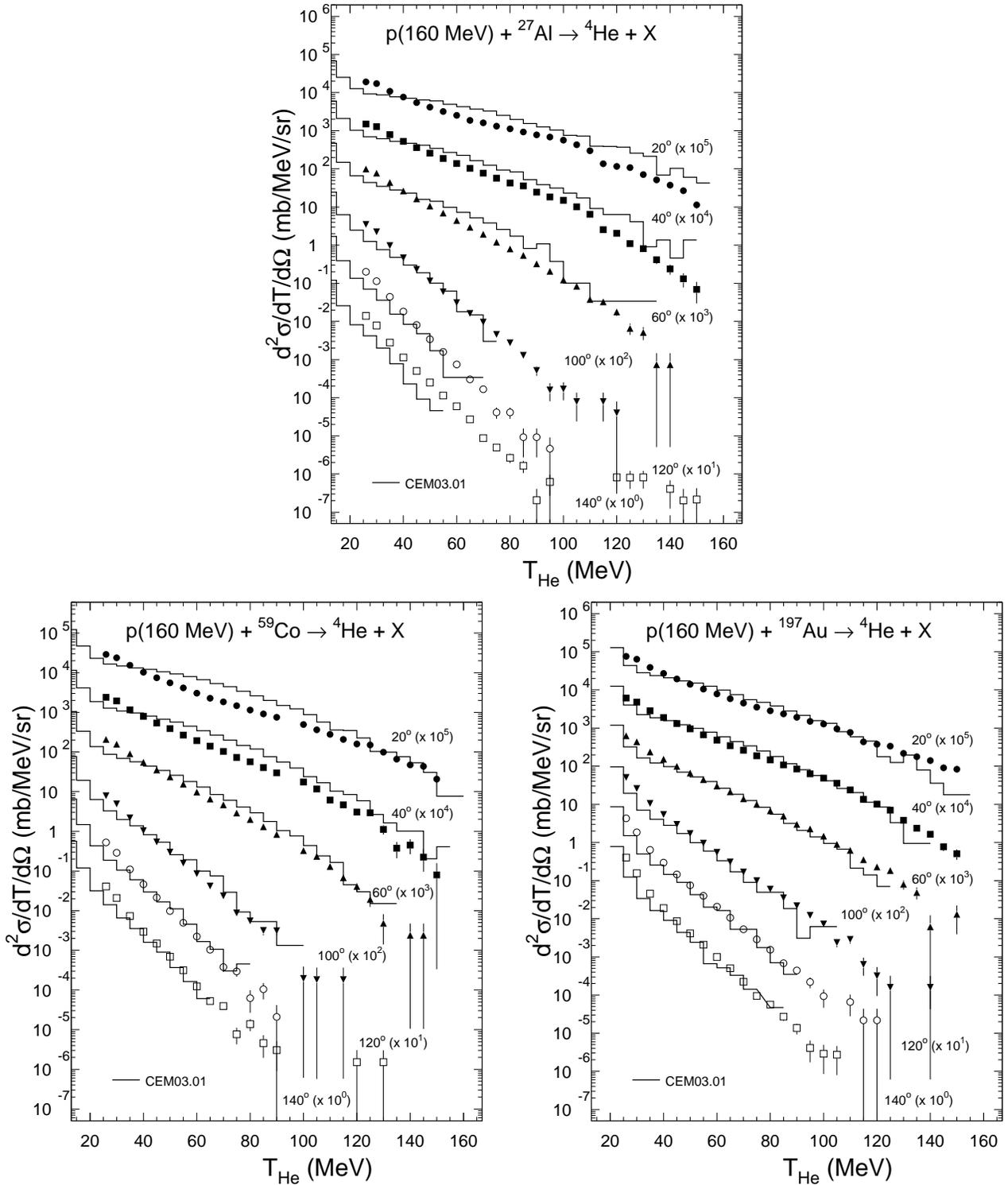}
\caption{CEM03.01 calculated
double-differential spectra of $^4$He
at 20, 40, 60, 100, 120, and 140
degrees from 160 MeV $p$ + $^{27}$Al, $^{59}$Co, and $^{197}$Au
compared with the Cowley {\it et al.} data \cite{Cowley96}.
}
\end{figure}

\clearpage

\begin{figure}[ht]                                                   %Fig.\ 22

%\vspace*{-10mm}
\centering

\hspace*{-5mm} 
\includegraphics[height=170mm,angle=-0]{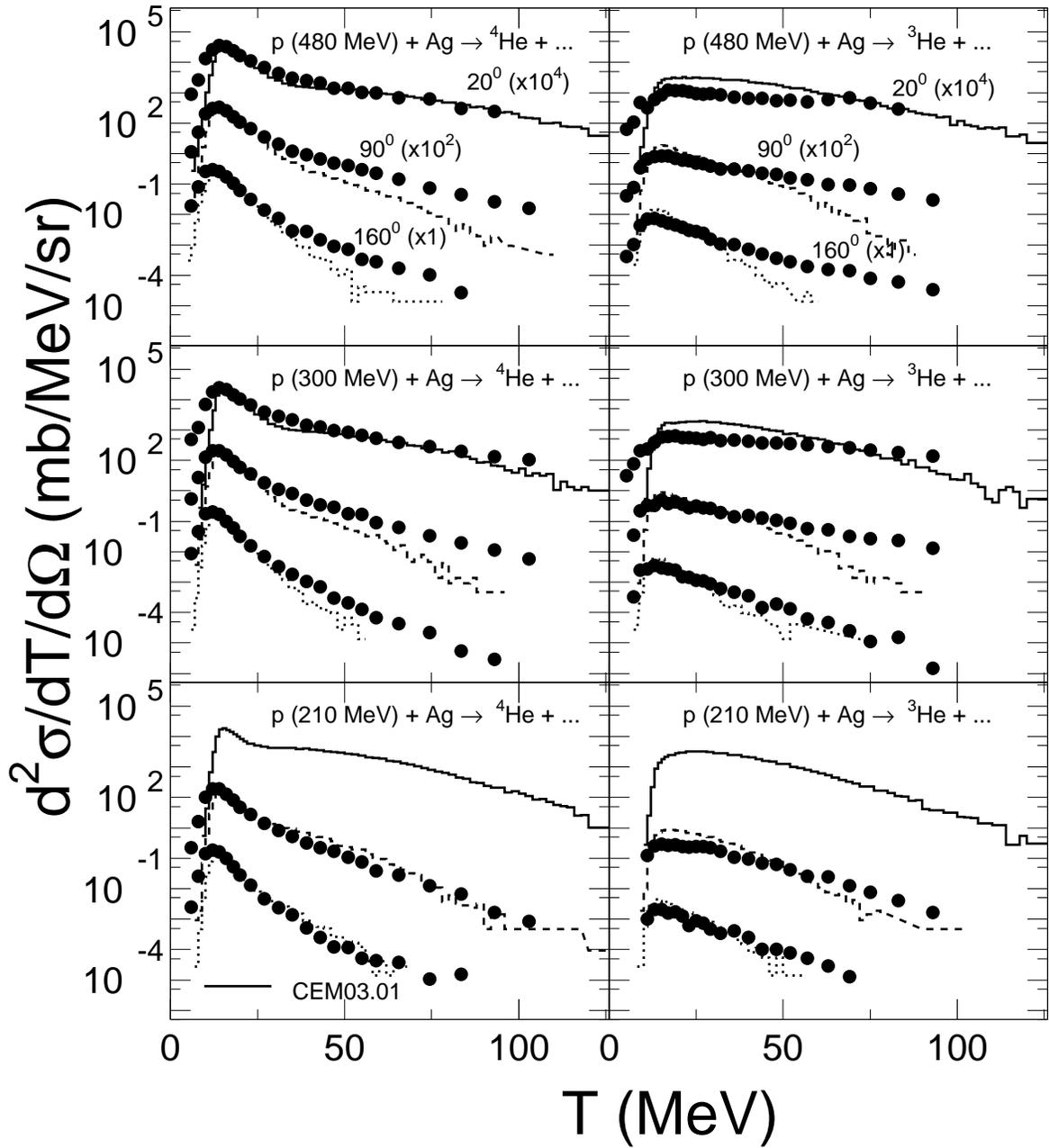}
\caption{
CEM03.01 calculated
double-differential spectra of $^3$He and  $^4$He
at 20, 90, and 160
degrees from 210, 300, and 480 MeV proton-silver
interactions
compared with the Green and Korteling data \cite{Green78}.
}
\end{figure}

\clearpage

\begin{figure}[ht]                                                   %Fig.\ 23

%\vspace*{-10mm}
\centering

\hspace*{-5mm} 
\includegraphics[height=200mm,angle=-0]{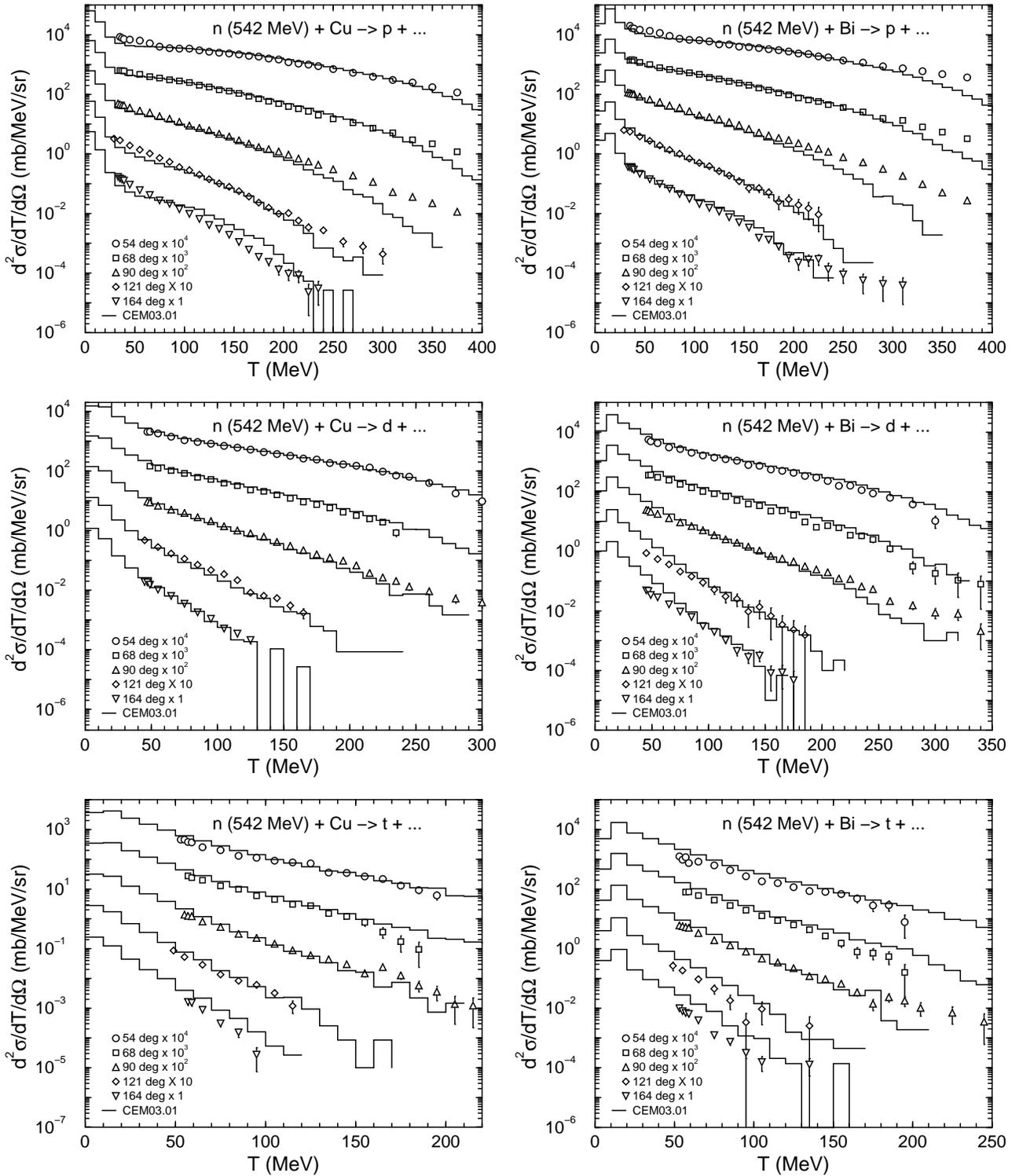}
\caption{
CEM03.01 calculated
double-differential spectra of 
$p$, $d$, and $t$ 
 at 54, 68, 90, 121, and 164 
degrees from interactions of
542 MeV neutrons with copper and bismuth
compared with the measurements by Franz {\it et al.} \cite{Franz90}.
}

\end{figure}

\clearpage

\begin{figure}[ht]                                                   %Fig.\ 24

%\vspace*{-10mm}
\centering

\hspace*{-5mm} 
\includegraphics[height=150mm,angle=-0]{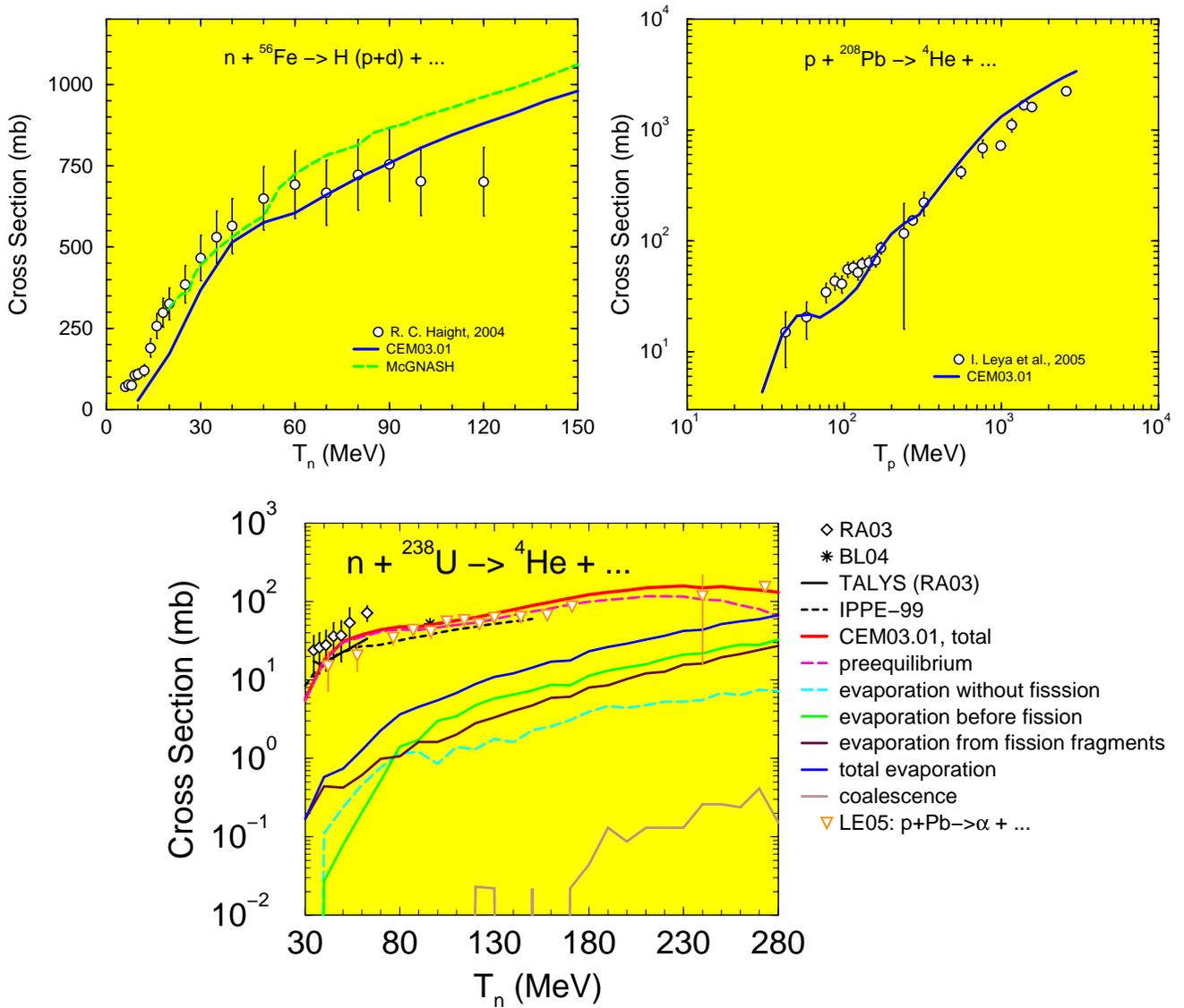}
\caption{
Examples of H ($p+d$) and $^4$He total production cross 
sections calculated by CEM03.01 compared with experimental data 
\cite{Haight04}--\cite{Leya05}
and results by TALYS from \cite{Raeymackers03},
McGNASH \cite{Talou06}, and IPPE-99 \cite{Ignatyuk00}.
For the production of $^4$He from $n + ^{238}$U (bottom plot),
we show the total cross section as predicted by CEM03.01 with
a thick red solid line, as well as contributions to
it from preequilibrium emission, from events with evaporation
that are not followed by fission, from evaporation before fission,
evaporation from fission fragments,
total evaporation (both with and without fission), and from
coalescence, shown with different color lines as indicated in the legend.}
\end{figure}

\clearpage

%\vspace{4mm}
{\large\bf 6.  Evaporation} \\

CEM03.01 and LAQGSM03.01 and their later versions
use an extension of the Generalized Evaporation Model
(GEM) code GEM2 by Furihata \cite{Furihata1}--\cite{Furihata3}
%$^{37,38}$ 
after the preequilibrium stage of reactions to describe
evaporation of nucleons, complex particles, and 
light fragments heavier than $^4$He (up to $^{28}$Mg)
from excited compound nuclei and to describe their
fission, if the compound nuclei are heavy enough 
to fission ($Z \ge 65$).
The GEM is an extension by Furihata
of the Dostrovsky evaporation model \cite{Dostrovsky}
%^{39}$ 
as implemented in LAHET \cite{LAHET}
%$^{14}$ 
to include up to 66 types of particles and fragments that
can be evaporated from an excited compound nucleus plus a modification
of the version of Atchison's fission model \cite{RAL}--\cite{RAL2007}
used in LAHET. Many 
of the parameters were adjusted by Furihata for a better description of 
fission reactions when using it in conjunction with the extended 
evaporation model. 

A very detailed description of the GEM, together with a large amount
of results obtained for many reactions
using the GEM coupled either with the Bertini or ISABEL INC models
in LAHET may be found in \cite{Furihata1,Furihata2}.
%$^{37,38}$. 
Therefore, we present here
only the main features of the GEM, following mainly
\cite{Furihata2}
%$^{38}$ 
and using as well useful information obtained in private communications 
with Dr.\ Furihata.

Furihata did not change in the GEM the general algorithms used in LAHET
to simulate evaporation and fission. The decay widths of evaporated 
particles and 
fragments are estimated using the classical Weisskopf-Ewing statistical
model \cite{weisewing}. 
%$^{40}$. 
In this approach, the decay probability $P_j$ for the 
emission of a particle $j$ from a parent compound nucleus $i$ with 
the total kinetic energy in the center-of-mass system between $\eps$ 
and $\eps + d \eps$ is
\beq
P_j(\eps) d \eps = g_j \sigma_{inv} (\eps )
{{\rho_d ( E -Q -\eps)}\over{\rho_i (E)}} \eps d \eps ,
\eeq
where $E$ [MeV] is the excitation energy of the parent nucleus $i$ with mass
$A_i$ and charge $Z_i$, and $d$ denotes a daughter nucleus with mass $A_d$ 
and charge $Z_d$ produced after the emission of ejectile $j$ with mass $A_j$ 
and charge $Z_j$ in its ground state. $\sigma_{inv}$ is the cross section for 
the inverse reaction, $\rho_i$ and $\rho_d$ are the level densities 
[MeV]$^{-1}$
of the parent and the daughter nucleus, respectively. 
$g_j = (2S_j+1) m_j/\pi^2 \hbar^2$, 
where $S_j$ is the spin and $m_j$ is the reduced mass
of the emitted particle $j$. The $Q$-value is calculated using the excess mass
$M(A,Z)$ as $Q=M(A_j,Z_j)+M(A_d,Z_d) - M(A_i,Z_i)$. In GEM2, four mass tables 
are used to calculate $Q$ values, according to the following priorities, where
a lower priority table is only used outside the range of validity of the higher
priority one:
(1) the Audi-Wapstra mass table
%$^{18}$
\cite{Audi95},
(2) theoretical masses calculated by M\"oller {\it et al.}
%$^{19}$
\cite{moller95},
(3) theoretical masses calculated by Comay {\it et al.}
%$^{41}$
\cite{Comay},
(4) the mass excess calculated using the old Cameron formula
%$^{42}$
\cite{Cameron57}.
As does LAHET, GEM2 uses Dostrovsky's formula
%$^{39}$ 
\cite{Dostrovsky}
to calculate
the inverse cross section $\sigma_{inv}$ for all emitted particles and 
fragments
\begin{equation}
\sigma_{inv} (\eps) = \sigma_{g} \alpha  \left(
1 + {\beta \over \eps} \right) \mbox{ ,}
%\label{a2}
\end{equation}
which is often written as
$$
\sigma_{inv} (\eps) =
\cases{\sigma_g c_n (1 + b/ \eps)& for neutrons \cr
\sigma_g c_j (1- V/ \eps)& for charged particles ,\cr}
$$
where $\sigma_g = \pi R^2_b$ [fm$^2$] is the geometrical cross section, and
\beq
V = k_j Z_j Z_d e^2 / R_c
\eeq
is the Coulomb barrier in MeV. 

One important new ingredient in GEM2 in comparison with LAHET, which
considers evaporation of only 6 particles (n, p, d, t, $^3$He, and $^4$He),
is that Furihata includes the possibility of evaporation of up to 66 types 
of particles and fragments and incorporates into GEM2 several alternative
sets of parameters
$b$, $c_j$, $k_j$, $R_b$, and $R_c$ for each particle type.

The 66 ejectiles considered by GEM2 for evaporation 
are selected to satisfy the following criteria:
(1) isotopes with $Z_j \leq 12$;
(2) naturally existing isotopes or isotopes near the stability line;
(3) isotopes with half-lives longer than 1 ms. All the 66 ejectiles
considered by GEM2 are shown in Table 1.
%n, p, d, t, $^{3,4,6,8}$He, $^{6--9}$Li, $^{7,9--12}$Be, $^{8,10--13}$B,
%$^{10--16}$C, $^{12--17}$N, $^{14--20}$O, $^{17--21}$F, $^{18--24}$Ne,
%$^{21--25}$Na, and $^{22--28}$Mg.

%\newpage
\begin{center}
Table 1. The evaporated particles considered by GEM2

\vspace{2mm}
\begin{tabular}{rlllllll}
\hline\hline 
 $Z_j$\hspace{2mm} & \multicolumn{7}{l} {Ejectiles} \\
\hline
0\hspace{2mm}  & n       &         &         &         &         &         &         \\
1\hspace{2mm}  & p       &\hspace{1mm}   d     &\hspace{1mm}   t     &         &         &         &         \\
2\hspace{2mm}  &$^{3 }$He&\hspace{1mm}$^{4 }$He&\hspace{1mm}$^{6 }$He&\hspace{1mm}$^{8 }$He&         &         &         \\
3\hspace{2mm}  &$^{6 }$Li&\hspace{1mm}$^{7 }$Li&\hspace{1mm}$^{8 }$Li&\hspace{1mm}$^{9 }$Li&         &         &         \\
4\hspace{2mm}  &$^{7 }$Be&\hspace{1mm}$^{9 }$Be&\hspace{1mm}$^{10}$Be&\hspace{1mm}$^{11}$Be&\hspace{1mm}$^{12}$Be&         &         \\
5\hspace{2mm}  &$^{8 }$B &\hspace{1mm}$^{10}$B &\hspace{1mm}$^{11}$B &\hspace{1mm}$^{12}$B &$\hspace{1mm}^{13}$B &         &         \\
6\hspace{2mm}  &$^{10}$C &\hspace{1mm}$^{11}$C &\hspace{1mm}$^{12}$C &\hspace{1mm}$^{13}$C &\hspace{1mm}$^{14}$C &\hspace{1mm}$^{15}$C &\hspace{1mm}$^{16}$C \\
7\hspace{2mm}  &$^{12}$N &\hspace{1mm}$^{13}$N &\hspace{1mm}$^{14}$N &\hspace{1mm}$^{15}$N &\hspace{1mm}$^{16}$N &\hspace{1mm}$^{17}$N &         \\
8\hspace{2mm}  &$^{14}$O &\hspace{1mm}$^{15}$O &\hspace{1mm}$^{16}$O &\hspace{1mm}$^{17}$O &\hspace{1mm}$^{18}$O &\hspace{1mm}$^{19}$O &\hspace{1mm}$^{20}$O \\
9\hspace{2mm}  &$^{17}$F &\hspace{1mm}$^{18}$F &\hspace{1mm}$^{19}$F &\hspace{1mm}$^{20}$F &\hspace{1mm}$^{21}$F &         &         \\
10\hspace{2mm} &$^{18}$Ne&\hspace{1mm}$^{19}$Ne&\hspace{1mm}$^{20}$Ne&\hspace{1mm}$^{21}$Ne&\hspace{1mm}$^{22}$Ne&\hspace{1mm}$^{23}$Ne&\hspace{1mm}$^{24}$Ne\\
11\hspace{2mm} &$^{21}$Na&\hspace{1mm}$^{22}$Na&\hspace{1mm}$^{23}$Na&\hspace{1mm}$^{24}$Na&\hspace{1mm}$^{25}$Na&         &         \\
12\hspace{2mm} &$^{22}$Mg&\hspace{1mm}$^{23}$Mg&\hspace{1mm}$^{24}$Mg&
\hspace{1mm}$^{25}$Mg&\hspace{1mm}$^{26}$Mg&\hspace{1mm}$^{27}$Mg&
\hspace{1mm}$^{28}$Mg\\
\hline\hline 
\end{tabular}
\end{center}

GEM2 includes several options for the parameter set in expressions (40,41):

1) The ``simple'' parameter set is given as $c_n = c_j =k_j = 1$,
$b=0$, and $R_b  = R_c = r_0 (A^{1/3}_j + A^{1/3}_d)$ [fm];
users need to input $r_0$.

2) The ``precise'' parameter set is used in GEM2 as the default,
and we use this set in our present work.

A) For all light ejectiles up to $\alpha$ ($A_j \leq 4$), the
parameters determined by Dostrovsky {\it et al.}\ \cite{Dostrovsky}
%$^{39}$ 
are used in GEM2, namely:
$c_n = 0.76 + c_a A_d^{-1/3}$, $b=(b_a A^{-2/3}_d - 0.050)/
(0.76 + c_a A^{-1/3}_d)$
(and $b=0$ for $A_d \geq 192$), where $c_a = 1.93$ and $b_a = 1.66$,
$c_p = 1 + c$, $c_d = 1 + c/2$, $c_t = 1 + c/3$, $c_{^3He} = c_\alpha = 0$,
$k_p = k$, $k_d = k + 0.06$, $k_t = k + 0.12$,
$k_{^3He} = k_\alpha - 0.06$, where $c$, $k$, and $k_\alpha$ are listed
in Table 2 for a set of $Z_d$. Between the $Z_d$ values listed in Table 2,
$c$, $k$, and $k_\alpha$ are interpolated linearly.The nuclear distances 
are given by $R_b = 1.5 A^{1/3}$ for neutrons and protons, and
$1.5 (A^{1/3}_d + A^{1/3}_j)$ for d, t, $^3$He, and $\alpha$.

\begin{center}
Table 2. $k$, $k_\alpha$, and $c$ parameters used in GEM2

\vspace{2mm}
\begin{tabular}{|c|c|c|c|}
\hline \hline
\hspace{5mm} $Z_d$\hspace{5mm} &\hspace{5mm} $k$\hspace{5mm} &\hspace{5mm}
 $ k_\alpha$\hspace{5mm} &\hspace{5mm} $c$\hspace{5mm}  \\
\hline
$\leq 20$ & 0.51 & 0.81 & 0.0 \\
30 & 0.60 & 0.85 & -0.06 \\
40 & 0.66 & 0.89 & -0.10 \\
$\geq 50$ & 0.68 & 0.93 & -0.10\\
\hline \hline
\end{tabular}
\end{center}

The nuclear distance for the Coulomb barrier is expressed as
$R_c = R_d + R_j$, where $R_d = r^c_0 A^{1/3}$, $r^c_0 = 1.7$,
and $R_j=0$ for neutrons and protons, and $R_j = 1.2$
for d, t, $^3$He, and $^4$He.  We note that several of these parameters 
are similar to the original values published by
Dostrovsky {\it et al.}\ \cite{Dostrovsky}
%$^{39}$ 
but not exactly the same. Dostrovsky {\it et al.}\ \cite{Dostrovsky}
had $c_a = 2.2$, 
$b_a = 2.12$, and $r^c_0 = 1.5$. Also, for the $k$, $k_\alpha$, and $c$
parameters shown in Table 2, they had slightly different values, shown 
in Table 3. 

\begin{center}
Table 3. $k_p$, $c_p$, $k_\alpha$, and $c_\alpha$ parameters from 
Ref.\ \cite{Dostrovsky}

\vspace{2mm}
\begin{tabular}{|c|c|c|c|c|}
\hline \hline
\hspace{5mm} $Z_d$\hspace{5mm} &\hspace{5mm} $k_p$\hspace{5mm} &\hspace{5mm}
 $c_p$\hspace{5mm} & $ k_\alpha$\hspace{5mm} &\hspace{5mm} $c_\alpha$\hspace{5mm}  \\
\hline
10        & 0.42 & 0.50 & 0.68 & 0.10\\
20        & 0.58 & 0.28 & 0.82 & 0.10\\
30        & 0.68 & 0.20 & 0.91 & 0.10 \\
50        & 0.77 & 0.15 & 0.97 & 0.08 \\
$\geq 70$ & 0.80 & 0.10 & 0.98 & 0.06\\
\hline \hline
\end{tabular}
\end{center}

B) For fragments heavier than $\alpha$ ($A_j \geq 4$),
the ``precise'' parameters 
of GEM2 use values by Matsuse {\it et al.}\ \cite{Matsuse82}, namely:
$c_j = k = 1$,
$R_b = R_0(A_j) + R_0(A_d) + 2.85$ [fm],
$R_c = R_0(A_j) + R_0(A_d) + 3.75$ [fm], where
$R_0(A) = 1.12A^{1/3} - 0.86 A^{-1/3}$.

3) The code GEM2 
contains two other options for the parameters of the inverse cross sections.

A) A set of parameters due to Furihata for light ejectiles in combination
with Matsuse's parameters for fragments heavier than $\alpha$. Furihata 
and Nakamura 
determined $k_j$ for p, d, t, $^3$He, and $\alpha$ as follows
\cite{Furihata3}:
%$^{44}$
$$k_j = c_1 \log (Z_d) + c_2 \log (A_d) + c_3 .$$
The coefficients $c_1$, $c_2$, and $c_3$ for each ejectile are shown 
in Table 4.

\begin{center}
Table 4. $c_1$, $c_2$, and $c_3$ for p, d, t, $^3$He, and $\alpha$ from
%$^{44}$
\cite{Furihata3}\\
\vspace{2mm}
\begin{tabular}{|c|c|c|c|}
\hline \hline
\hspace{5mm} Ejectile\hspace{5mm} &\hspace{5mm} $c_1$\hspace{5mm} &\hspace{5mm}
 $ c_2$\hspace{5mm} &\hspace{5mm} $c_3$\hspace{5mm}  \\
\hline
p        & 0.0615 & 0.0167 & 0.3227\\
d        & 0.0556 & 0.0135 & 0.4067\\
t        & 0.0530 & 0.0134 & 0.4374 \\
$^3$He   & 0.0484 & 0.0122 & 0.4938 \\
$\alpha$ & 0.0468 & 0.0122 & 0.5120\\
\hline \hline
\end{tabular}
\end{center}

When these parameters are chosen in GEM2, the following nuclear
radius $R$ is used in the calculation of $V$ and $\sigma_g$:
$$
R =\cases{0&for  $A=1$ ,\cr
   1.2&for $2\leq A \leq 4$ ,\cr
   2.02&for $5\leq A \leq 6$ ,\cr
   2.42&for $A=7$ ,\cr
   2.83&for $A=8$ ,\cr
   3.25&for $A=9$ ,\cr
   1.414A^{1/3}_d + 1&for $A \geq 10$ .\cr}
$$
B) The second new option in GEM2 is to use 
Furihata's parameters for light ejectiles up to $\alpha$ and the
Botvina {\it et al.}\ \cite{Botvina87}
parameterization for inverse cross sections
for heavier ejectiles. Botvina {\it et al.}\ \cite{Botvina87}
found that
$\sigma_{inv}$ can be expressed as
\beq
\sigma_{inv} = \sigma_g \cases{(1 - V / \eps)&for  $\eps \geq V + 1$ [MeV],\cr
\exp[{\alpha (\eps - V - 1)] / (V+1)}&for $\eps < V + 1$ [MeV],\cr}
\eeq
where
$$ \alpha = 0.869 + 9.91 / Z_j ,$$
$$ V = {Z_j Z_d \over {r_0^b (A^{1/3}_j + A^{1/3}_d)}} ,$$
$$r_0^b = 2.173 {{1+6.103 \times 10^{-3}Z_jZ_d}
\over{1 + 9.443 \times 10^{-3}Z_j Z_d}} \mbox{ [fm].}$$

The expression of $\sigma_{inv}$ for $\eps < V + 1$ shows the fusion reaction
in the sub-barrier region. When using Eq.\ (42) instead of Eq.\ (40), the total
decay width for a fragment emission can not be calculated analytically.
Therefore, the total decay width must be calculated numerically and takes 
much CPU time.

The total decay width $\Gamma_j$ is calculated by integrating Eq.\ (39) with 
respect to the total kinetic energy $\eps$ from the Coulomb barrier $V$ 
up to the
maximum possible value, $(E-Q)$. The good feature of Dostrovsky's 
approximation for the inverse cross sections, Eq.\ (40),  
is its simple energy dependence that allows the analytic integration of 
Eq.\ (39).
By using Eq.\ (40) for $\sigma_{inv}$, the total decay width for the particle 
emission is
\beq
\Gamma_j = {{g_j\sigma_g \alpha}\over{\rho_i(E)}}
\int^{E-Q}_V \eps \Bigl(1 + {\beta\over\eps} \Bigr)
\rho_d (E -Q-\eps) d \eps .
\eeq
The level density $\rho(E)$ is calculated in GEM2 according to the Fermi-gas
model using the expression
%$^{46}$
\cite{GilbertCameron}
\beq
\rho(E) = { \pi \over {12} } 
{ {\exp ({2\sqrt{a(E-\delta)}}) } \over {a^{1/4}(E-\delta)^{5/4}} } ,
\eeq
where $a$ is the level-density parameter and $\delta$ is the pairing energy
in MeV. As does LAHET, GEM2 uses the $\delta$ values evaluated by Cook
{\it et al.}\ \cite{Cook67}.
For those values not evaluated by Cook {\it et al.},
$\delta$'s from Gilbert and Cameron
\cite{GilbertCameron}
%$^{46}$ 
are used instead. The simplest option for the level-density parameter 
in GEM2 is $a = A_d /8$ [MeV$^{-1}$], but the
default is the Gilbert-Cameron-Cook-Ignatyuk (GCCI) parameterization from
LAHET \cite{LAHET}:
%$^{14}$
\beq
a = \tilde a {{1 - e^{-u}} \over {u}} +
a_I\Biggl( 1 - {{1 - e^{-u}} \over {u}}\Biggr),
\eeq
where $u=0.05(E-\delta)$, and
$$a_I = (0.1375 - 8.36 \times 10^{-5} A_d) \times A_d ,$$
$$
\tilde a = \cases{ A_d/8&for  $Z_d < 9$ or $N_d < 9$,\cr
 A_d(a' + 0.00917 S)&for others.\cr}
$$
For deformed nuclei with 
$54 \leq Z_d \leq 78$, 
$86 \leq Z_d \leq 98$,
$86 \leq N_d \leq 122$,
or $130\leq N_d \leq 150$,
$a' = 0.12$ while $a' = 0.142$ for other nuclei. The shell corrections $S$ 
is expressed as a sum of separate contributions from neutrons and protons,
{\it i.e.} $S = S(Z_d) + S(N_d)$ from \cite{GilbertCameron,Cook67}
%$^{46,47}$ 
and are tabulated in \cite{Furihata1}. 
%$^{38}$.

The level density is calculated using Eq.\ (44) only for high excitation
energies, $E \geq E_x$, where $E_x = U_x + \delta$ and $U_x = 2.5 + 150/A_d$
(all energies are in MeV). At lower excitation energies,
the following \cite{GilbertCameron}
%$^{46}$ 
is used for the level density:
\beq
\rho(E) = {\pi \over {12} } {1 \over T} \exp((E-E_0)/T) ,
\eeq
where $T$ is the nuclear temperature defined as $1/T = \sqrt{a/U_x} - 
1.5/U_x$. To provide a smooth connection of Eqs.\ (44) and (46) at $E=E_x$,
$E_0$ is defined as 
$E_0 = E_x - T(\log T - 0.25 \log a - 1.25 \log U_x + 2 \sqrt{aU_x})$.

For $E-Q-V < E_x$, substituting Eq.\ (46) into Eq.\ (44) we can calculate the
integral analytically, if we neglect the dependence of the level-density 
parameter $a$ on $E$:
\beq
\Gamma_j = { {\pi g_j \sigma_g \alpha} \over {12 \rho_i (E)} } \nonumber \\ 
\{I_1(t,t) + (\beta + V) I_0(t)\} , \nonumber
\eeq
where $I_0(t)$ and $I_1(t,t_x)$ are expressed as
\bea
I_0(t) &=& e^{-E_0/T}  (e^t - 1) , \nonumber \\
I_1(t,t_x) &=& e^{-E_0/T}  T\{(t-t_x + 1) e^{t_x} -t -1\} ,  \nonumber 
\eea
where $t = (E-Q-V)/T$ and $t_x = E_x/T$.
For $E-Q-V \geq E_x$, the integral of Eq.\ (43) cannot be solved analytically 
because of the denominator in Eq.\ (44). However, it is approximated as
\beq
\Gamma_j = { {\pi g_j \sigma_g \alpha} \over {12 \rho_i (E)} }
[ I_1(t,t_x) + I_3(s,s_x)e^s + (\beta + V) 
 \{ I_0(t_x) - I_2(s,s_x) e^s \} ] ,
\eeq
where $I_2(s,s_x)$ and $I_3(s,s_x)$ are given by
$$
I_2(s,s_x) = 2 \sqrt{2} \{s^{-3/2} + 1.5 s^{-5/2} + 3.75 s^{-7/2} 
- (s^{-3/2}_x + 1.5 s^{-5/2}_x + 3.75s^{-7/2}_x) e^{s_x -s} \} , 
$$
\bea
I_3(s,s_x) 
&=& (\sqrt{2}a)^{-1} [ 2 s^{-1/2} + 4 s^{-3/2} + 13.5s^{-5/2} 
+ 60.0s^{-7/2} + 325.125s^{-9/2}  \nonumber \\
&-& \{(s^2-s^2_x)s^{-3/2}_x + (1.5s^2 + 0.5 s^2_x)s^{-5/2}_x 
+ (3.75s^2 + 0.25s^2_x) s^{-7/2}_x 
+ (12.875s^2  \nonumber \\
&+& 0.625s^2_x)s^{-9/2}_x 
+ (59.0625s^2 + 0.9375s^2_x) s^{-11/2}_x 
+ (324.8s^2_x + 3.28s^2_x ) s^{-13/2}_x \} e^{s_x - s} ], \nonumber 
\eea
with $s=2\sqrt{a(E-Q-V-\delta)}$ and $s_x = 2 \sqrt{a(E_x-\delta)}$.   

The particle type $j$ to be evaporated is selected in GEM2 by the 
Monte-Carlo
method according to the probability distribution calculated as 
$P_j = \Gamma_j / \sum_j \Gamma_j$, where $\Gamma_j$ is given by
Eqs.\ (47) or (48). The total kinetic energy $\eps$ of the emitted
particle $j$ and the recoil energy of the daughter nucleus is chosen 
according to the probability distribution given by Eq.\ (39). The angular
distribution of ejectiles is simulated to be isotropic in the 
center-of-mass system.

According to Friedman and Lynch \cite{Friedman83},
%$^{48}$,
it is important to include excited states in the particle emitted via the 
evaporation process along with evaporation of particles in their ground states,
because it greatly enhances the yield of heavy particles.
Taking this into consideration, GEM2 includes evaporation of complex 
particles and
light fragments both in the ground states and excited states. An excited
state of a fragment is included in calculations if its half-life
$T_{1/2}(s)$
satisfies the following condition:
\beq
{ T_{1/2} \over {\ln 2} } > { \hbar \over {\Gamma_j^*} } ,
\eeq
where $\Gamma_j^*$ is the decay width of the excited particle (resonance).
GEM2 calculates  $\Gamma_j^*$ in the same manner as for a ground-state particle 
emission.
The $Q$-value for the resonance emission is expressed as $Q^* = Q + E^*_j$,
where $E^*_J$ is the excitation energy of the resonance. The spin state of
the resonance $S^*_j$ is used in the calculation of $g_j$, instead of the
spin of the ground state $S_j$. GEM2 uses the ground state masses $m_j$ for
excited states because the difference between the masses is negligible.

Instead of treating a resonance as an independent particle, GEM2 simply 
enhances the
decay width $\Gamma_j$ of the ground state particle emission as follows:
\beq
\Gamma_j = \Gamma^0_j + \sum_{n} \Gamma^n_j ,
\eeq
where $\Gamma^0_j$ is the decay width of the ground state particle
emission, and $\Gamma^n_j$ is that of the $n$th excited state 
of the particle $j$ emission which satisfies Eq.\ (49).

The total-kinetic-energy distribution of the excited particles is 
assumed to be the same as that of the ground-state particle.
$S^*_j$, $E^*_j$, and $T_{1/2}$ used in GEM2 are extracted from the
Evaluated Nuclear Structure Data File (ENSDF) database 
maintained by the National Nuclear Data Center at Brookhaven National 
Laboratory \cite{ENSDF}.
%$^{49}$. 

Note that when including evaporation of up to 66 particles in GEM2,
its running time increases significantly compared to the case when
evaporating only 6 particles, up to $^4$He. The major particles emitted 
from
an excited nucleus are n, p, d, t, $^3$He, and $^4$He. For most
cases, the total emission probability of particles heavier than $\alpha$
is negligible compared to those for the emission of light ejectiles. 
Our detailed study of different reactions (see, {\it e.g.}, \cite{TRAMU}
and references therein) shows that
if we study only nucleon and complex-particle spectra or only
spallation and fission products and are not interested in light fragments, 
we can consider evaporation of only 6 types of particles in GEM2 and save
much time, getting results very close
to the ones calculated with the more time consuming ``66" option.
In CEM03.01 and LAQGSM03.01, we have introduced an input parameter called
{\bf nevtype} that defines the number of types of particles to
be considered at the evaporation stage. The index of each
type of particle that can be evaporated corresponds to the particle
arangement in Table 1, with values, {\it e.g.}, of 1, 2, 3, 4, 5, and 6  
for n, p, d, t, $^3$He, and $^4$He, with succeeding values
up to 66 for $^{28}$Mg. All 66 particles that can
possibly evaporate are listed in CEM03.01 and
LAQGSM03.01 together with their
mass number, charge, and spin values in the {\bf block data bdejc}.
For all ten examples of inputs and outputs of CEM03.01 included in
Appendices 1 and 2 of the CEM03.01 User Manual \cite{CEM03.01}, 
whose results are plotted in the figures in Appendix 3 of \cite{CEM03.01},
we have performed calculations taking into account only 6 types
of evaporated particles ({\bf nevtype = 6}) as well as with the ``66"
option ({\bf nevtype = 66}) and we provide the corresponding computing time
for these examples in the captions to the appropriate figures shown in 
Appendix 3 of Ref.\ \cite{CEM03.01}. 
The ``6" option can be up to  several times faster than the 
``66" option,
providing meanwhile almost the same results. Therefore we recommend
that users of CEM03.01 and LAQGSM03.01 
use 66 for the value of the input parameter
{\bf nevtype} only when they are interested in all fragments heavier than
$^4$He; otherwise, we recommend the default value of 6 for {\bf nevtype}, 
saving computing time. Alternatively, users
may choose intermediate values of {\bf nevtype}, for example 9 if one wants
to calculate the production of $^6$Li, or 14 for modeling the production
of $^9$Be and lighter fragments and nucleons only, while still saving 
computing time compared to running the code with the maximum value of 66. 

Examples of calculation by CEM03.01 the reactions
800 MeV $p$ + $^{197}$Au and 1 GeV $p$
+ $^{56}$Fe using the ``6'' and ``66'' options are shown in 
Figs.\ 25 and 26.

\newpage

\begin{figure}[ht]                                                 %Fig. 25
\centering
\vspace*{-3mm}
\includegraphics[width=146mm,angle=-0]{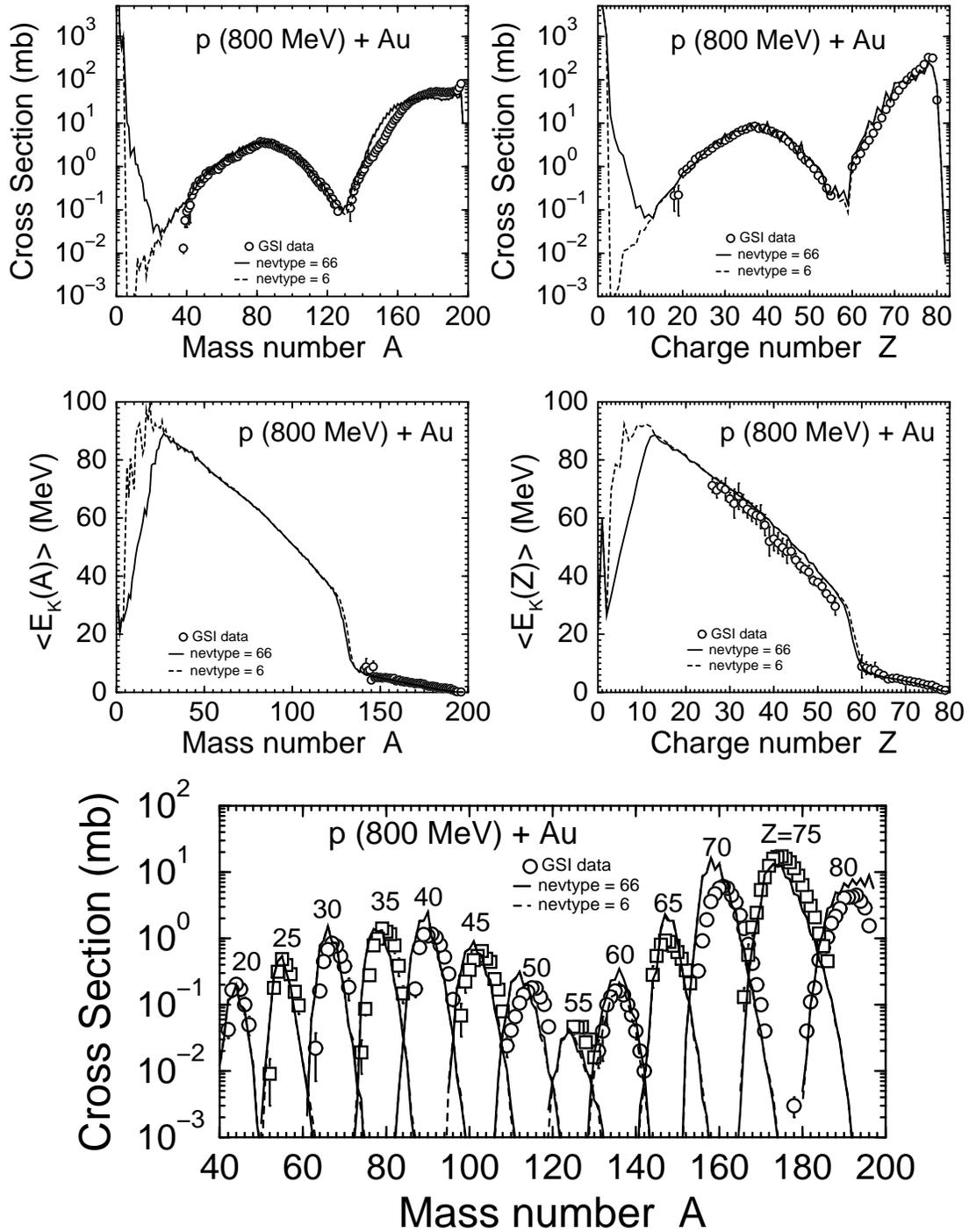}
\caption{
The measured \cite{Rejmund01} mass and charge 
distributions of the product yields from the reaction 
800 MeV/A $^{197}$Au+p and of the mean kinetic 
energy of these products, and the mass distributions
of the cross sections for the production of 
thirteen elements with the charge Z from 20 to 80 (open symbols)
compared with CEM03.01 results.
The results shown in this figure are for ten million 
simulated inelastic events. 
The {\bf nevtype=66} option requires
53 hr 34 min 31 sec of computing time on a SunBlade 100, 500 MHz
computer, while the {\bf nevtype=6} option requires only 12 hr 48 min 3 sec,
providing almost the same results for the spallation and fission products.
The fragment ($2<Z<13$, $6<A<29$) results are very different, therefore
we need to use the option {\bf nevtype=66} when we are interested in
fragment production.
}

\vspace*{-50mm}
\end{figure}

\newpage
\begin{figure}[ht]                                                 %Fig. 26
\centering
\vspace*{-3mm}
\includegraphics[width=150mm,angle=-0]{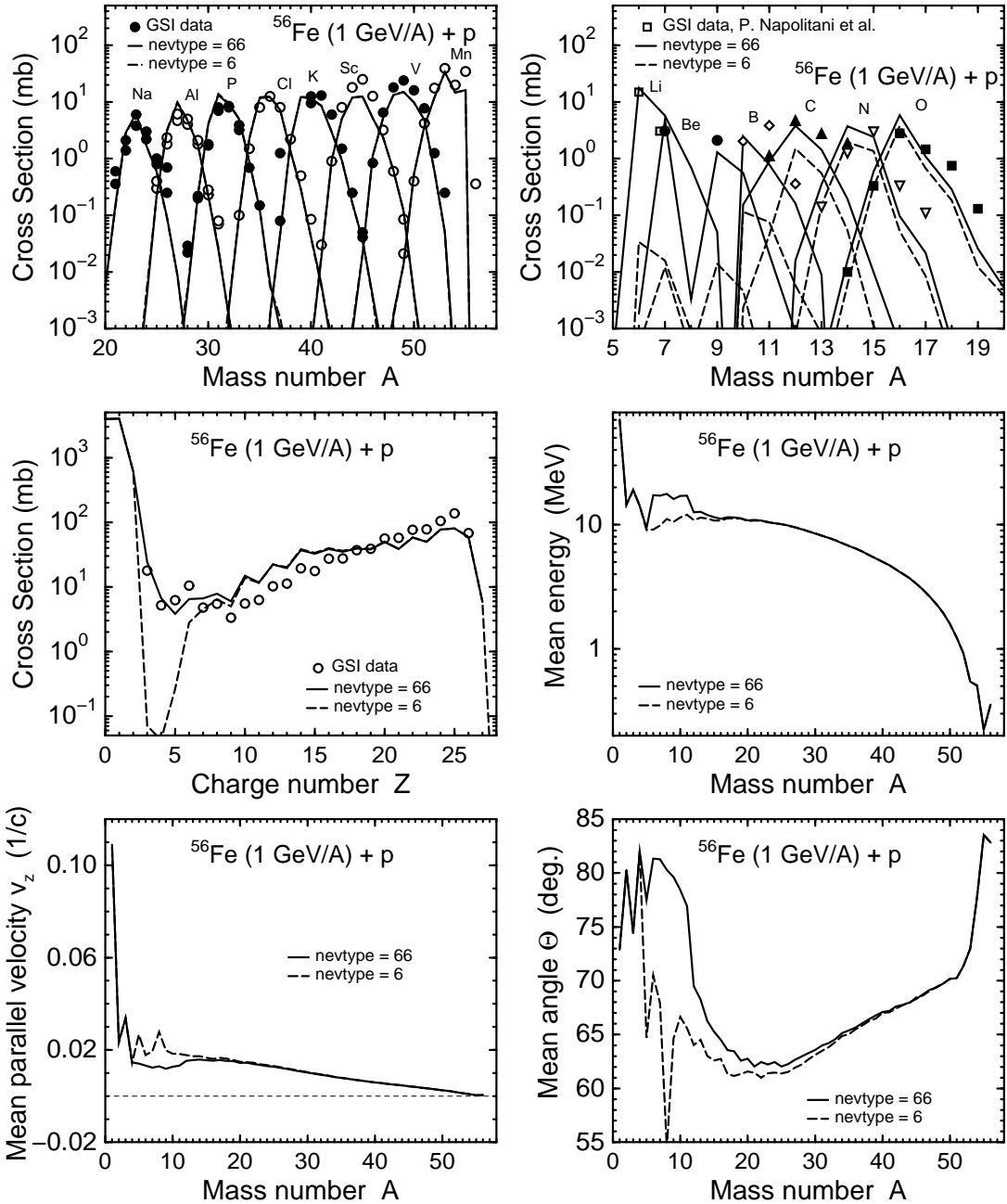}
\caption{
Experimental mass distributions 
of the yields of eight isotopes from Na to Mn  \cite{Villagrasa}
and of all light fragments from Li to O \cite{Napolitani}
from the reaction 1 GeV/A $^{56}$Fe+p and 
the charge distribution of the product yield
compared with CEM03.01 results.
The results shown in this figure are for ten million 
simulated inelastic events. 
The {\bf nevtype=66} option requires
27 hr 38 min 37 sec of computing time on a SunBlade 100, 500 MHz
computer, while the {\bf nevtype=6} option requires only 8 hr 27 
min 48 sec,
providing almost the same results for the spallation products.
The yields of light fragments, especially of Li and Be, differ by several
orders of magnitude, therefore
we need to use the option {\bf nevtype $>$ 6} when we are interested in 
light-fragment production.
}

\vspace*{-50mm}
\end{figure}

\clearpage

Let us note that increasing the number of types of particles that can 
be evaporated from 6, employed by all old versions of CEM and LAQGSM which
use our own evaporation model based on the Weisskopf-Ewing 
statistical theory of particle evaporation~\cite{weisewing} 
and consider some newer features according to
Ref.\ \cite{iljinov92}
(see details, {\ e.g.}, in \cite{CEM95}),
to 66 in the 03.01 and later versions, where we replaced
our evaporation model with a modification of GEM2 by Furihata
\cite{Furihata1}--\cite{Furihata3} as described above, together
with considering the Fermi breakup model to disintegrate 
exited nuclei with $A < 13$ in the new versions of our codes,
allows us to improve considerably
the description of fragment yields from various nuclear reactions.
Two examples with such results are shown in Figs.\ 27 and 28.

Fig.\ 27 shows an example of an excitation function of interest
to space applications used to estimate the risk assessment for 
electronic devices (SEUs and MBUs), namely $^{28}$Si(p,x)$^7$Be,
calculated with an already quite old version of CEM, CEM2k \cite{CEM2k},
and with our recent versions CEM03.01 \cite{CEM03.01} and
CEM03.02 \cite{CEM03.02} compared with 
%available 
experimental data from our T-16 compilation \cite{T16comp}.
CEM2k, which does not consider evaporation of up to 66
types of particles and the Fermi breakup
of light excited nuclei, underestimates these data by 
%a factor of 
two orders of magnitude, providing only a tiny production
of $^7$Be, only at high incident energies as final residual 
nuclei produced at the end of the reaction after  
INC, preequilibrium, and evaporation.
The new versions of our codes describe this reaction
much better than CEM2k,
though there is still room for improvement and we plan a
further improvement of the description of fragment production
by our codes.

\begin{figure}[ht]                                                 %Fig. 27
\centering
\includegraphics[width=140mm,angle=-0]{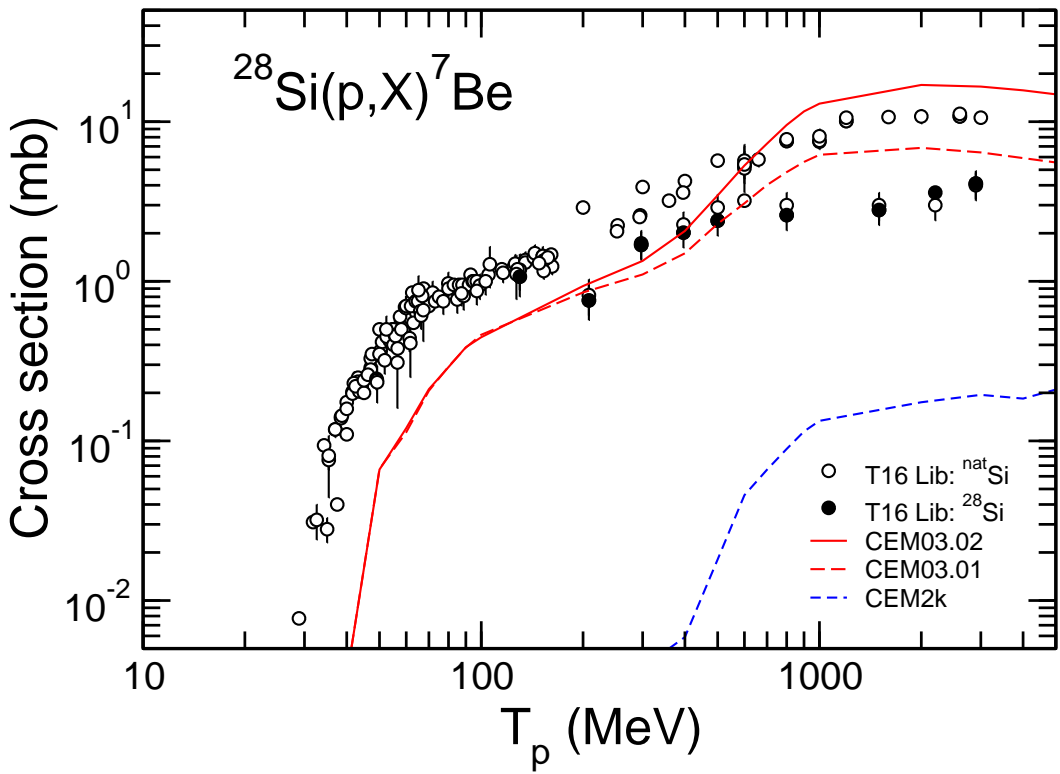}
\caption{Comparison of the 
$^{28}$Si(p,X)$^7$Be excitation function
calculated by CEM2k \cite{CEM2k},
CEM03.01 \cite{CEM03.01},
and CEM03.02 \cite{CEM03.02} 
with 
%available 
experimental data 
from our T-16 compilation \cite{T16comp}.
}
\end{figure}

Fig.\ 28 compares results from CEM03.01 \cite{CEM03.01}
and from its ``S'' and ``G'' modifications
described briefly in Section 10, CEM03.S1 and CEM03.G1 \cite{01s1g1},
for the total production cross sections of Li and Be isotopes
produced in interactions of 1.2 GeV protons with thirteen target nuclei
from Al to Th, measured recently at the Cooler Synchrotron
Facility COSY of the Forschungszentrum J\"{u}lich \cite{Herbach06}.

\begin{figure}[ht]                                                 %Fig.\ 28
\vspace*{3mm}
\centering
\hspace*{-5mm}
\includegraphics[height=92mm,angle=-0]{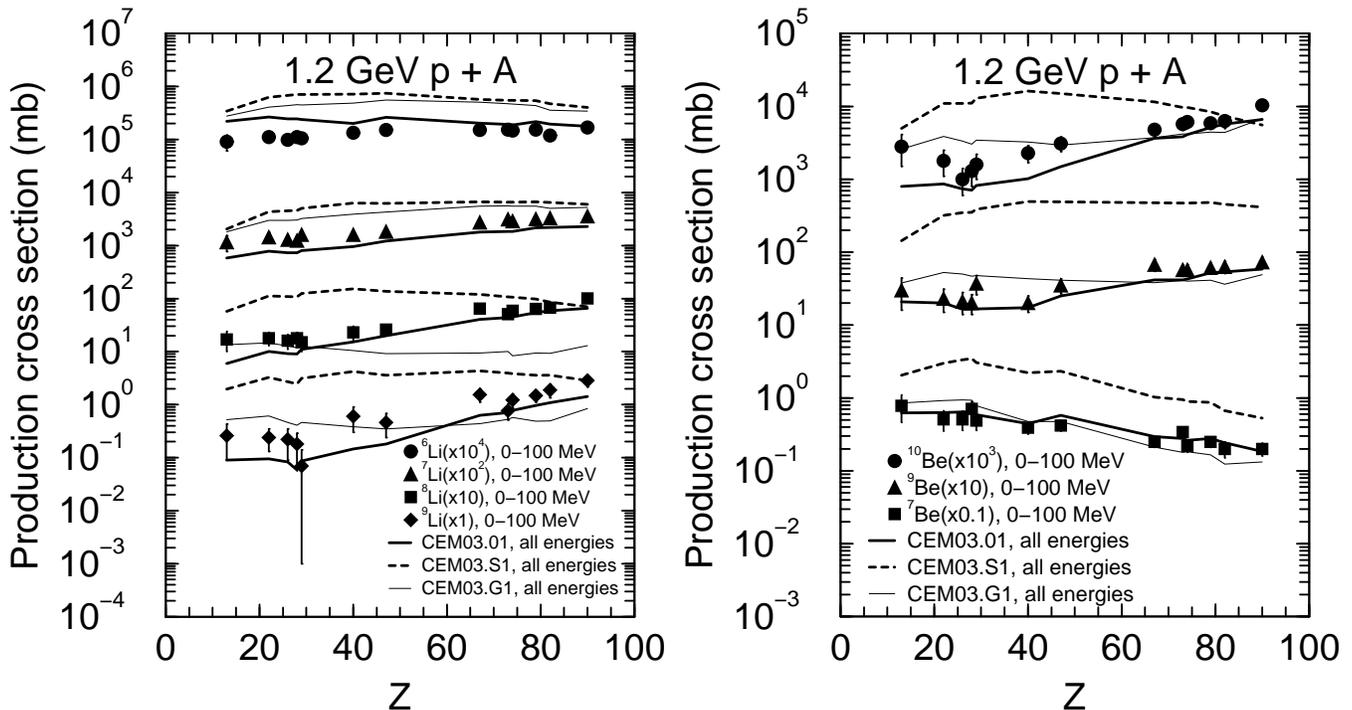}

\caption{
Comparison of measured \cite{Herbach06} (symbols)
production cross sections of
Li and Be isotopes with kinetic energies
below 100 MeV for 1.2 GeV proton-induced reactions
on targets between Al and Th with results
from CEM03.01, CEM03.S1, and CEM03.G1 (lines), as indicated.
}
\end{figure}
%\clearpage            % Use to start references on new page.

We see that on the whole, with only a few exceptions,
all versions of our current codes describe
reasonably the shape and the absolute values of the
measured total production cross sections for all these
fragments,
while the old versions would produce such fragments only as 
residual nuclei and would underestimate the data by orders 
of magnitude, as discussed above.
% discussing the results presented in Fig.\ 27. 
More details on comparisons of our results with the NESSI data
 \cite{Herbach06} may be found in Ref.\ \cite{01s1g1}.

However, a reasonable description by our current codes of the
experimental integrated fragment yields demonstrated above
does not yet assure a similar agreement with the fragment spectra,
where we still have some serious problems to solve.
Fig.\ 29 shows examples of some problems we still have 
in CEM03.03, CEM03.01 \cite{CEM03.01},
as well as in the ``S'' and ``G'' versions CEM03.S1 and CEM03.G1
\cite{01s1g1} 
in the description of 
%$^3$He, 
$^6$Li %, $^7$Li, $^7$Be, $^9$Be, $^{10}$Be, and $^{10}$B
spectra 
at 20, 45, 60, 90, and 110 degrees
from proton-aluminum interactions at 200 MeV measured
recently in Ref.\ \cite{Machner06}
(we got very similar results
also for the $^7$Li, $^7$Be, $^9$Be, $^{10}$Be, and $^{10}$B
spectra). 
We see that on the whole,  with only a few exceptions,
all these recent versions of our codes describe
reasonably the low-energy part of the measured spectra,
where fragments are produced mostly via evaporation.
But all codes fail to reproduce correctly the
high-energy tails of the spectra of all the fragments
heavier than $^4$He.
% (the spectra
%of $^4$He are described well by all models and the $^3$He spectra
%are reproduced not so well, but still reasonably).
Note that the authors of this experiment \cite{Machner06}
have measured similar fragment spectra also from $^{59}$Co and 
$^{197}$Au, and the problems with their correct description by
our codes for these systems are only graver.
We believe that we have these problems
(just like other models have; as of today, we do not know any
models or codes able to reproduce well 
all these data) 
because the high-energy fragments from these reactions are produced 
via preequilibrium mechanisms of nuclear reactions,
not considered by our current models for fragments heavier than $^4$He.
Our models consider preequilibrium emission of particles 
only up to $^4$He, 
this is probably the reason why they describe quite well the
$^4$He spectra, and not so well but still reasonably the $^3$He spectra,
but fail completely to reproduce the high energy tails of all 
heavier fragments. The incident energy of bombarding protons
in these reactions is too low to produce fragments via multigragmentation;
this is the reason why the ``S'' version does not work well here.
The ``G'' version describes production of fragments 
from these reactions only via fission-like binary decays;
this mechanism contributes only to the evaporation part
of the spectra, therefore it also fails to
reproduce the high-energy results
(see more details on the ``G'' and ``S'' versions 
of our codes in Section 10). We plan to address 
preequilibrium emission of fragments heavier than $^4$He
in the future, with a hope to solve these problems.\\

%\vspace{4mm}
{\large\bf 7.  Fission} \\

The fission model used in GEM2 is based on Atchison's model 
\cite{RAL,RAL1}
%$^{12,13}$
as implemented in LAHET \cite{LAHET},
%$^{14}$,
often referred in the literature as the Rutherford Appleton Laboratory (RAL)
fission model, which is where Atchison developed it. In GEM2 there are two 
choices of parameters for the fission model: one of them is the original 
parameter set by Atchison \cite{RAL,RAL1}
%$^{12,13}$ 
as implemented in LAHET \cite{LAHET},
%$^{14}$, 
and the other is a parameter set developed by Furihata
\cite{Furihata1,Furihata2}.\\
%$^{37,38}$.\\

%\vspace{4mm}
{\large\it  7.1. Fission Probability}\\

The Atchison fission model is designed to describe only fission of
nuclei with $Z \geq 70$. It assumes that fission competes only with
neutron emission, {\it i.e.}, from the widths $\Gamma_j$ of n, p, d, 
t, $^3$He, and $^4$He,
the RAL code calculates the probability of evaporation of any 
particle. When a charged particle is selected to be evaporated, 
no fission competition is taken into account. When a neutron is
selected to be evaporated, the code does not actually simulate its evaporation,
instead it considers that fission may compete,
and chooses either fission or evaporation of a neutron according to
the fission probability $P_f$. This quantity is treated by the RAL code 
differently
for the elements above and below $Z=89$. The reasons Atchison split the 
calculation of the fission probability $P_f$ are: (1) there is
very little experimental information on fission in the region $Z=85$ to 88,
(2) the marked rise in the fission barrier for nuclei with $Z^2/A$ below
about 34 (see Fig.\ 2 in \cite{RAL1})
%$^{13}$) 
together with the disappearance of asymmetric 
mass splitting, indicates that a change in the character of the fission
process occurs. If experimental information were available, a split between
regions around $Z^2/A \approx 34$ would be more sensible \cite{RAL1}.
%$^{13}$.

1) $70 \leq Z_j \leq 88$. For fissioning nuclei with $70 \leq Z_j \leq 88$,
GEM2 uses the original Atchison calculation of the neutron emission
width $\Gamma_n$ and fission width $\Gamma_f$ to estimate the fission
probability as
\beq
P_f={\Gamma_f \over {\Gamma_f +\Gamma_n}}={1 \over {1+ \Gamma_n/\Gamma_f} }.
\eeq
Atchison uses \cite{RAL,RAL1}
%$^{12,13}$ 
the Weisskopf and Ewing statistical model \cite{weisewing}
%$^{40}$
with an energy-independent pre-exponential factor for the level density 
(see Eq.\ (44)) and Dostrovsky's \cite{Dostrovsky}
%$^{39}$ 
inverse cross section for neutrons
and estimates the neutron width $\Gamma_n$ as

\newpage
\begin{figure}[ht]     

\vspace*{-10mm}                                            %Fig. 29
\centering
\includegraphics[width=145mm,angle=-0]{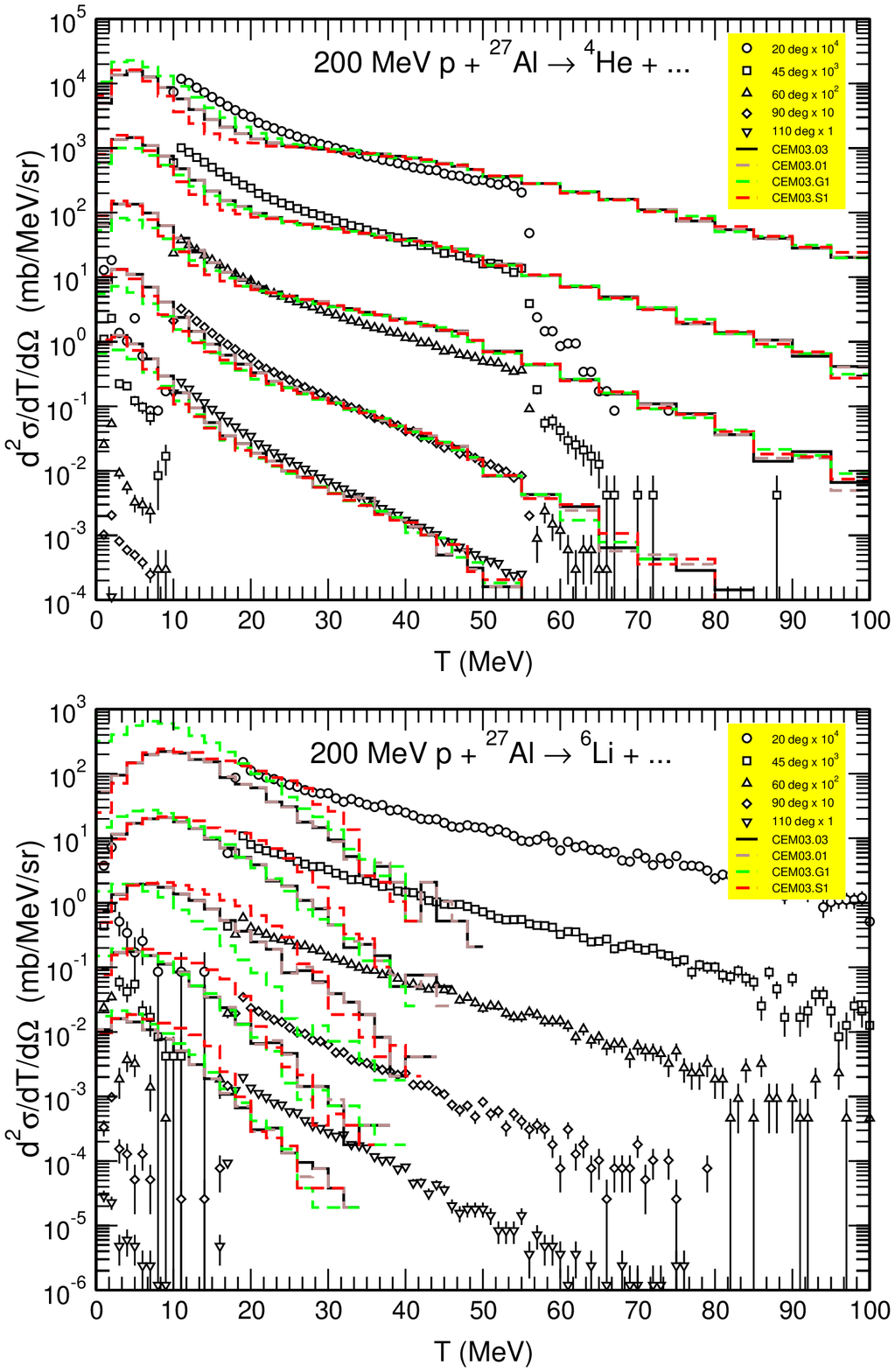}
\caption{Experimental
%$^3$He, 
$^4$He and %, 
$^6$Li
%, $^7$Li, $^7$Be, $^9$Be, $^{10}$Be, and $^{10}$B
spectra 
at 20, 45, 60, 90, and 110 degrees
from proton-aluminum interactions at 200 MeV  \cite{Machner06}
compared with results from CEM03.03, CEM03.01,
CEM03.S1, and CEM03.G1, as indicated.
}
\end{figure}

\clearpage

\beq
\Gamma_n = 0.352 \bigl(1.68 J_0 + 1.93A_i^{1/3}J_1 
+ A_i^{2/3}(0.76J_1 - 0.05 J_0)\bigr),
\eeq
{\noindent
where $J_0$ and $J_1$ are
}
 functions of the level-density parameter $a_n$
and $s_n (=2\sqrt{a_n(E-Q_n-\delta)})$, 
$$J_0 = { {(s_n-1) e^{s_n} + 1} \over {2 a_n} },$$
$$J_1 = { {(2s_n^2 - 6s_n + 6) e^{s_n} + s_n^2 -6} \over {8a_n^2} }.$$
Note that the RAL model uses
a fixed value for the level-density parameter $a_n$, namely
\beq
a_n = (A_i - 1) /8,
\eeq
and this approximation is kept in GEM2 when
calculating the fission probability according to Eq.\ (51), although it 
differs from
the GCCI parameterization (45) used in GEM2 to calculate particle
evaporation widths.
The fission width for nuclei with $70 \leq Z_j \leq 88$ is calculated in 
the RAL model and in the GEM as
\beq
\Gamma_f = { {(s_f - 1)e^{s_f} + 1} \over a_f },
\eeq
where $s_f = 2 \sqrt{a_f (E-B_f - \delta)}$ and the level-density parameter
in the fission mode $a_f$ is fitted by Atchison to describe the measured
$\Gamma_f / \Gamma_n$ to be \cite{RAL1}:
%$^{13}$
\beq
a_f = a_n \Bigl(1.08926 + 0.01098 ( \chi - 31.08551)^2\Bigr),
\eeq
and $\chi = Z^2/A$.
The fission barriers $B_f$ [MeV] are approximated by
\beq
B_f = Q_n + 321.2 - 16.7 { {Z^2_i} \over A} + 0.218 
\Biggl({ {Z^2_i} \over {A_i} }\Biggr)^2 .
\eeq
Note that neither the angular momentum nor the excitation energy of the 
nucleus are taken into account in finding the fission barriers.

2) $Z_j \geq 89$. For heavy fissioning nuclei with $Z_j \geq 89$
GEM2 follows the RAL model 
\cite{RAL,RAL1}
%$^{12,13}$ 
and does not calculate at all
the fission width $\Gamma_f$ and does not use Eq.\ (51) to estimate
the fission probability $P_f$. Instead, the following semi-empirical
expression obtained by Atchison
\cite{RAL,RAL1}
%$^{12,13}$ 
by approximating the experimental values of
$\Gamma_n / \Gamma_f$ published by
Vandenbosch and Huizenga \cite{VanHui}
%$^{50}$ 
is used to calculate the fission probability:
\beq
\log (\Gamma_n / \Gamma_f ) = C(Z_i) ( A_i - A_0(Z_i)),
\eeq
where $C(Z)$ and $A_0(Z)$ are constants depending on the nuclear 
charge $Z$ only. The values of these constants are those
used in the current version of LAHET \cite{LAHET}
%$^{14}$ 
and are tabulated in Table 5 
(note that some adjustments of these values have been done since
Atchison's papers
\cite{RAL,RAL1}
%$^{12,13}$ 
were published).  

In this approach the fission probability $P_f$ is
independent of the excitation energy of the fissioning nucleus
and its angular momentum.\\

\newpage
{\large\it
7.2. Mass Distribution}\\

The selection of the mass of the fission fragments depends on whether the
fission is symmetric or asymmetric. For a pre-fission nucleus with
$Z^2_i/A_i \leq 35$, only symmetric fission is allowed. For $Z^2_i/A_i > 35$,
both symmetric and asymmetric fission are allowed, depending on the
excitation energy of the fissioning nucleus. No new parameters were determined
for asymmetric fission in GEM2.

For nuclei with $Z^2_i/A_i > 35$, whether the fission is symmetric or not is 
determined by the asymmetric fission probability $P_{asy}$
\beq
P_{asy} = { {4870e^{-0.36E}} \over {1+4870e^{-0.36E} } } .
\eeq
\begin{center}
Table 5. $C(Z)$ and $A_0(Z)$ values used in GEM2

\vspace{2mm}
\begin{tabular}{|c|c|c|}
\hline \hline
\hspace{5mm} Z\hspace{5mm} &\hspace{5mm} $C(Z)$\hspace{5mm} &\hspace{5mm}
 $ A_0(Z)$\hspace{5mm} \\
\hline
89 & 0.23000 & 219.40\\
90 & 0.23300 & 226.90\\
91 & 0.12225 & 229.75\\
92 & 0.14727 & 234.04\\
93 & 0.13559 & 238.88\\
94 & 0.15735 & 241.34\\
95 & 0.16597 & 243.04\\
96 & 0.17589 & 245.52\\
97 & 0.18018 & 246.84\\
98 & 0.19568 & 250.18\\
99 & 0.16313 & 254.00\\
100& 0.17123 & 257.80\\
101& 0.17123 & 261.30\\
102& 0.17123 & 264.80\\
103& 0.17123 & 268.30\\
104& 0.17123 & 271.80\\
105& 0.17123 & 275.30\\
106& 0.17123 & 278.80\\
\hline \hline
\end{tabular}
\end{center}

\vspace*{5mm}
\noindent{\sf 7.2.a. Asymmetric fission.} 
For asymmetric fission, the mass of one of the 
post-fission fragments $A_1$ is selected from a Gaussian distribution of mean 
$A_f = 140$ and width $\sigma_M = 6.5$. The mass of the second fragment is
$A_2 = A_i -A_1$.\\

\noindent{\sf 7.2.b. Symmetric fission.} For symmetric fission, 
$A_1$ is selected from a Gaussian distribution of mean $A_f = A_i/2$ 
and two options for the width $\sigma_M$
as described below.

The first option for choosing $\sigma_M$ is the original Atchison 
approximation:
\beq
\sigma_M =
\cases{3.97+0.425(E-B_f)-0.00212(E-B_f)^2 ,& \cr 
25.27 ,&\cr}
\eeq
for $(E-B_f)$ below or above 100 MeV, respectively. In this expression all
values are in MeV and the fission barriers $B_f$ are calculated according
to Eq.\ (56) for nuclei with $Z_i \leq 88$. For nuclei with $Z_i > 88$,
the expression by Neuzil and Fairhall \cite{ Neuzil}
%$^{51}$ 
is used:
\beq
B_f = C - 0.36 (Z^2_i/A_i) ,
\eeq
where $C = 18.8, 18.1, 18.1,$ and $18.5$ [MeV] for odd-odd, even-odd, 
odd-even, and even-even nuclei, respectively.
 
The second option in GEM2 for $\sigma_M$ (used here) was found by 
Furihata
\cite{Furihata1,Furihata2}
 as:
\beq
\sigma_M = C_3 (Z^2_i/A_i)^2 + C_4 (Z^2_i/A_i) + C_5(E-B_f) + C_6 .
\eeq
The constants $C_3=0.122$, $C_4 =-7.77$, $C_5=3.32\times10^{-2}$, and
$C_6=134.0$ were obtained by fitting with GEM2 the recent Russian collection of
experimental fission-fragment mass distributions \cite{Rusanov}.
%$^{52}$. 
In this expression, the fission barriers $B_f$ by Myers and Swiatecki
\cite{Myers99}
%$^{53}$ 
are used. More details may be found in Ref.\ \cite{Furihata2}.\\
%$^{38}$\\

{\large\it 7.3. Charge Distribution}\\

The charge distribution of fission fragments is assumed to be a Gaussian 
distribution of mean $Z_f$ and width $\sigma_Z$. $Z_f$ is expressed as
\beq
Z_f = { {Z_i+Z'_1 -Z'_2} \over 2} ,
\eeq
where
\beq
Z_l' = {{65.5A_l} \over {131+A_l^{2/3}}}, \mbox{$l=1$ or 2}.
\eeq
The original Atchison model uses $\sigma_Z = 2.0$. An investigation
by Furihata  \cite{Furihata2}
%$^{38}$ 
suggests that $\sigma_Z = 0.75$ provides a better 
agreement with data; therefore $\sigma_Z = 0.75$ is used in GEM2
and in our code.\\

{\large\it 7.4. Kinetic Energy Distribution}\\

The kinetic energy of fission fragments [MeV] is determined by a
Gaussian distribution with mean $\eps_f$ and width $\sigma_{\eps_f}$.

The original parameters in the Atchison model are:\\
$$ \eps_f = 0.133Z_i^2/A_i^{1/3} - 11.4 ,$$
$$\sigma_{\eps_f} = 0.084 \eps_f .$$

Furihata's parameters in the GEM, which we also use, are:
\beq
\eps_f =
\cases{0.131 Z^2_i/A_i^{1/3} ,& \cr 
0.104Z^2_i/A_i^{1/3} + 24.3 ,&\cr}
\eeq
for $Z^2_i/A^{1/3}_i \leq 900$ and $900 < Z^2_i/A_i^{1/3} \leq 1800$,
respectively, according to Rusanov {\it et al.}\ \cite{Rusanov}.
%$^{52}$
By fitting the experimental data by Itkis {\it et al.} \cite{Itkis90},
%$^{54}$, 
Furihata found the following expression for $\sigma_{\eps_f}$
\beq
\sigma_{\eps_f} =
\cases{C_1 (Z^2_i/A^{1/3}_i - 1000) + C_2 ,& \cr 
C_2 ,&   \cr}
\eeq
for $Z^2_i/A_i^{1/3}$ above and below 1000, respectively, and the values
of the fitted constants are $C_1 = 5.70 \times 10^{-4}$ and
$C_2 = 86.5$.
The experimental data used by Furihata for fitting are the values
extrapolated to the nuclear temperature 1.5 MeV by
Itkis {\it et al.}\ \cite{Itkis90}.
%$^{54}$
More details may be found in \cite{Furihata2}.
%$^{38}$.

We note that Atchison has also modified his original version using 
recent data and published \cite{RAL3}
%$^{55}$ 
improved (and more complicated)
parameterizations for many quantities and distributions in his model,
but these modifications \cite{RAL3} 
%$^{55}$ 
have not been included either in LAHET or in GEM2.\\

{\large\it 7.5. Modifications to GEM2 in CEM03.01 and LAQGSM03.01}\\

First, we fixed several observed uncertainties and small errors
in the 2002 version of GEM2 Dr.\ Furihata kindly sent us. Then,
we extended GEM2 to describe fission of lighter nuclei, down to $Z \ge 65$,
and modified it \cite{fitaf}
so that it provides a good description of fission
cross sections when it is used after our INC and preequilibrium
models.

If we had merged GEM2 with the INC and preequilibrium-decay 
modules of CEM or of LAQGSM without any modifications, the new code
would not describe correctly fission cross sections (and the yields of
fission fragments). This is because Atchison
fitted the parameters of his RAL fission model when it followed 
the Bertini INC \cite{BertiniINC} which differs
from ours. In addition, Atchison did not model 
preequilibrium emission. Therefore, the distributions of fissioning
nuclei in $A$, $Z$, and excitation energy $E^*$ simulated by Atchison
differ significantly from the distributions we get; 
as a consequence, all the fission characteristics are also different.
Furihata used GEM2 coupled either with the Bertini INC 
\cite{BertiniINC} or with
the ISABEL \cite{ISABEL} INC code, which also differs from our INC, and did 
not include preequilibrium particle emission. Therefore the distributions
of fissioning nuclei simulated by Furihata differ from those in
our simulations, so the parameters adjusted by Furihata to work well 
with her INC are not appropriate for us. To get a good description 
of fission cross sections (and fission-fragment yields)
we have modified at least two parameters in GEM2 as used
in CEM03.01 and LAQGSM03.01
(see more details in \cite{SATIF6,SantaFe02,Pavia06}).

The main parameters that determine the fission cross sections
calculated by GEM2 are the level-density parameter in the
fission channel, $a_f$ (or more exactly, the ratio $a_f/a_n$
as calculated by Eq.\ (55)) for preactinides, and parameter
$C(Z)$ in Eq.\ (57) for actinides. The sensitivity of results to
these parameters is much higher than to either the fission-barrier 
heights used in a calculation or other parameters of the model.
Therefore we choose \cite{fitaf} to adjust only these two 
parameters in our merged 
%CEM2k+GEM2 
codes. We do not change
the form of systematics (55) and (57) derived by Atchison.
We only introduce additional coefficients both to $a_f$ and $C(Z)$,
replacing $a_f \to C_a \times a_f$ in Eq.\ (55) and 
$C(Z_i) \to C_c \times C(Z_i)$
in Eq.\ (57) and fit $C_a$ and $C_c$
to experimental proton-induced fission cross sections 
covered by Prokofiev's systematics \cite{Prokofiev}
for both CEM03.01 and LAQGSM03.01. 
No other parameters in GEM2 have been changed. 
For preactinides, we fit only $C_a$.
The values of $C_a$ found in our fit to Prokofiev's systematics 
are close to one and vary smoothly with the
proton energy and the charge or mass number of the target.
This result gives us some confidence in our procedure, and
allows us to interpolate the values of $C_a$
for nuclei and incident proton energies not analyzed by Prokofiev.
For actinides, as described in \cite{SATIF6,SantaFe02},
we have to fit both $C_a$
and $C_c$. The values of $C_a$ we find are also very close to one,
while the values of $C_c$ are more varied, but both of them
change smoothly with the proton energy and Z or A of the   
target, which again allows us to interpolate 
them for nuclei and energies outside Prokofiev's systematics.

We fix the fitted values of $C_a$ and $C_c$ in data blocks 
in our codes and use the routines 
{\bf fitafpa} and {\bf fitafac} to interpolate to nuclei not covered by 
Prokofiev's systematics. We believe that such a procedure 
provides a reasonably accurate fission cross-section calculation, 
at least for proton energies and target nuclei not too far 
from the ones covered by the systematics.

Let us mention that the situation with the fitting procedure of 
parameters $C_a$ and $C_c$ is quite tricky, as it should
be redone after all major improvements of other parts of
our codes describing INC, preequilibrium, or evaporation.
This is a major minus of such types of models like GEM2 that are 
based mostly on systematics of available experimental data 
rather than on fundamental physics. (This was the main reason we 
started to develop our own improved evaporation and fission   
models for our codes \cite{SantaFe02}
that would describe experimental data not worse than GEM2 
but would be based on physics rather than on systematics of available 
data; this work is not completed yet, so we have to use
GEM2 in our codes so far.)
Indeed, 
after making our major improvements to the INC and preequilibrium parts 
CEM and LAQGSM as described above, the mean values of the 
mass and charge numbers, $A$ and $Z$ of the excited compound nuclei
produced after the preequilibrium stage of nuclear reactions and their
mean excitation energy $E^*$ have changed slightly, which
affects the probability of heavy compound nuclei
(especially preactinides) to fission. This means that
the procedure of fitting the  $C_a$ and $C_c$ parameters
which we performed in \cite{fitaf} to provide the 
best description by CEM2k and LAQGSM
of fission cross sections was no longer correct.
We had to redo this fitting for the latest versions of our 
CEM03.01 and LAQGSM03.01, ensuring that they
describe as well as possible fission cross sections from various
reactions. Fig.\ 30 shows examples of fission cross sections
for proton-induced reactions on 
 $^{186}$W, $^{184}$W, $^{183}$W and $^{182}$W. 
The improved CEM03.01 reproduces the recent Uppsala 
measurements \cite{Eismont05} of proton-induced fission cross 
sections.

\begin{figure}[h]                                                 %Fig. 30
\centering
\includegraphics[width=100mm,angle=-0]{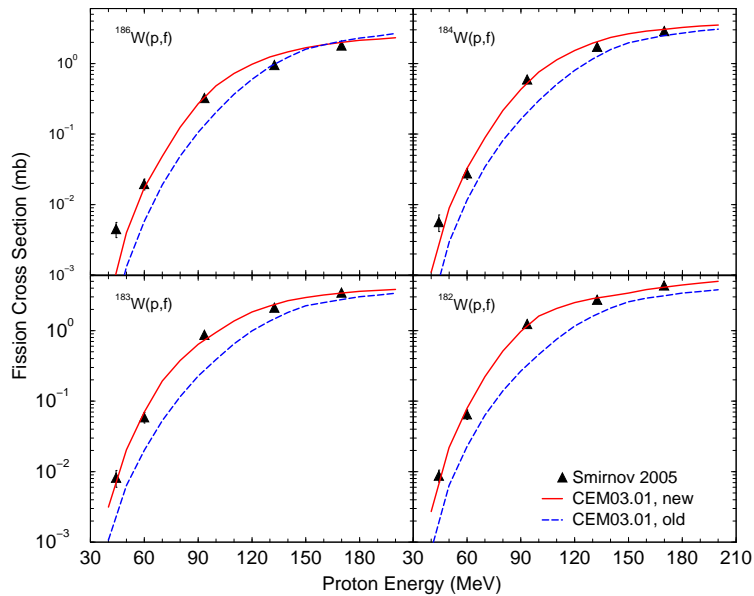}
\caption{Experimental \cite{Eismont05}
proton-induced fission cross sections of $^{186}$W, $^{184}$W, $^{183}$W,
and $^{182}$W compared with improved (red solid lines) and old
(blue dashed lines, from \cite{Eismont05}) CEM03.01 calculations.
}
\end{figure}

Another example of some problems of the fission model of GEM2 based on
systematics of measured data is shown in Fig.\ 31 for 
fragment mass distributions from several reactions
induced by bremsstrahlung with the maximum energy $E_0$ of 15, 20, 30,
and 70 MeV on $^{235}$U and $^{238}$U targets
measued in \cite{Jacobs79} compared with results by CEM03.02 \cite{CEM03.02}.
We see that the fission model of GEM2 used in our CEM03.02
describes surprisingly well these mass distributions of fission
fragments at $E_0 = 15$, 20, and 30 MeV, where experimentasl data
were available to fit the mass distribution of fission
fragments in 
the GEM2 systematics, but fails to reproduce correctly
such distribution at  $E_0 = 70$ MeV, requiring an additional
refitting of the GEM systematics at this energy to improve
the agreement with the data.

\begin{figure}[ht]                                                 %Fig. 31
\centering
\includegraphics[width=135mm,angle=-0]{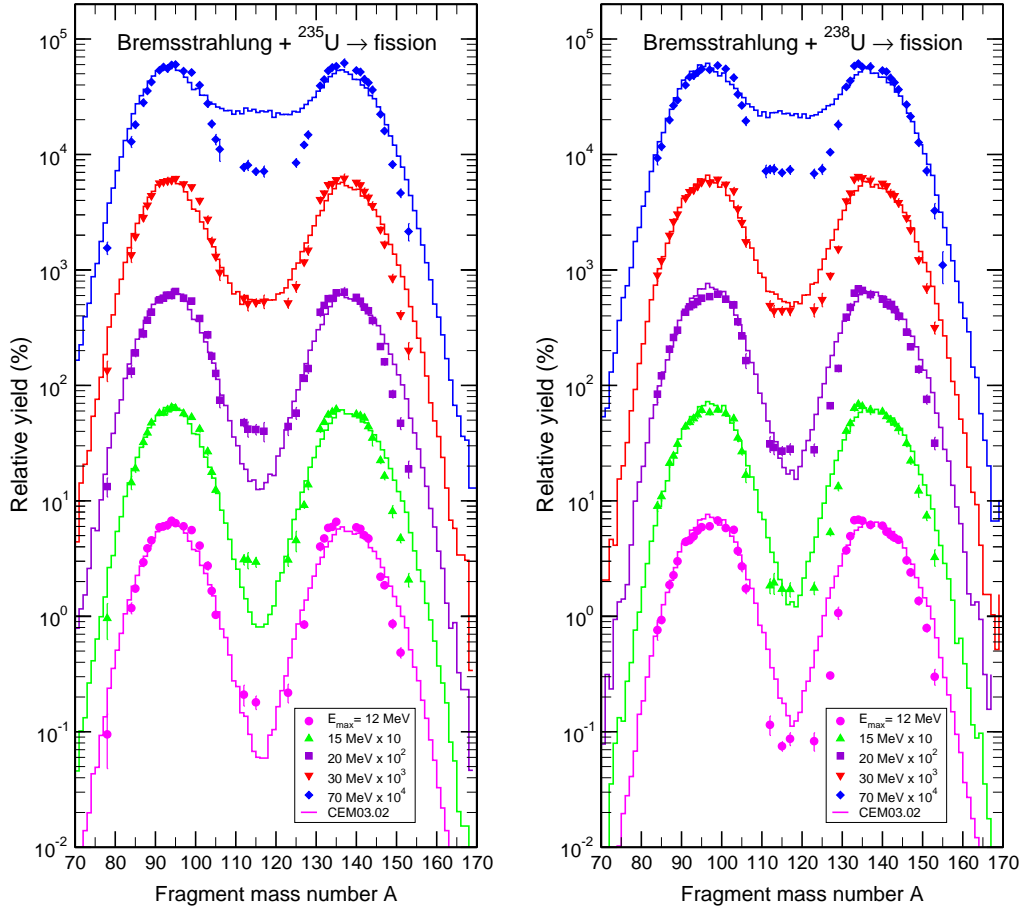}
\caption{
Experimental  \cite{Jacobs79} fission
fragment mass distributions (symbols) from reactions
induced by bremsstrahlung with the maximum energy $E_0$ of 15, 20, 30,
and 70 MeV on $^{235}$U and $^{238}$U targets
compared with results by CEM03.02 \cite{CEM03.02} (histogams).
}
\end{figure}

Finally, to conclude the fission Section, Figs.\ 32 and 33 shows examples
of spallation, fission, and fragmentation 
experimental  product cross sections
from 1 GeV $p$ + $^{238}$U \cite{Taieb03,Bernas03}
and 2 GeV $d$ + $^{208}$Pb \cite{Enqvist02}
measured recently at GSI in inverse kinematics 
compared with results by CEM03.01 ($p$ + U, Fig.\ 32) and
LAQGSM03.01 ($d$ + Pb, Fig.\ 33), respectively.
Our models describe quite well the fission-
fragment yields, as well as the spallation and fragmentation
product cross sections and agree with most of the GSI
data with an acuracy of a factor of two or better.
Similar agreements are obtained with other reactions
measured at GSI in inverse kinematics  
(see, {\it e.g.}, \cite{Mashnik04,Pavia06,TRAMU,Varenna03}).

\newpage
\begin{figure}[ht]                                                %Fig. 32

%\vspace*{-10mm}            
\centering
\includegraphics[width=155mm,angle=-0]{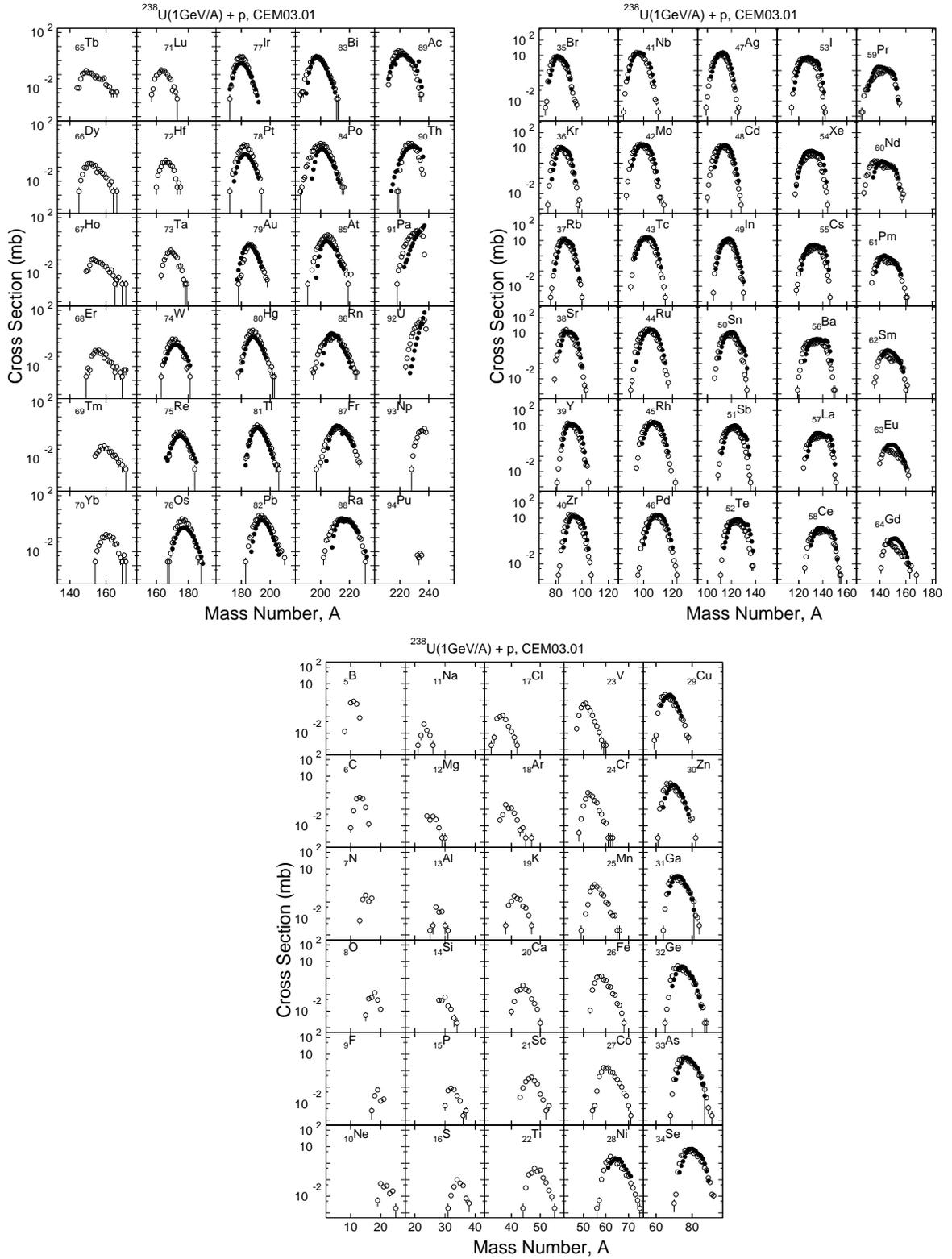}
\caption{
Comparison of measured \cite{Taieb03,Bernas03} 
spallation, fission, and fragmentation cross
sections for the reaction $^{238}$U(1 GeV/A) + p (filled circles)
with our CEM03.01 results (open circles). Experimental data for
isotopes from B to Co, from Tb to Ta, and for Np and Pu
are not available so we present only our predictions. 
}

%\vspace*{-50mm}
\end{figure}

\clearpage

\begin{figure}[ht]                                                 %Fig. 33
\centering
\includegraphics[width=155mm,angle=-0]{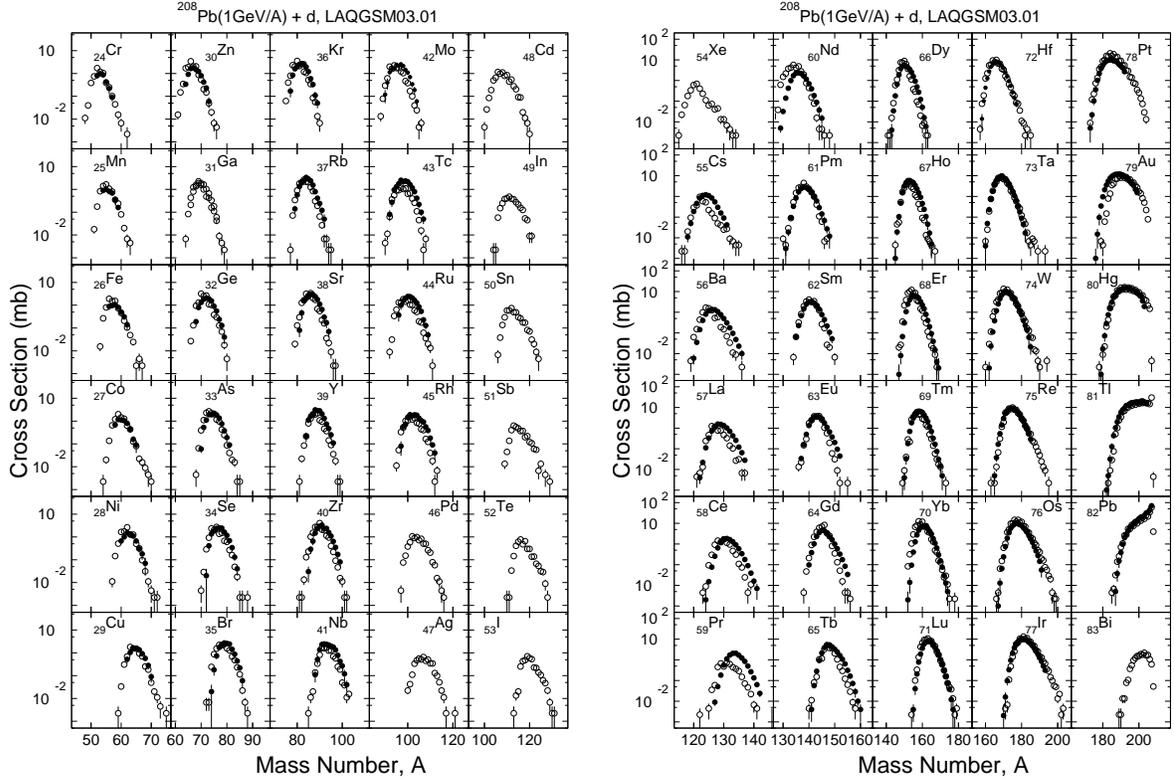}
\caption{
Comparison of measured \cite{Enqvist02} 
spallation and fission-product cross
sections for the reaction $^{208}$Pb(1 GeV/A) + d (filled circles)
with our LAQGSM03.01 results (open circles).
}
\end{figure}

{\large\bf 8.  The Fermi Breakup Model} \\

After calculating the coalescence stage of a reaction, CEM03.01 
and LAQGSM03.01
move to the description of the last slow stages of the interaction,
namely to preequilibrium decay and evaporation, with a possible competition
of fission. But as mentioned above,
if the residual nuclei have atomic numbers with  $A < 13$,
 CEM03.01 
and LAQGSM03.01 use the Fermi breakup model \cite{Fermi:50}
to calculate their further disintegration instead of using
the preequilibrium and evaporation models.
The newer 03.02 versions of our codes use the Fermi breakup model also
during the preequilibrium and/or evaporation stages of reactions,
when the residual nucleus has an atomic number with $A < 13$.

Finally, the latest 03.03 
versions of our codes use the Fermi breakup model also
to disintegrate the unstable fission fragments with  $A < 13$ that 
can be produced in very rare cases of very asymmetric fission.

All formulas and details of the algorithms used in the version of the
Fermi breakup model developed in the former group of 
Prof.~Barashenkov
at Joint Institute for Nuclear Research (JINR), Dubna, Russia and
used by our codes
% in CEM03.01 
may be found in \cite{GEANT4}.
%(true, with some misprints).
All the information
needed to calculate the breakup of an excited nucleus is its excitation
energy $U$ and the mass and charge numbers $A$ and $Z$. The total energy
of the nucleus in the rest frame will be
$E = U + M(A,Z)$, where $M$ is the mass of the nucleus.
The total probability per unit time for a nucleus to break up into $n$
components in the final state ({\it e.g.}, a possible residual nucleus,
nucleons, deuterons, tritons, alphas, {\it etc.}) is given by
\begin{equation}
W(E,n) = (V/\Omega)^{n-1} \rho_n(E) ,
\end{equation}
where $\rho_n$ is the density of final states, $V$ is the volume 
of the decaying system and $\Omega = (2\pi \hbar)^3$
is the normalization volume. The density $\rho_n(E)$ can be defined as a
product of three factors:
\begin{equation}
\rho_n(E) = M_n(E) S_n G_n .
\end{equation}
The first one is the phase space factor defined as
\begin{equation}
M_n(E) = \int_{-\infty}^{+\infty} \cdots \int_{-\infty}^{+\infty}
\delta \left( \sum_{b=1}^{n} \vec p_b \right) \delta \left( E - \sum_{b=1}^n
\sqrt{p^2 + m_b^2} \right) \prod_{b=1}^n d^3 p_b ,
\end{equation}
where $\vec p_b$ are fragment momenta. The second one is the spin factor
\begin{equation}
S_n = \prod_{b=1}^n (2 s_b +1) ,
\end{equation}
which gives the number of states with different spin orientations. The last
one is the permutation factor
\begin{equation}
G_n = \prod_{j=1}^k {1 \over n_j !} ,
\end{equation}
which takes into account identical particles in the final state
($n_j$ is the number of components of $j$-type particles and $k$ is defined
by $n = \sum\nolimits_{j=1}^k n_j$). For example, if we have in the final
state six particles ($n = 6$) and two of them are alphas, three are nucleons,
and one is a deuteron, then $G_6 = 1/ (2! 3! 1!) = 1/12$.
For the non-relativistic case, the integration in Eq.~(68) can be evaluated
analytically (see, {\it e.g.}, \cite{GEANT4}) and the probability for
a nucleus to disintegrate into $n$ fragments with masses $m_b$,
where $b = 1, 2, 3, \dots, n$ is
\begin{equation}
W(E,n) = S_n G_n \left( {V \over \Omega} \right)^{n-1}
\left( {1 \over \sum\nolimits_{b=1}^n m_b}
\prod_{b=1}^n m_b \right)^{3/2}
{ (2\pi)^{3(n-1)/2} \over \Gamma (3(n-1)/2) } E^{(3n-5)/2},
\end{equation}
where $\Gamma(x)$ is the gamma function.

The angular distribution of $n$ emitted fragments is assumed to be
isotropic in the c.m.\ system of the disintegrating nucleus and their
kinetic energies are calculated from momentum-energy conservation.
The Monte-Carlo method is used to randomly select the decay channel
according to probabilities defined by Eq.\ (71). Then, for a given channel,
the code
calculates kinematic quantities for each fragment according to the
$n$-body phase space distribution using Kopylov's method \cite{Kopylov:70}.
Generally, the Fermi breakup model
considers formation of fragments only in their ground and those
low-lying states which are stable for nucleon emission. However,
as already mentioned in Section 2,
several unstable fragments with large lifetimes:
$^5$He, $^5$Li, $^8$Be, $^9$B, {\it etc.} are considered as well
by the initial version of the Fermi breakup model code as
described in Ref.\ \cite{GEANT4}.
The randomly chosen channel will be allowed to decay only
if the total kinetic energy $E_{kin}$ of all fragments at the moment of
breakup is positive, otherwise a new simulation will be performed and
a new channel will be selected. The
total kinetic energy $E_{kin}$ can be calculated according to the equation:
\begin{equation}
E_{kin} = U + M(A,Z) - E_{Coulomb} - \sum_{b=1}^n (m_b + \epsilon_b ) ,
\end{equation}
where $m_b$ and $\epsilon_b$ are masses and excitation energies of
the fragments, respectively, and $E_{Coulomb}$ is the Coulomb barrier
for the given channel. It is approximated by
\begin{equation}
E_{Coulomb} = {3 \over 5} {e^2 \over r_0}
\left( 1 + {V \over V_0} \right)^{-1/3}
\left( {Z^2 \over A^{1/3}} - \sum_{b=1}^n {Z_b^2 \over A_b^{1/3}} \right) ,
\end{equation}
where $A_b$ and $Z_b$ are the mass number and the charge of the $b$-th
particle of a given channel, respectively.
$V_0$ is the volume of the system corresponding to normal
nuclear density and $V = k V_0$ is the decaying system volume
(we assume $k=1$ in our codes).

Thus, the Fermi breakup model we use
has only one free parameter, $V$ or $V_0$,
the volume of the decaying system, which is estimated as follows:
\begin{equation}
V = 4\pi R^3 /3 = 4 \pi r_0^3 A /3 ,
\end{equation}
where we use $r_0 = 1.4$ fm.

There is no limitation on the number $n$ of fragments a nucleus
may break up into in our version of the breakup model,
in contrast to implementations in other codes, such as
$n \le 3$ in MCNPX, or $n \le 7$ in LAHET.

In comparison with its initial version as described in  \cite{GEANT4},
%[137], 
the Fermi breakup model used in
CEM03.02 and LAQGSM03.02 has been modified \cite{CEM03.02}
%[123] 
to decay the unstable light fragments
that were produced by the original code. As mentioned above, 
the initial routines that describe the
Fermi breakup model were written more than twenty years ago 
in the group of Prof.\ Barashenkov
at JINR, Dubna, and unfortunately had some problems. First, 
those routines allowed in rare cases
production of some light unstable fragments like 
$^5$He, $^5$Li, $^8$Be, $^9$B, {\it etc.} as a result of a 
breakup of some light excited nuclei. Second,  they
allowed very rarely even production of ``neutron stars'' 
(or ``proton stars''),{\it i.e.}, residual ``nuclei'' 
produced via Fermi breakup that consist of only neutrons (or
only protons). Lastly, those routines could even crash the 
code, due to cases of division by 0. All
these problems of the Fermi breakup model routines were 
addressed and solved in CEM03.02 \cite{CEM03.02};
%[123]; 
the changes were then put in LAQGSM03.02 \cite{CEM03.02}.
%[123]. 
Several bugs are also fixed.
However, even after solving these problems and after 
implementing the improved Fermi
breakup model into CEM03.02 and LAQGSM03.02 \cite{CEM03.02},
%[123], 
our event generators still could
produce some unstable products via very asymmetric fission, 
when the excitation energies of those
fragments were below 3 MeV so they were not checked and 
disintegrated with the Fermi breakup model. 
Our analysis \cite{LAQGSM03.03} had shown
that such events could occur
very rarely, in less than 0.0006\% of all simulated events,
so that production of such
unstable nuclides affects by less than 0.0006\% the other correct 
cross sections calculated by our codes.
However, these unstable nuclides are not physical and 
should be eliminated. This was
the reason why a universal checking of all unstable light 
products has been incorporated into CEM03.03 and
LAQGSM03.03. Such unstable products are forced to disintegrate 
via Fermi breakup independently of their excitation energy. 
The latest versions of the CEM03 and LAQGSM03 event
generators do not produce any such unstable products.

Examples of several results from CEM03.02 compared with results from
CEM03.01 and from the older versions 
CEM95 and CEM2k and with experimental data for a number of
reactions relevant to the Fermi breakup model
are presented Figs.\ 34 to 36.
Fig.\ 34 presents the mass  
distribution of the product yields from the reaction 
730 MeV p +  $^{27}$Al
calculated with CEM03.01 without considering the Fermi breakup mode
during the preequilibrium and evaporation stages of reactions
and with CEM03.02 that does consider this mode during these stages
and uses an extended version of the model
without unstable products, as described above.
CEM03.01 provides a small yield of unphysical
unstable $^5$He (just like the older versions CEM95 and CEM2k do),
while the newer version
does not produce such unstable nuclides.
The results from CEM03.02 also agree a little better with
available experimental data (red squares on Fig.\ 34) than
those from CEM03.01. Similar results are obtained for
several other reactions on other light targets.

\begin{figure}[ht]                                                 %Fig. 34
\centering
\hspace*{-5mm}
\includegraphics[width=120mm,angle=-90]{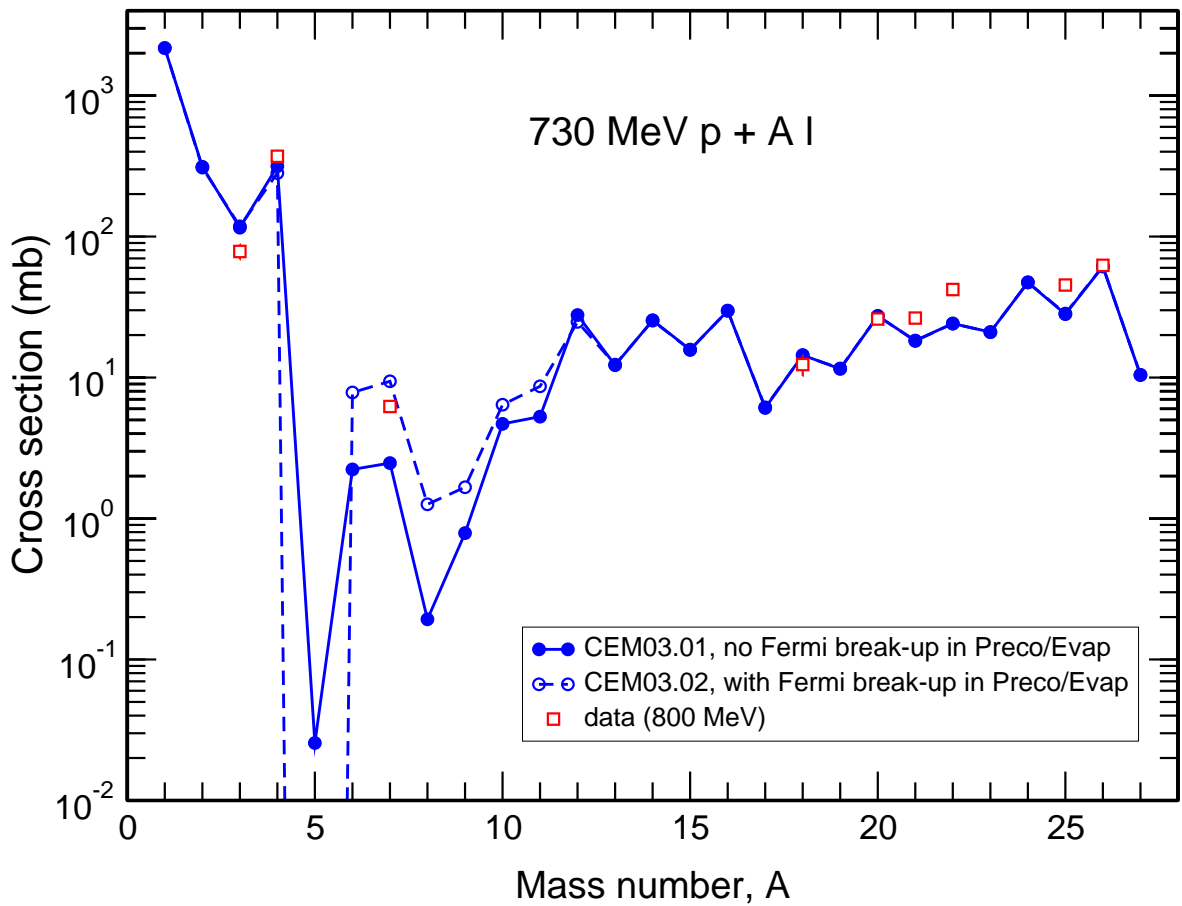}
\caption{Mass  
distribution of the product yields from the reaction 
730 MeV p +  $^{27}$Al
calculated with CEM03.01 without considering the Fermi breakup mode
during the preequilibrium and evaporation stages of reactions
(solid circles connected with a solid line) and 
with the improved version of the code 
CEM03.02 that does consider the Fermi breakup mode
during the preequilibrium and evaporation stages of reactions
(open circles connected with a dashed line), compared with experimental 
data available
at a nearby energy of 800 MeV from the T-16 Lib compilation 
\cite{T16comp} (open red squares).
}
\end{figure}

Our results show that CEM03.02 and LAQGSM03.02
do not predict unstable unphysical 
nuclides and describe the yields of most products a little better
than the 03.01
or older versions of CEM and LAQGSM. The question about
how the spectra of different particles from
different nuclear reactions are described by the 03.02 version
in comparison with older versions of our codes is not obvious. 
We studied this question on several reactions and 
Fig.\ 35 presents examples of particle spectra calculated
with CEM03.01 and CEM03.02 for the reaction
730 MeV p + Al. We see that spectra of n, p, d, t, $^3$He,
and $^4$He calculated by CEM03.02 that uses the Fermi breakup
model to describe disintegration of exited nuclei with $A < 13$ 
are almost the same as such spectra calculated by CEM03.01
which uses the preequilibrium and evaporation models
to describe such reactions when $A > 12 $ after the
INC stage of reactions. This result is very interesting from a
physical point of view; it is not trivial at all and could not be
forecast easily in advance. We see that spectra of these particles
predicted by different models are almost the same. In other words,
the theoretical spectra of particles do not depend much on the
models we use to calculate them, but depend mostly on the 
final phase space calculated by these models: If the phase spaces
calculated by different models are correct and near to each other,
than the spectra of secondary particles calculated
by these models would be also near to each other and would be
not very sensitive to the
dynamics of reactions considered (or not) by the models we use.

\begin{figure}[ht]                                                 %Fig. 35
\centering
\hspace*{-5mm}
\includegraphics[width=120mm,angle=-90]{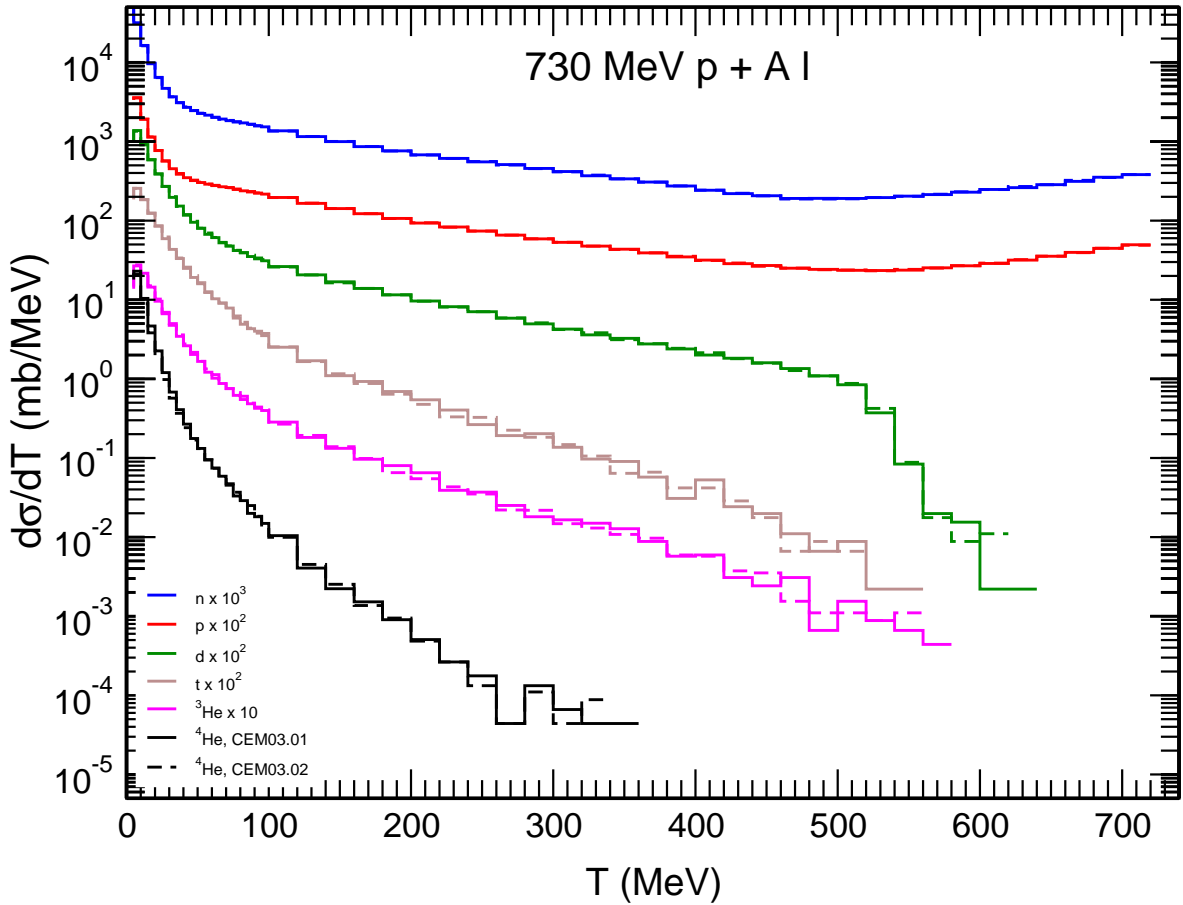}
\caption{
Angle-integrated energy spectra of secondary n, p, d, t, $^3$He, and $^4$He
from
the reaction 
730 MeV p +  $^{27}$Al
calculated with CEM03.01 without considering the Fermi breakup mode
during the preequilibrium and evaporation stages of reactions
(solid histograms) and 
with CEM03.02 that does consider the Fermi breakup mode
during the preequilibrium and evaporation stages of reactions
(dashed histograms).
}
\end{figure}

Fig.\ 36 shows examples of double differential
 experimental neutron spectra
 at 15, 30, 60, 90, 120, and 150 deg from 
3.0, 1.5, 0.8, and 0.597 GeV 
p + Al (symbols) compared with results by
CEM03.01 without considering the Fermi breakup mode
during the preequilibrium and evaporation stages of reactions
(dashed histograms),  
with CEM03.02 that does consider this mode
during the preequilibrium and evaporation stages of reactions
(solid histograms),
and with the old version of the code, CEM95 \cite{CEM95}.
%[8]. 
We see that the 
results from CEM03.02 and CEM03.01 agree much better with the data
at neutron energies of 20--100 MeV (where the contribution from
the preequilibrium mode is the most important) than the results
from CEM95 do. Results from CEM03.02 for these neutron spectra 
are nearly identical to those from CEM03.01, just as observed
for the energy spectra of n, p, d, t, $^3$He, and $^4$He shown
in Fig.\ 35.

\begin{figure}[ht]                                                 %Fig. 36
\centering
%\hspace*{-5mm}
\includegraphics[width=170mm,angle=-0]{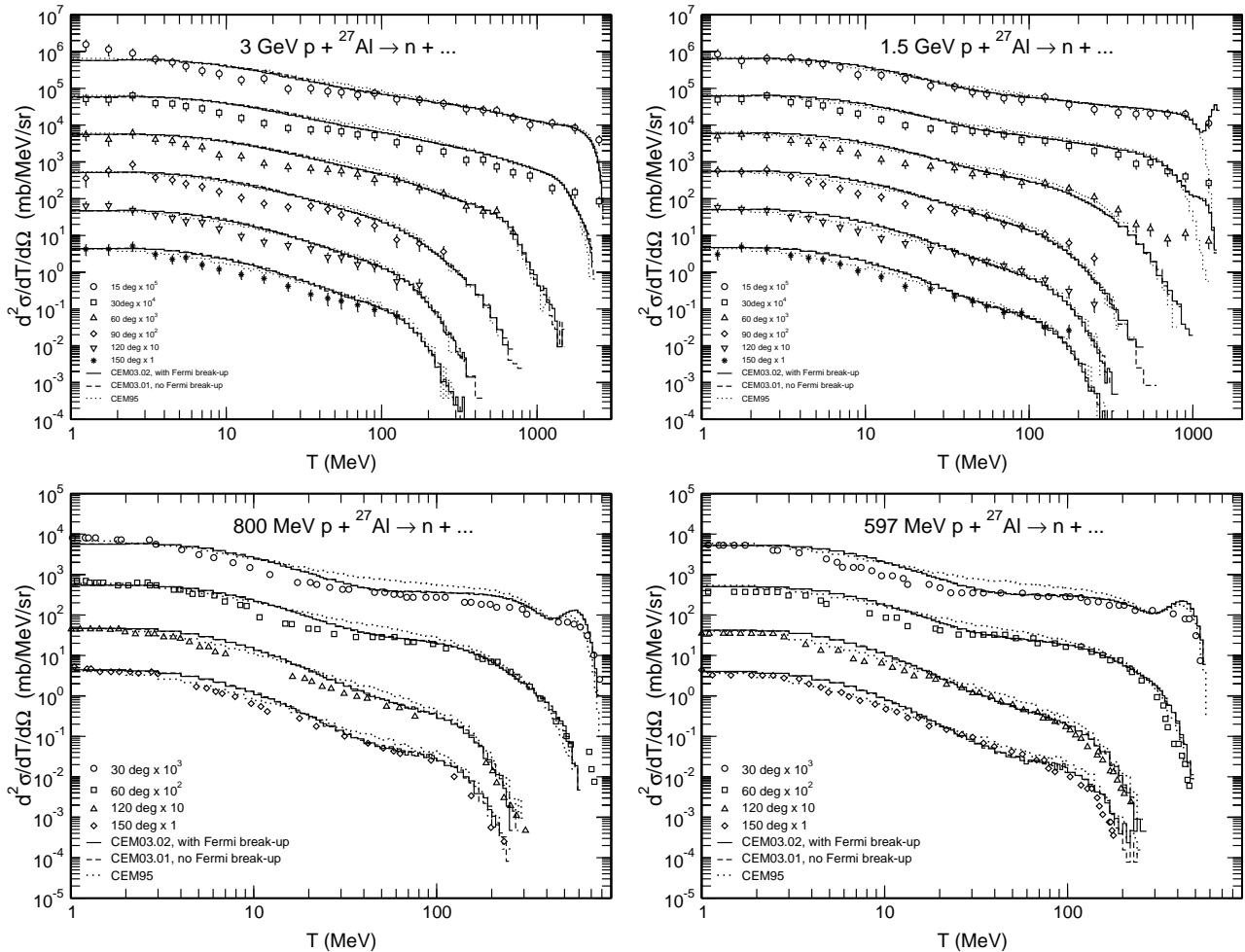}
\caption{
Experimental %[20,31,32]
\cite{Ishibashi97}--\cite{Amian92}
%\cite{Amian93}
double differential neutron spectra at 15, 30, 60, 90, 120, and 150 deg from 
3.0, 1.5, 0.8, and 0.597 GeV 
p + Al (symbols) compared with results from
CEM03.01 without considering the Fermi breakup mode
during the preequilibrium and evaporation stages of reactions
(dashed histograms),  
with CEM03.02 which does consider this mode
during the preequilibrium and evaporation stages of reactions
(solid histograms),
and with the old version of the code, CEM95 \cite{CEM95}.
}
\end{figure}

{\large\bf 9. Reactions Involving Pions and Photons; Excitation Functions}\\

Protons from the beam and energetic secondary particles
generate in
ADS targets and surrounding shielding not only nucleons,
complex particles, and fragments heavier than $^4$He
(in addition to residual nuclei), but also pions
of different energies. The transport codes used in ADS 
applications should be able to describe correctly
their spectra and yields, to 
transport them, and to consider their further interactions with nuclei. 
We have shown that our event generators describe reasonably well 
spectra and yields of nucleon and complex particles
from various reactions, and a little worse but still reasonably
the yields and low energy part of spectra of fragments heavier than $^4$He, 
with still retaining some unsolved problems for the high energy part
of fragment spectra. 
Here, we demonstrate that our codes describe reasonably well also
pion spectra from different reactions as well as
various pion-induced reactions.

Fig.\ 37 shows an example of $\pi^+$ spectra from 562.5 MeV n + Cu
calculated by CEM03.01 compared with the LANL measurements
by Brooks {\it et. al.} \cite{Brooks91}.
We see a reasonably good agreement of CEM03.01 calculations
with these pion spectra. Similar results are obtained with
our codes for other pion spectra (and for the integrated pion yield)
measured in various reactions induced by different projectiles
at different energies (see, {\it e.g.}, \cite{JNRS05,LAQGSM03.01,Pavia06}
and Fig.\ 12 above).

%\newpage
\begin{figure}[ht]                                                 %Fig. 37
\centering
\includegraphics[width=100mm,angle=-90]{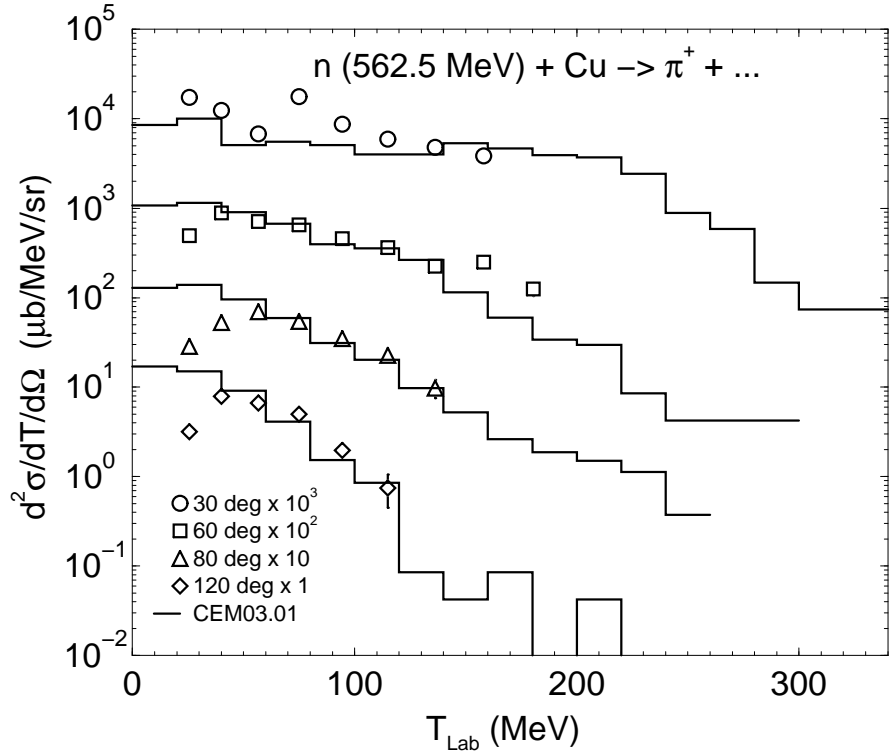}
\caption{
Experimental $\pi^+$ spectra from 562.5 MeV n + Cu \cite{Brooks91}
compared with CEM03.01 results \cite{CEM03.01}. 
}
\end{figure}

Fig.\ 38 shows an example of neutron spectra from 1.5 GeV $\pi^+$ + Fe
calculated by CEM03.01 compared with 
experimental data by Nakamoto {\it et. al.} \cite{Nakamoto97}.
We see again a reasonably good agreement of our results with these
measured data.  We obtain similar results for other pion-induced 
reactions.

\begin{figure}[ht]                                                 %Fig. 38
\centering
\includegraphics[width=100mm,angle=-90]{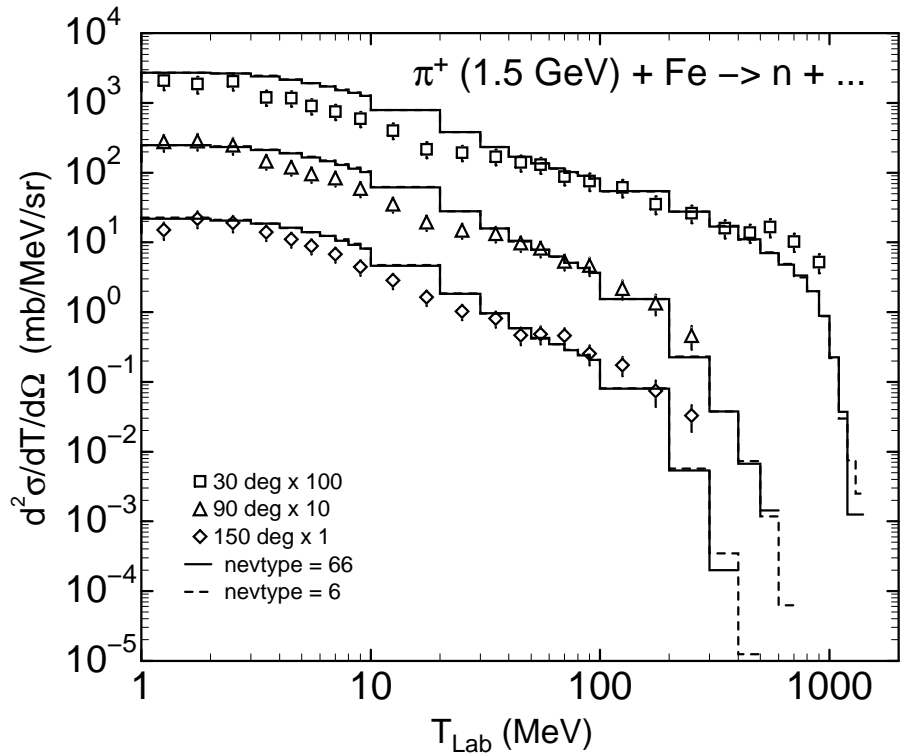}
\caption{
Experimental neutron spectra from 1.5 GeV $\pi^+$ + Fe \cite{Nakamoto97}
compared with CEM03.01 results \cite{CEM03.01}. 
The results shown in this figure are for one million simulated
inelastic events. The {\bf nevtype=66} option (see Section 6)
requires
3 hr 30 min 52 sec of computing time on a SunBlade 100, 500 MHz
computer, while the {\bf nevtype=6} option requires only 1 hr 51 min 58 sec,
providing almost the same results.
}
\end{figure}

The secondary neutral pions produced in ADS targets and shielding
materials would decay later into two photons, as their mean life
is of only $(8.4 \pm 0.6) \times 10^{-17}$ s. That is, 
the transport codes should transport the produced photons 
and our event generators should be able to describe 
properly their further interactions with various nuclei.

If the photon energy is above several GeV, such 
photonuclear interactions are calculated in transport codes
using our high-energy event generator LAQGSM03.01.
%, depending on the mass number of the nucleus, as discussed above. 
Several examples of
results by LAQGSM03.01 for photonuclear reactions
are shown in Fig.\ 14 of Section 3.2; more results 
for other high-energy photonuclear reactions
may be found in Refs.\ \cite{LAQGSM03.01,Varenna06}.

Interaction of photons of lower energies with nuclei
are calculated in transport codes using our low-energy
event generator CEM03.01.
Figs.\ 39 and 40 show examples of proton spectra from
300 MeV $\gamma + ^{64}$Cu and isotopic yields of 
products produced by bremsstrahlung
reactions on $^{197}$Au and $^{209}$Bi at $E_0 = 1$ GeV
calculated by CEM03.01 compared with experimental data
\cite{SCH82,Sakamoto03}.
Figs.\ 39 and 40 demonstrate that      
CEM03.01 allows us to describe reasonably well
many photonuclear reactions 
needed for ADS and other applications, as well as to analyze 
mechanisms of photonuclear reactions for fundamental
studies (see \cite{JNRS05} for more details).

It is well known that the most difficult characteristics of 
nuclear reactions to be described by any theoretical model are excitation
functions, {\it i.e.}, cross sections for the production of a given
isotope as functions of the energy of the projectile.
We have studied with different versions of our models
several thousands of excitations functions for various reactions
(see, {\it e.g.}, 
\cite{Titarenko07,Titarenko07a,Mashnik97,BigMedicalRep,Titarenko07b,p_FePRC}
and references therein); therefore, we limit ourself here
to analysis of only a few recent measurements published only
recently \cite{Ammon08} and calculated with CEM03.03
just the day before our present Advanced Workshop.

Figs.\ 41 to 43 show excitation functions for the production of
$^3$He, $^4$He, $^{21}$Ne, and $^{38}$Ar from interactions
of protons with Iron and Nickel measured recently by 
Ammon {\it et al.} \cite{Ammon08} 
compared with our CEM03.03 results, with results by TALYS \cite{TALYS} 
and by INCL/ABLA \cite{INCL,ABLA} from \cite{Ammon08},
with systematics
by Konobeev and Korovin \cite{Konobeev93} (for 
$^3$He and $^4$He from p+Fe), 
with previous measurements by other authors,
and with our predictions for the p+Fe excitation
functions we published in 1997 \cite{Mashnik97}.
On the whole, CEM03.03 is able to reproduce
quite well these new measurements.

%\newpage

\begin{figure}[ht]                                                 %Fig.39
\centering
\includegraphics[width=100mm,angle=-90]{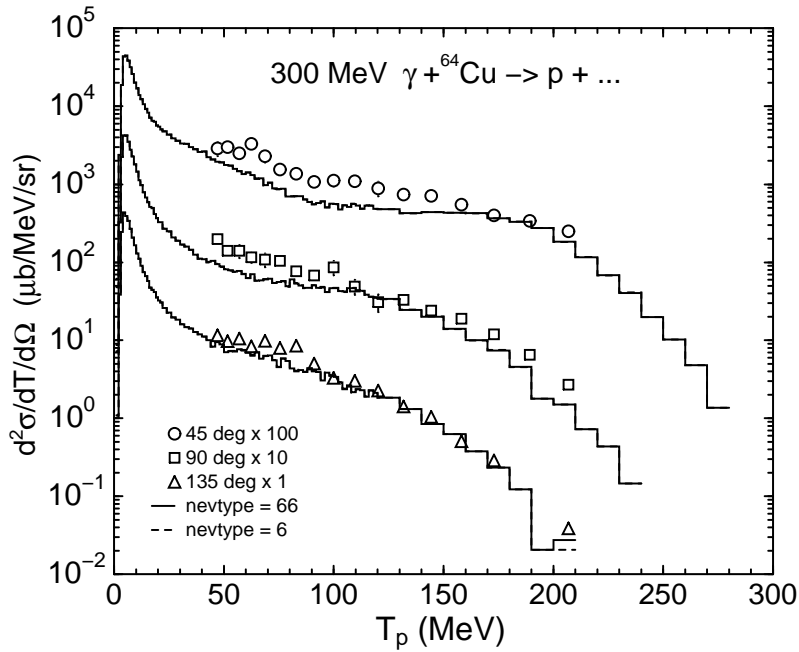}
\caption{
Proton spectra at 45$^{\circ}$, 90$^{\circ}$, and 135$^{\circ}$
from the reaction
300 MeV $\gamma$ + Cu. Symbols are experimental data from \cite{SCH82} and
histograms are CEM03.01 results \cite{CEM03.01}.
}
\end{figure}

It is interesting to observe on the right panels
of Figs.\ 41 and 42 that the CEM95 \cite{CEM95} 
version of CEM predicted these p+Fe excitation functions reasonably
well: the old CEM95 predictions \cite{Mashnik97} agree quite
well with the recently measured data \cite{Ammon08}.

We observe some disagreements of our results
with the new data, especially 
for the production of $^{21}$Ne at low energies from both Fe and Ni,
indicating us that there are still problems to be solved in CEM03.03
and we have to improve further our models
to describe properly this type of reaction. We see
that TALYS \cite{TALYS} and INCL/ABLA \cite{INCL,ABLA} also
encounter difficulties in reproducing these data,
indicating similar problems occur with other models.\\

{\large\bf 10. CEM03.S1, CEM03.G1, LAQGSM03.S1, and LAQGSM03.G1}\\

From the results presented already and in the cited references,
we can conclude that CEM03.01 and LAQGSM03.01 
and their later 03.02 and 03.03 versions are able to predict 
well a large variety of nuclear reactions of interest to ADS. 
Therefore, they can be employed with confidence as reliable event generators
in transport codes used as ``workhorses'' for ADS 
and other applications.
This is true only if we are not interested in the
production of intermediate-mass fragments from not too heavy 
targets at not too high energies, targets that do not fission from an 
``orthodox'' point of view, therefore not producing 
such fragments.
To be able to predict production of intermediate-mass fragments
from not too heavy targets at such intermediate energies with our
codes, we still need to solve some problems, just as do authors of 
other similar Monte-Carlo codes.

Such a problem is presented in the low-energy parts of 
the Fe(p,x)$^{21}$Ne and  Ni(p,x)$^{21}$Ne
excitation functions shown in Figs.\ 42 and 43;
two more examples,
%of such problems 
for other reactions, are 

%\newpage

%\begin{figure}[ht]                                                 %Fig.40
%\centering

\newpage
%%%%%%%%%%%%%%%%%%%%%%%%%%%%%%%%%%%%%%%%%%%%%%%%%%%%%%%%%%%%%%      Fig.40
\vspace{-5mm}
\begin{figure}[!ht]
\begin{center}

\includegraphics[width=13.0cm]{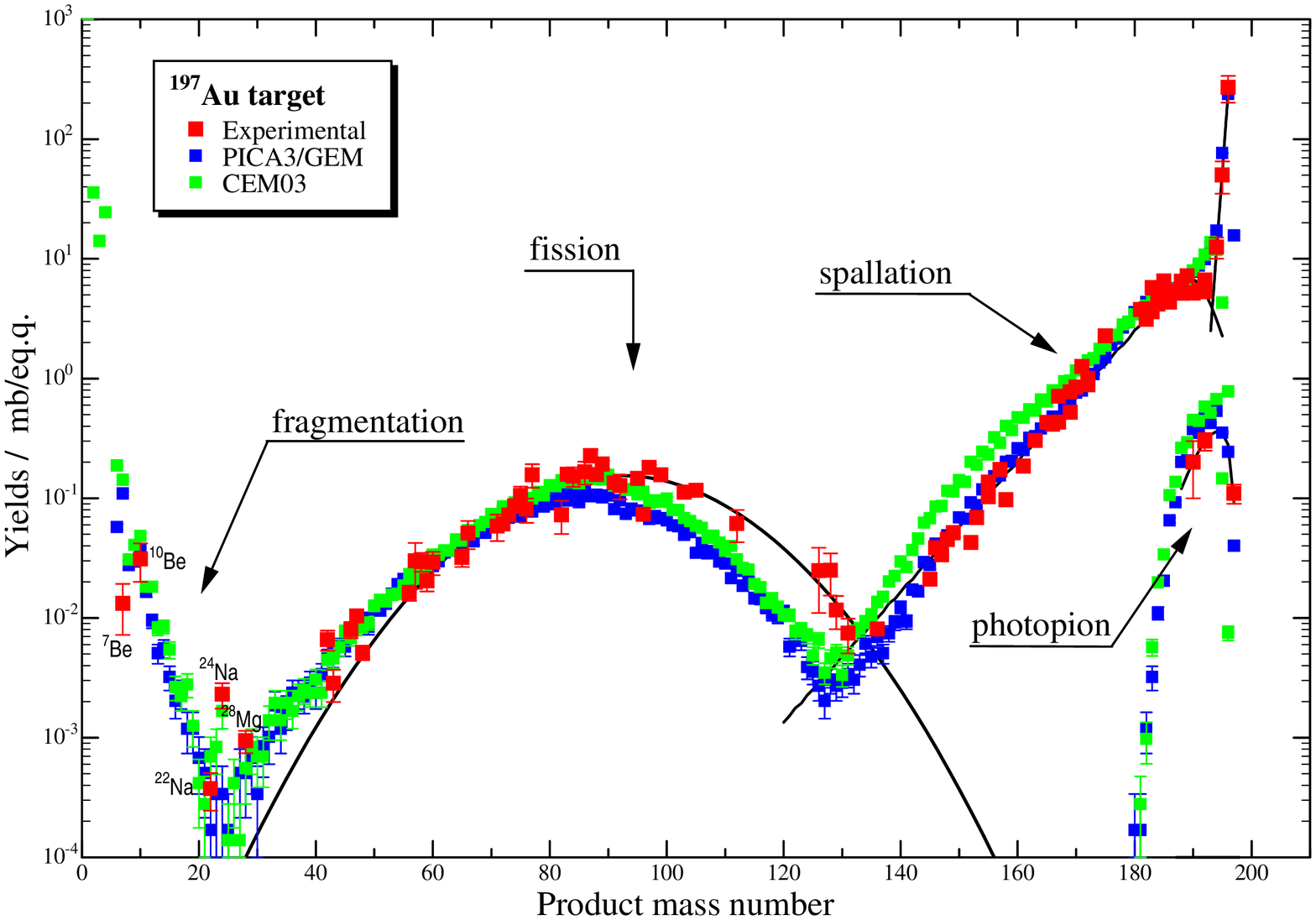}
\includegraphics[width=13.0cm]{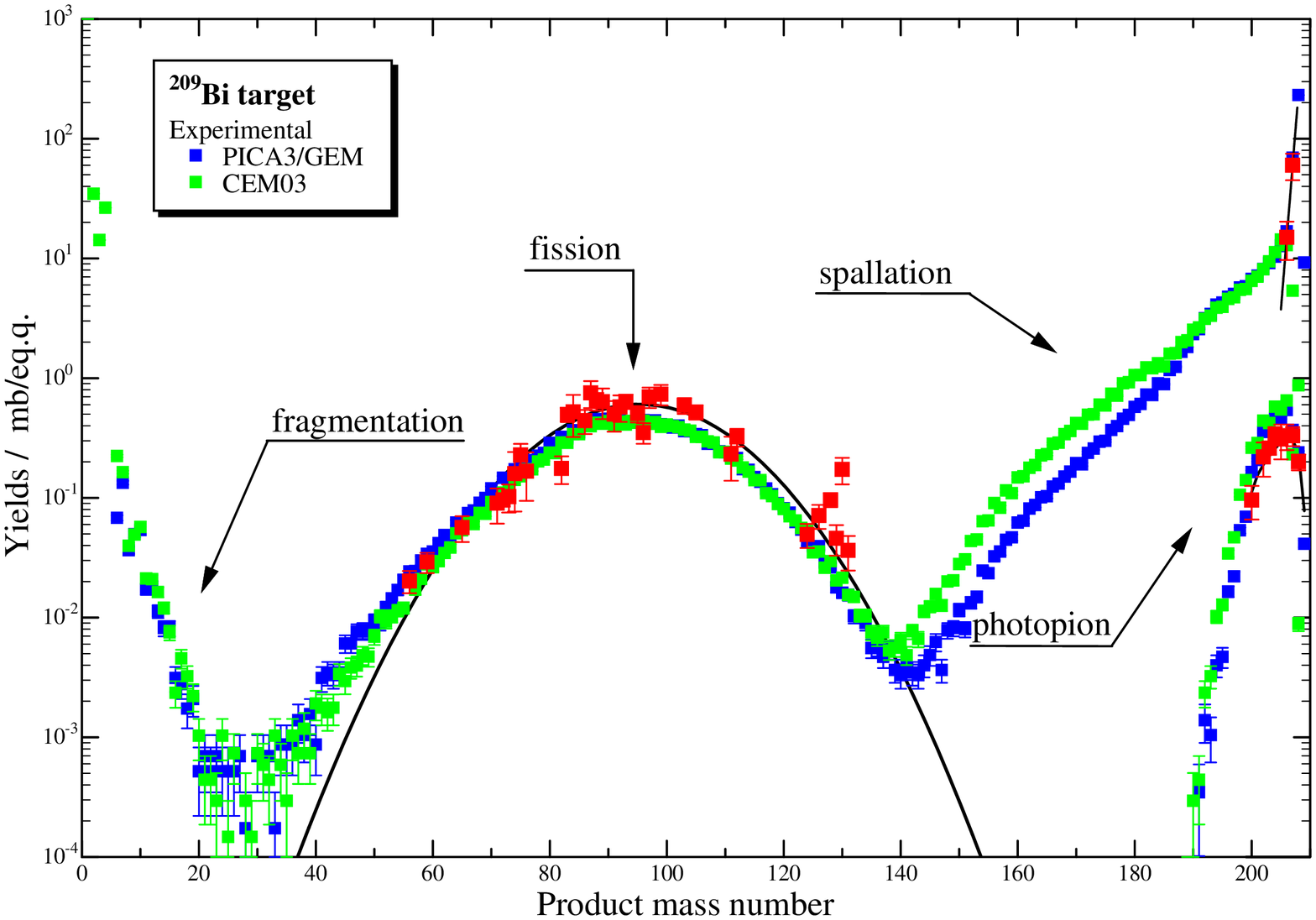}
%\epsfig{figure=Bi1000v2.eps,width=125mm,height=85mm,angle=-0}
%\includegraphics[width=100mm,angle=-0]{Au1000v2.eps}
%
%\vspace*{-2mm}
%\epsfig{figure=Bi1000v2.eps,width=125mm,height=85mm,angle=-0}
%\includegraphics[width=100mm,angle=-0]{Bi1000v2.eps}
%
%\vspace*{-2mm}
\caption{
Comparison of CEM03.01 results (green symbols)
for the isotopic yields of products produced by bremsstrahlung
reactions on $^{197}$Au and $^{209}$Bi at $E_0 = 1$ GeV
with experimental data (red symbols)
from the review %[15]
\cite{Sakamoto03}
and calculations by PICA3/GEM (blue symbols);
the PICA3/GEM results are from several publications and
are presented in Fig.\ 18 of %[15]
\cite{Sakamoto03} 
with the corresponding citations. The mass yields for the fission
products shown by black curves represent approximations based on 
experimental data by Prof.\ Sakamoto's group.
This figure was done for us by  Dr.\  Hiroshi Matsumura
by adding our CEM03.01 results to Fig.\ 18 of the review %[15].
\cite{Sakamoto03}.
}
\end{center}
\end{figure}
\clearpage

\begin{figure}[ht]                                                 %Fig.41
\centering

%\hspace*{-20mm}
\includegraphics[width=100mm,angle=90]{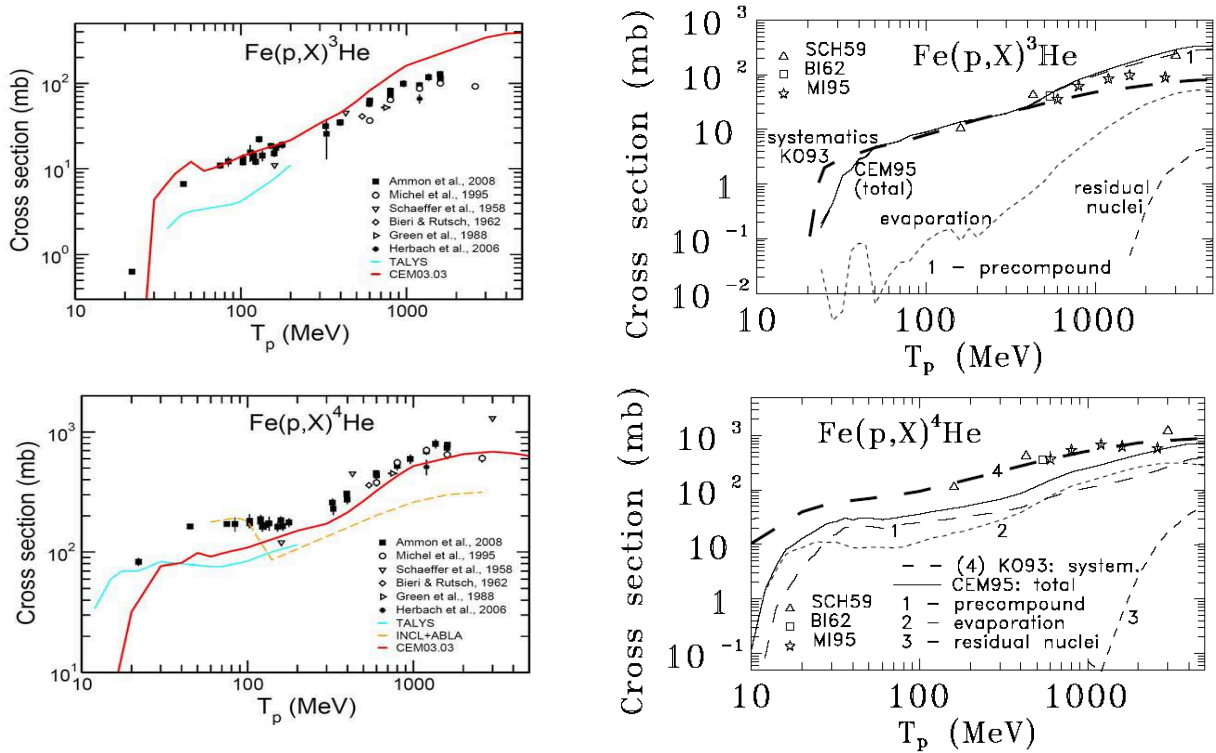}
\caption{
Excitation functions for the production of $^3$He and $^4$He
from p+Fe.
{\bf Left panel:}
Recent measurements by Ammon {\it et al.} \cite{Ammon08} (filled squares)
compared with our CEM03.03 results (red lines), with
results by TALYS \cite{TALYS} (blue line) from  \cite{Ammon08},
and with previous measurements shown with
different symbols, as indicated (see references to old
data in \cite{Ammon08}).
{\bf Right panel:}
The same excitation functions as on the left,
but predicted twelve years ago \cite{Mashnik97}
with the CEM95 version of CEM (solid lines),
compared with the experimental data available at that time
(see references in \cite{Mashnik97} and \cite{Ammon08}), 
and with systematics
by Konobeev and Korovin \cite{Konobeev93} (thick long-dashed lines).
Contributions from preequilibrium emission,
evaporation, and from residual nuclei to the total CEM95
yields are shown by thin dashed lines, as indicated.
}
\end{figure}

{\noindent
indicated in Figs.\ 44 and 45.  
}
Fig.\ 44 shows the 
mass distribution of the product yields from the
reaction 660 MeV p + $^{129}$I 
measured recently at JINR, Dubna \cite{Adam04} 
compared with our calculations with three recent versions of CEM.
The standard version of CEM03.01
does not describe production of isotopes with mass number 
$26 < A < 63$ from this reaction
observed in the experiment \cite{Adam04}. These
products are too heavy to be evaporated from compound nuclei
and the target is too light to fission in GEM2, therefore not
producing these isotopes as fission fragments (CEM03.01 and 
LAQGSM03.01 consider only ``conventional" fission of preactinides
and actinides and do not consider at all fission of nuclei with
$Z < 65$).

Fig.\ 45 shows 
a comparison \cite{ND2004} of the  mass distributions 
of the yields of eight isotopes from Na to Mn produced in the 
reactions 1500, 1000, 750, 500, and 300 MeV/A $^{56}$Fe + p
measured recently at GSI \cite{Villagrasa}
with three recent versions of LAQGSM.
We see that the standard version of LAQGSM03, referred as ``03.01'', 
fails to reproduce correctly 
production of fission-like heavy fragments from reactions 
with a medium-mass nuclear target at intermediate energies
(see the solid lines on the bottom two panels of Fig.\ 45), 
just as do all previous versions, the standard CEM03.01 and 
its predecessors, and most other currently 
available models (see, {\it e.g.}, \cite{Villagrasa,Baznat04}).
As already mentioned above, such nuclear targets are considered too
light to fission in conventional codes (including GEM2 and all models
currently employed in large-scale transport codes). Similarly, 
%{\noindent
the fragments are too light
to be produced as spallation residues at these intermediate energies
and too heavy to be produced via standard evaporation models.
%}

\begin{figure}[ht]                                                 %Fig.42
\centering

%\hspace*{-20mm}
\includegraphics[width=100mm,angle=90]{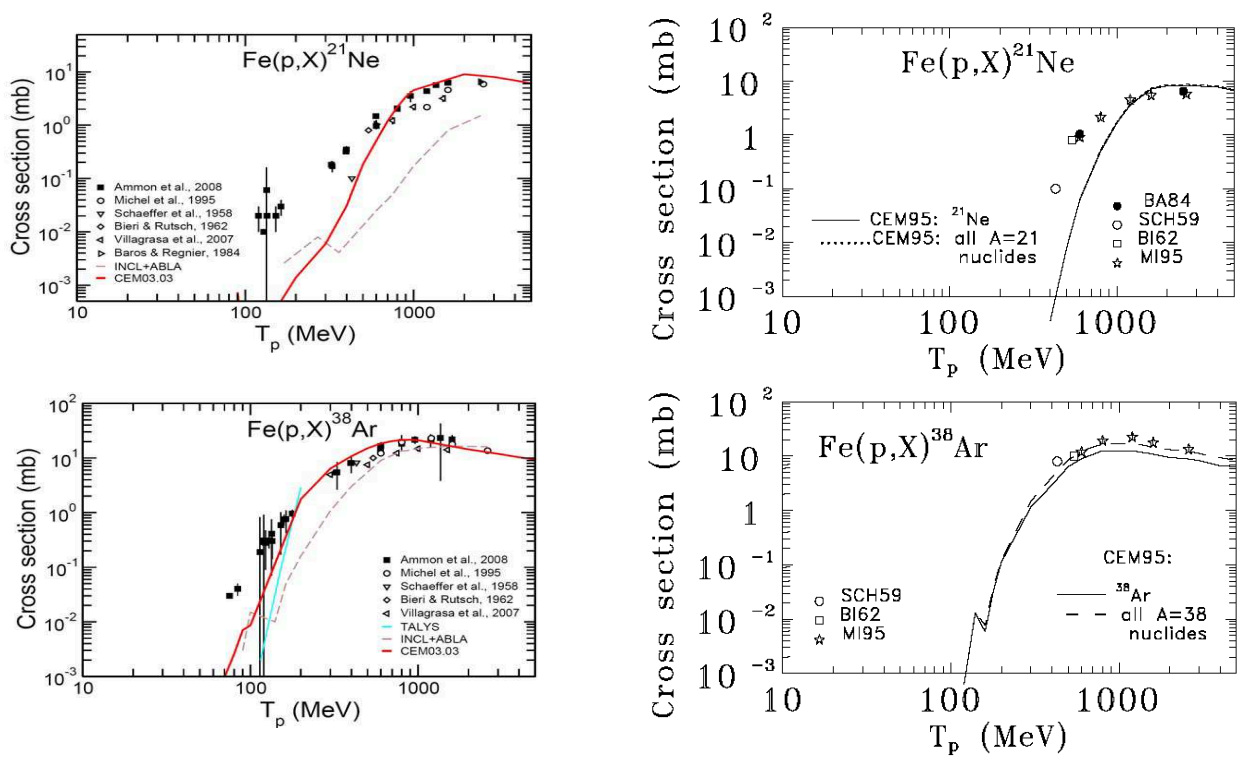}
\caption{
Excitation functions for the production of $^{21}$Ne and $^{38}$Ar
from p+Fe.
{\bf Left panel:}
Recent measurements by Ammon {\it et al.} \cite{Ammon08} (filled squares)
compared with our CEM03.03 results (red lines), with
results from TALYS \cite{TALYS} (blue solid line),
and from INCL/ABLA \cite{INCL,ABLA} (dashed brown line)
from \cite{Ammon08},
and with previous measurements shown with
different symbols, as indicated (see references to old 
data in \cite{Ammon08}).
{\bf Right panel:}
The same excitation functions as on the left,
but predicted twelve years ago \cite{Mashnik97}
with the CEM95 version of CEM,
compared with experimental data available at that time 
(see references in \cite{Mashnik97} and \cite{Ammon08}),
as indicated.
}
\end{figure}

We note that the problem of a reliable description
of intermediate-mass fragments from not too
heavy targets is still unresolved not 
only in our standard versions of CEM and LAQGSM, but also in most
other similar Monte-Carlo codes. Fig.\ 46 and 47 prove this with an 
example of comparison  \cite{p_FePRC}
the mass distributions of $^{56}$Fe(p,x) 
reaction products measured at ITEP \cite{p_FePRC} and GSI
\cite{Villagrasa} at 300 and 1000 MeV energies with results from
15 different different code systems:  
MCNPX (INCL, CEM2k, BERTINI, ISABEL), 
LAHET (BERTINI, ISABEL), CEM03 (.01, .G1, .S1), 
LAQGSM03 (.01, .G1, .S1), CASCADE-2004, LAHETO, and BRIEFF
(see references and details in \cite{p_FePRC}). 
We see that all these models give a sufficiently good description 
of the mass yields of the products close to target nucleus 
mass ($A > 35$--40). 
In the mass range $A<30$, however, a good description of the 
measured product nuclide yields is only given by the models 
that, apart from the conventional evaporation of 
nucleons and complex particles up to $^4$He,
allow for evaporation of heavy clusters (the CEM and LAQGSM 
versions). 

We also note that none of these models gives a good quantitative 
description of the whole set of experimental data.
However, as can be seen from Fig.\ 47, while still having these
problems, our CEM and LAQGSM codes provide the lowest
mean deviation factors between the calculated and measured
\cite{Villagrasa,p_FePRC} cross sections, averaged over all
energies of the beams and detected products in comparison with
other models (see more details in  \cite{p_FePRC}).

\begin{figure}[ht]                                                 %Fig.43
\centering

%\hspace*{-20mm}
\includegraphics[width=160mm,angle=0]{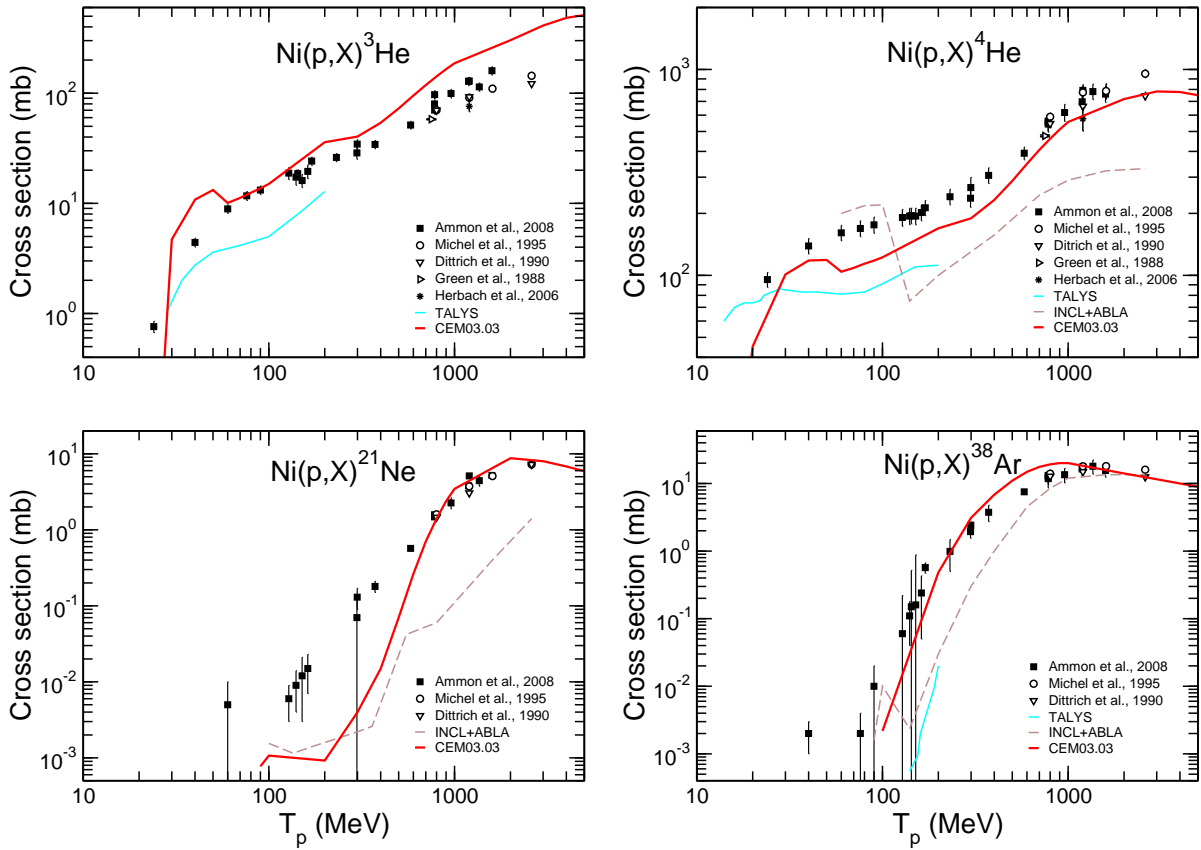}
\caption{
Excitation functions for the production of 
$^3$He, $^4$He,
$^{21}$Ne and $^{38}$Ar
from p+Ni.
Recent measurements by Ammon {\it et al.} \cite{Ammon08} (filled squares)
compared with our CEM03.03 results (red lines), with
results from TALYS \cite{TALYS} (blue solid line),
and from INCL/ABLA \cite{INCL,ABLA} (dashed brown line)
from \cite{Ammon08},
and with previous measurements shown with
different symbols, as indicated (see references to old 
data in \cite{Ammon08}).
}
\end{figure}

We have addressed this problem 
in two different ways \cite{ND2004,01s1g1,Baznat04}:

1) By implementing into CEM03.01 and LAQGSM03.01 the
 Statistical Multifragmentation Model (SMM) by Botvina {\it et al.} 
\cite{Botvina87}, \cite{SMM}--\cite{SMM08},
to consider multifragmentation as a mode competitive
to evaporation of particles and light fragments, when the
excitation energy $E^*$ of a compound nucleus produced after the
preequilibrium stage of a reaction is above $2\times A$ MeV. 
This way, we 
have produced the ``S" version of our codes (``S" stands for SMM),
CEM03.S1 and LAQGSM03.S1.

CEM03.S1 and LAQGSM03.S1 are exactly the same as CEM03.01 and 
LAQGSM03.01, but
consider also multifragmentation of excited nuclei produced 
after the preequilibrium stage of
reactions, when their excitation energy is above $2\times A$ MeV,
using SMM.
In SMM, within the total accessible phase space, a 
micro-canonical ensemble of all breakup
configurations, composed of nucleons and excited 
intermediate-mass fragments governs the
disassembly of the hot remnant. The probability of 
different channels is proportional to their
statistical weight. Several different breakup 
partitions of the system are possible. 
When after the preequilibrium stage of a reaction
$E^* > 2\times A$ MeV and we ``activate'' SMM to calculate
multifragmentation in the 03.S1 codes, the competitive evaporation
process are calculated also with a version of the
evaporation model by Botvina {\it et al.} from SMM, 
rather than using GEM2, as we do always in the standard
03.01 versions and in 03.S1 when $E^* \leq 2\times A$ MeV
and SMM is not invoked.

A very detailed description of SMM together with many 
results obtained for many reactions may be found in 
\cite{Botvina87}, \cite{SMM}--\cite{SMM08}
and references therein; 
many %useful 
details on SMM are presented
in the first paper of Ref.\ \cite{01s1g1}, therefore 
we do not repeat this here.

2) By replacing the Generalized Evaporation Model 
GEM2 by Furihata \cite{Furihata1}--\cite{Furihata3}
used in CEM03.01
and LAQGSM03.01 with the fission-like binary-decay model GEMINI
of Charity \cite{GEMINI}--\cite{GEMINI08}
which considers production of all possible fragments. 
This way, we have produced the ``G" version of our codes 
(``G" stands for GEMINI), CEM03.G1 and LAQGSM03.G1. 

CEM03.G1 and LAQGSM03.G1 are exactly the same as CEM03.01 and 
LAQGSM03.01, but use GEMINI instead of using GEM2.
Within GEMINI, a special treatment based on the Hauser-Feshbach 
formalism is used to calculate emission of the lightest particles, 
from neutrons and protons up to beryllium isotopes. 
The formation of heavier nuclei than beryllium is modeled 
according to the transition-state formalism developed 
by Moretto \cite{Moretto75}. 
All asymmetric divisions of the decaying compound nuclei are 
considered in the calculation of the probability of successive 
binary-decay configurations. GEMINI is described in details 
in the original publications  \cite{GEMINI}--\cite{Moretto75};
many details on GEMINI may be found also in the first paper 
of Ref.\ \cite{01s1g1}, therefore we do not elaborate here.

\begin{figure}[ht]                                                 %Fig.44
\centering

%\hspace*{-20mm}
\includegraphics[width=110mm,angle=-90]{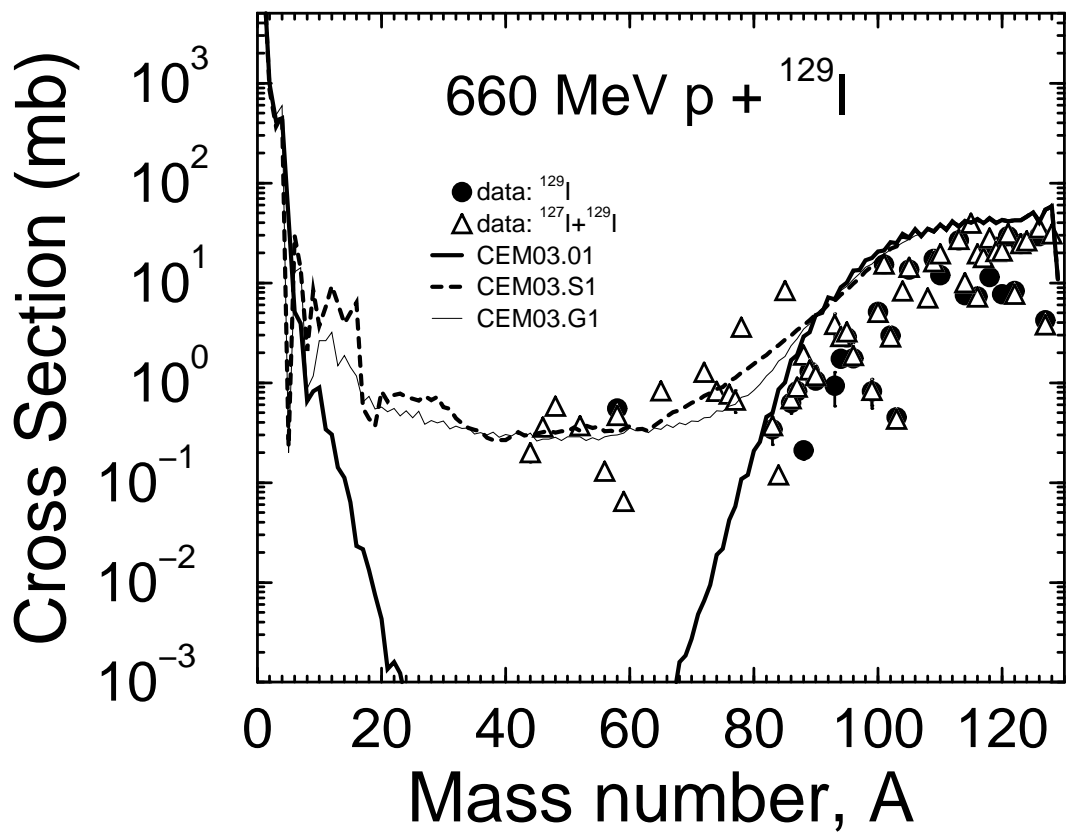}
\caption{Experimental \cite{Adam04}
mass number distribution of the product yield
from 660 MeV p + $^{129}$I compared with
calculations by
 CEM03.01
(thick solid line), CEM03.S1 (thick dashed line),
and CEM03.G1 (solid thin line), as indicated
}
\end{figure}

%\newpage
\begin{figure}[ht]                                                 %Fig.45
\centering

\vspace*{-15mm}
\includegraphics[width=170mm,angle=-0]{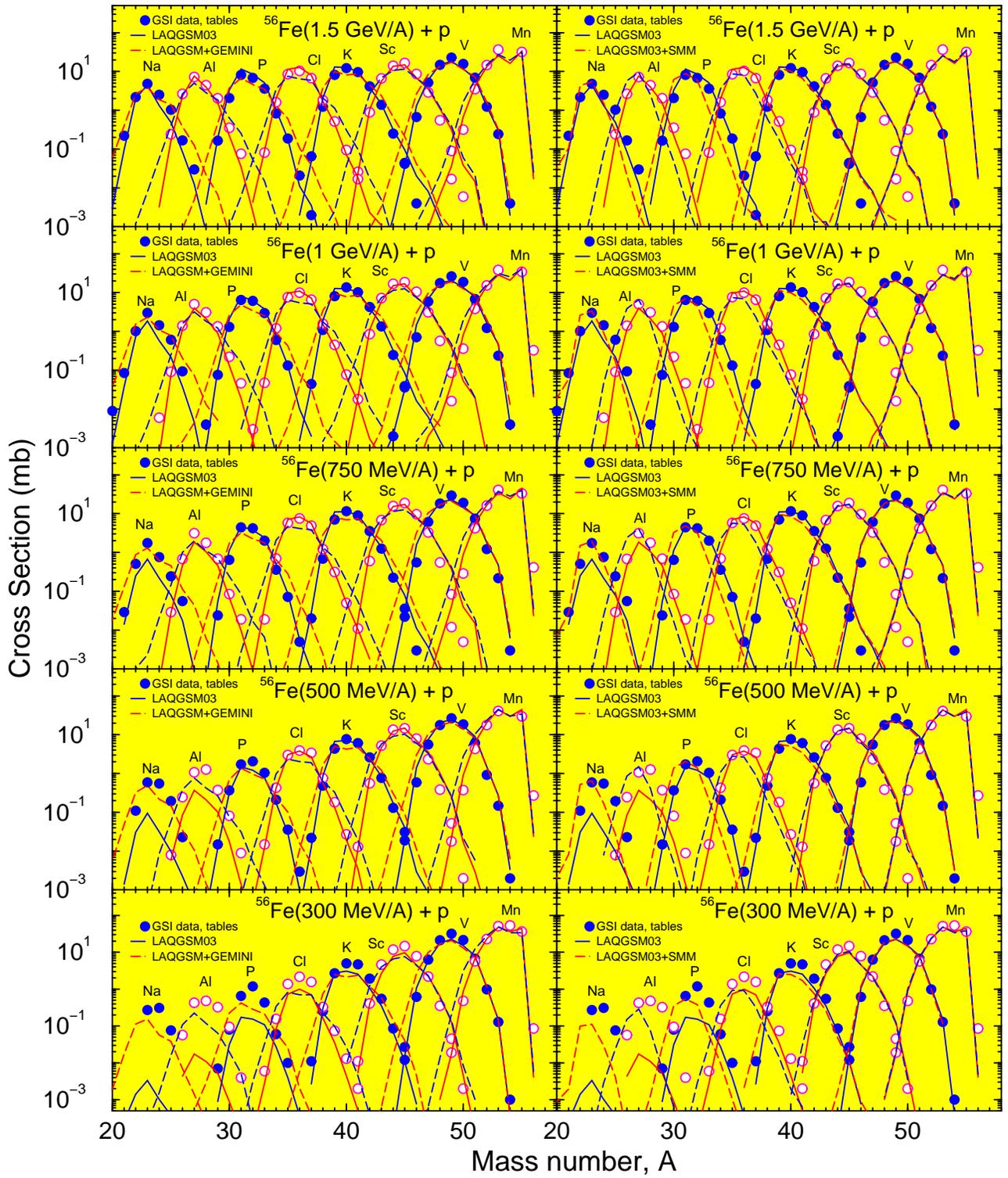}
\caption{
Experimental \cite{Villagrasa} mass distributions 
of the yields of eight isotopes from Na to Mn produced in the 
reactions 1500, 1000, 750, 500, and 300 MeV/A $^{56}$Fe + p
compared with LAQGSM03 (solid lines on both panels), 
LAQGSM03+GEMINI (dashed lines, left panel), and
LAQGSM03+SMM (dashed lines, right panel) results, respectively.
}
\end{figure}

\clearpage

%\newpage
\begin{figure}[ht]                                                 %Fig.46
\centering

%\vspace*{-30mm}
\includegraphics[width=150mm,angle=-0]{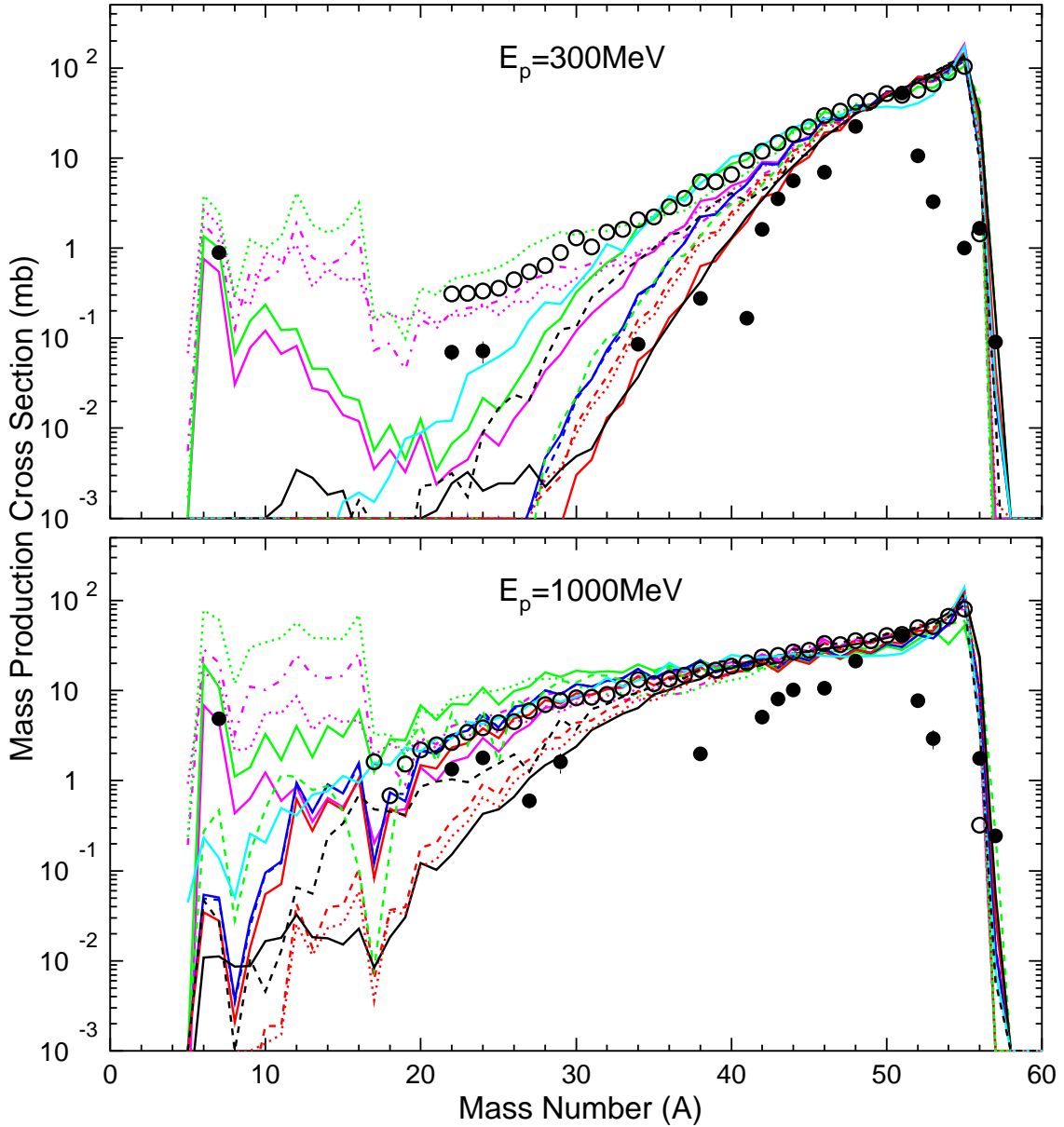}
\caption{
Mass distributions of $^{56}$Fe(p,x) reaction products measured
at ITEP (filled circles) \cite{p_FePRC} and GSI
(open circles) \cite{Villagrasa} for 300  and 1000 MeV
energies compared with results by %15 
14
codes shown
with different lines as following:
INCL/MCNPX (solid black), BRIEFF1.5.4g (dashed black), 
CEM03.01 (solid green), CEM2k/MCNPX (dashed green), 
CEM03.G1 (dotted green), %CEM03.S1 (dashed-dotted green), 
BERTINI (MCNPX - solid blue, LAHET -  dashed blue), 
ISABEL (MCNPX - solid red, LAHET -  dashed red, 
LAHETO - dotted red), LAQGSM03.01 (solid magenta), 
LAQGSM03.G1 (dotted magenta), LAQGSM03.S1 (dashed-dotted magenta), 
CASCADE-2004 (cyan). 
This figure is adopted from Ref.\ \cite{p_FePRC}, where
more details on this comparison may be found.
}
\end{figure}

\clearpage

%\newpage
\begin{figure}[ht]                                                 %Fig.47
\centering

%\vspace*{-30mm}
\includegraphics[width=90mm,angle=-90]{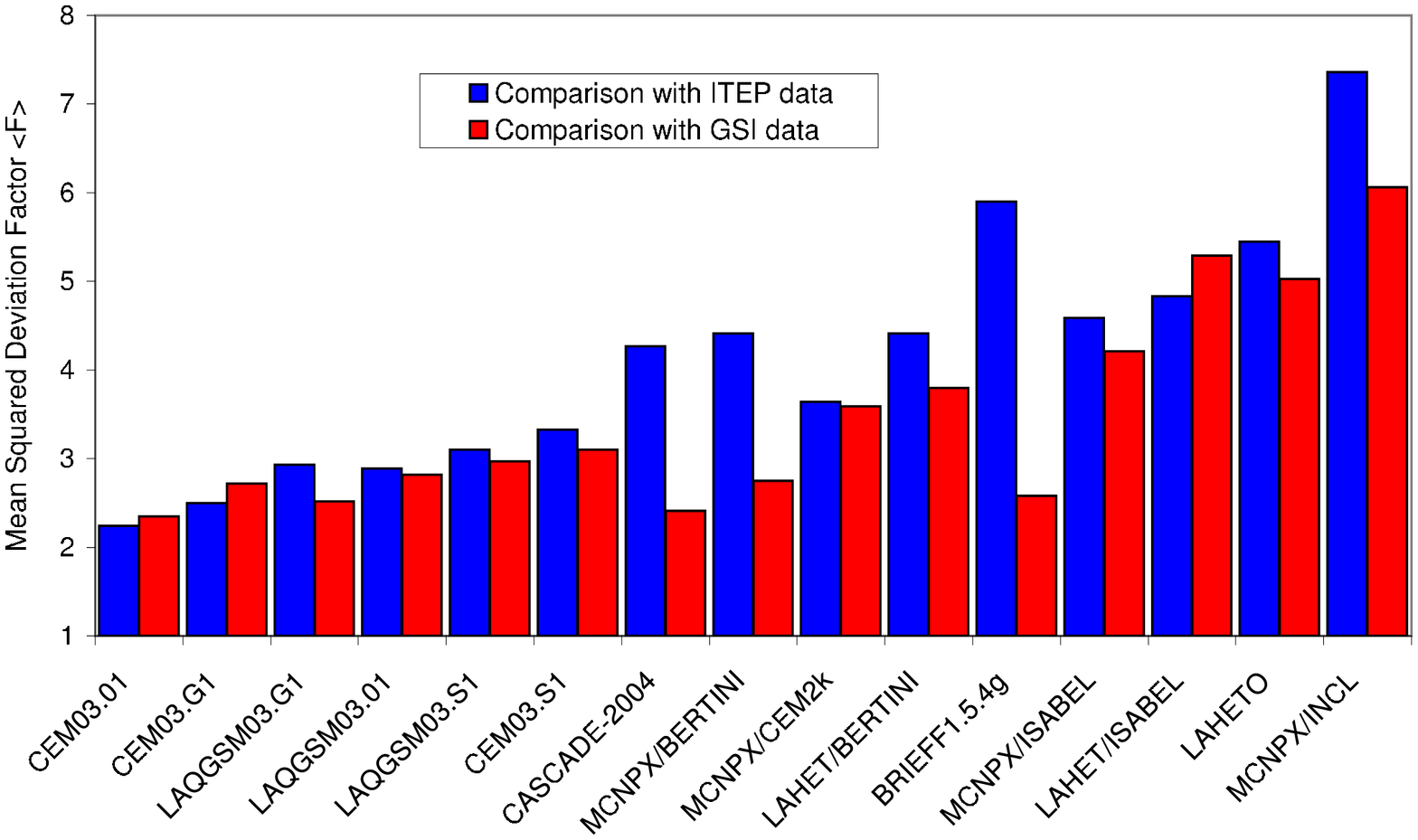}
\caption{
The predictive power of each tested code for the data shown in
Fig.\ 46: the mean-square deviation 
factors between calculations and measured
cross sections averaged over all energies 
and all products (see details in \cite{p_FePRC}).
}
\end{figure}

%\clearpage

Results by ``S'' and ``G'' versions of our codes are shown
together with standard 03.01 results 
in Figs.\ 44, 45, and 46
(see also Figs.\ 28 and 29 in Section 6).
We see that both ``S'' and ``G'' versions reproduce almost
equally well
the yields of intermediate-mass fragments from these reactions,
that can not be described by the standard CEM03.01 and LAQGSM03.01.

From one point of view,
this is an achievement, as, with the ``S'' and 
``G'' versions, we are able to describe reasonably well the 
reactions where the standard 03.01 versions do not work. 
On the other hand,
this also makes the situation more intricate, as from these 
(and other similar) results 
we can not choose easily between the ``S'' and ``G'' versions,
that makes more difficult to determine the real mechanisms of
such reactions.
We think that for such 
intermediate-energy proton-induced reactions the 
contribution of multifragmentation to the production of heavy 
fragments should not be very significant due to the relatively
low excitation energies involved. Such fragments are more 
likely to be produced via the fission-like binary decays 
modeled by GEMINI. Multifragmentation processes are
important and should be considered in reactions involving
higher excitation energies; at excitations probably
higher than the 2 MeV/nucleon considered here. 
We conclude \cite{ND2004} that it is impossible to make a
correct choice between fission-like and fragmentation reaction
mechanisms involved in these (or other similar)
reactions merely by comparing model results with the 
measurements of only product cross sections; addressing 
this question will probably require analysis of two- or multi-particle 
correlation measurements. Our conclusion  \cite{ND2004}
was confirmed by the recent coincidence measurements and calculations 
of residual and light particles in 1 GeV/A $^{56}$Fe + p
by Gentil {\it et al.} \cite{Gentil08,Leray_NUFRA07}.

Several more examples of the difficult situation with 
determination of the mechanisms of fragment production from 
reactions induced by intermediate-energy projectiles
using the ``S'' and ``G'' versions of our codes
are presented below in Figs.\ 48 to 51.

Fig.\ 48 shows data
recently measured at GSI \cite{Napolitani07}
on mass- and charge-product yield distributions
and mean kinetic energies of all products as a
function of the product mass number from the
reaction 1 GeV/nucleon $^{136}$Xe + p compared with calculations by 
the standard CEM03.03 and LAQGSM03.03, as well
as by their ``S'' versions.
More results for this reaction may be found in Ref.\ \cite{Mashnik_NUFRA07}.
We see that the standard 03.03 versions of both CEM and
LAQGSM codes fail to reproduce production of intermediate
mass fragments with $20 < A < 70$ from this reaction.
With the ``S'' versions, we performed calculations
for several different
values of the excitations energy $E^*$ of the nuclei produced
after the preequilibrium stage of reactions
when we start to consider
multifragmentation as a competitive to evaporation
mechanism of nuclear reactions, namely, at
$E^* > 2$, 4, 4.5, and 5 MeV/nucleon
in the case of CEM03.S1, and 
$E^* > 1.5$, 2, and 4 MeV/nucleon 
for LAQGSM03.S1. We see that the best agreement
of results by LAQGSM03.S1 (solid red lines in Fig.\ 48)
with the GSI data  \cite{Napolitani07} were achieved
for $E^* > 2$ MeV/nucleon, 
just as Dr.\ Botvina suggested to us and how it was implemented
in both LAQGSM03.S1 and CEM03.S1 versions of our codes.
But in the case of CEM03.S1 (solid blue lines in Fig.\ 48), 
we got the best agreement with the data for 
a higher energy, namely $E^* > 4.5$ MeV/nucleon.

%\newpage
\begin{figure}[ht]                                                 %Fig.48
\centering

%\vspace*{-30mm}
\includegraphics[width=160mm,angle=-0]{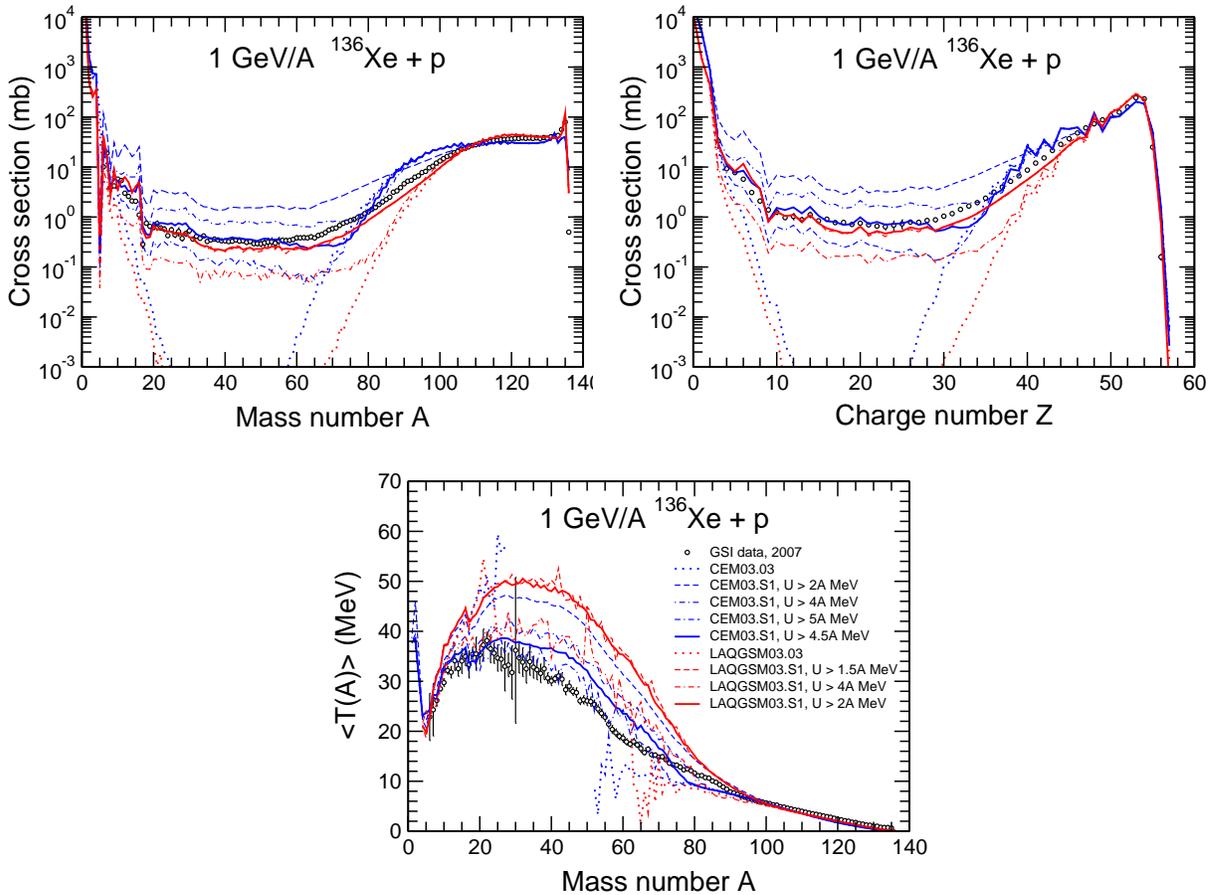}
\caption{Mass- and charge-product yield distributions
and mean kinetic energy of all products as 
functions of the product mass number from the
reaction 1 GeV/nucleon $^{136}$Xe + p \cite{Napolitani07} 
(open circles) recently measured at GSI
compared with calculations by 
the standard CEM03.03 and LAQGSM03.03 (dotted lines), 
as well as by their ``S'' versions (solid and dot-dashed lines)
for different
values of the excitations energy $E^*$ of the nuclei produced
after the preequilibrium stage of reactions above which
we start to consider multifragmentation as a competitive 
reaction mechanism, namely, at $E^* > 2$, 4, 4.5, and 5 
MeV/nucleon in the case of CEM03.S1, and $E^* > 1.5$, 2, and 4
MeV/nucleon for LAQGSM03.S1, as indicated.
}
\end{figure}

The fact that we get 
different ``best'' values of $E^*$ when we need to ``activate''
multifragmentation as a competition to the simple evaporation
mechanism in different codes is not contradictory and does imply
anything physical: The INC
of CEM is completely different from the one of LAQGSM, therefore
the mean mass $<A>$ and charge number $<Z>$ of residual nuclei 
produced after the preequilibrium stage of a reaction 
%(the same in CEM and LAQGSM)
with an excitation $E^*$ higher than a certain values, {\it e.g.}, 
2 MeV/nucleon, should be also different, as should be
the distributions of such nuclei with respect to their $E^*$.
This results in turn in different fragments produced via 
multifragmentation from this reaction as predicted by CEM03.S1 and
LAQGSM03.S1. If we would use another INC, {\ e.g.}, INCL \cite{INCL},
followed or not by preequilibrium emission, we would get 
still another ``best'' value of $E^*$ when we need to start considering 
multifragmentation, as another INC would predict other values of
$<A>$, $<Z>$, and distributions of $E^*$. 
This makes the situation of considering
multifragmentation by different codes more intricate:
the condition of its ``activation'' seems to be model-dependent.

   This situation becomes even more unclear if we refer to the results
shown in Fig.\ 49; the same data \cite{Napolitani07}
can be reproduced, even a little better, with the ``G'' versions
of our codes, without considering multifragmentation at all.

%\newpage
\begin{figure}[ht]                                                 %Fig.49
\centering

%\vspace*{-30mm}
\includegraphics[width=160mm,angle=-0]{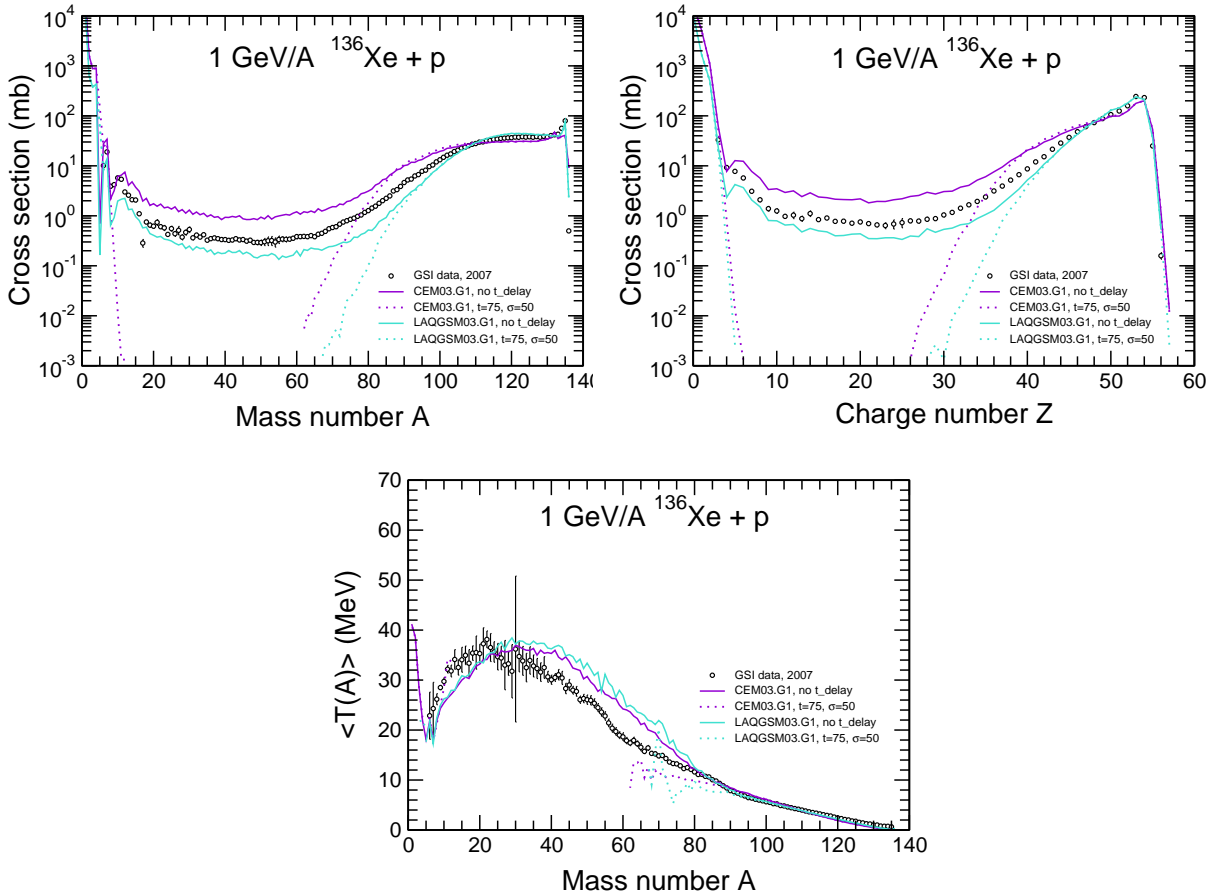}
\caption{
The same experimental data \cite{Napolitani07}
as in previous figure, but compared with results
of calculations by CEM03.G1 (violet lines)
and LAQGSM03.G1 (turquoise lines),
as indicated.
}
\end{figure}

%\clearpage
As described above,
CEM03.G1 and LAQGSM03.G1 are exactly the same as CEM03.01 and 
LAQGSM03.01, only replacing GEM2 \cite{Furihata1}--\cite{Furihata3}
with GEMINI \cite{GEMINI}--\cite{GEMINI08}.
As can be seen from Figs.\ 28, 29, 49, 50,
and from results presented in Refs.\ \cite{01s1g1,Baznat04,Mashnik_NUFRA07},   the ``G''

\newpage
\begin{figure}[ht]                                                 %Fig.50
\centering

%\vspace*{-2mm}
\includegraphics[width=150mm,angle=-0]{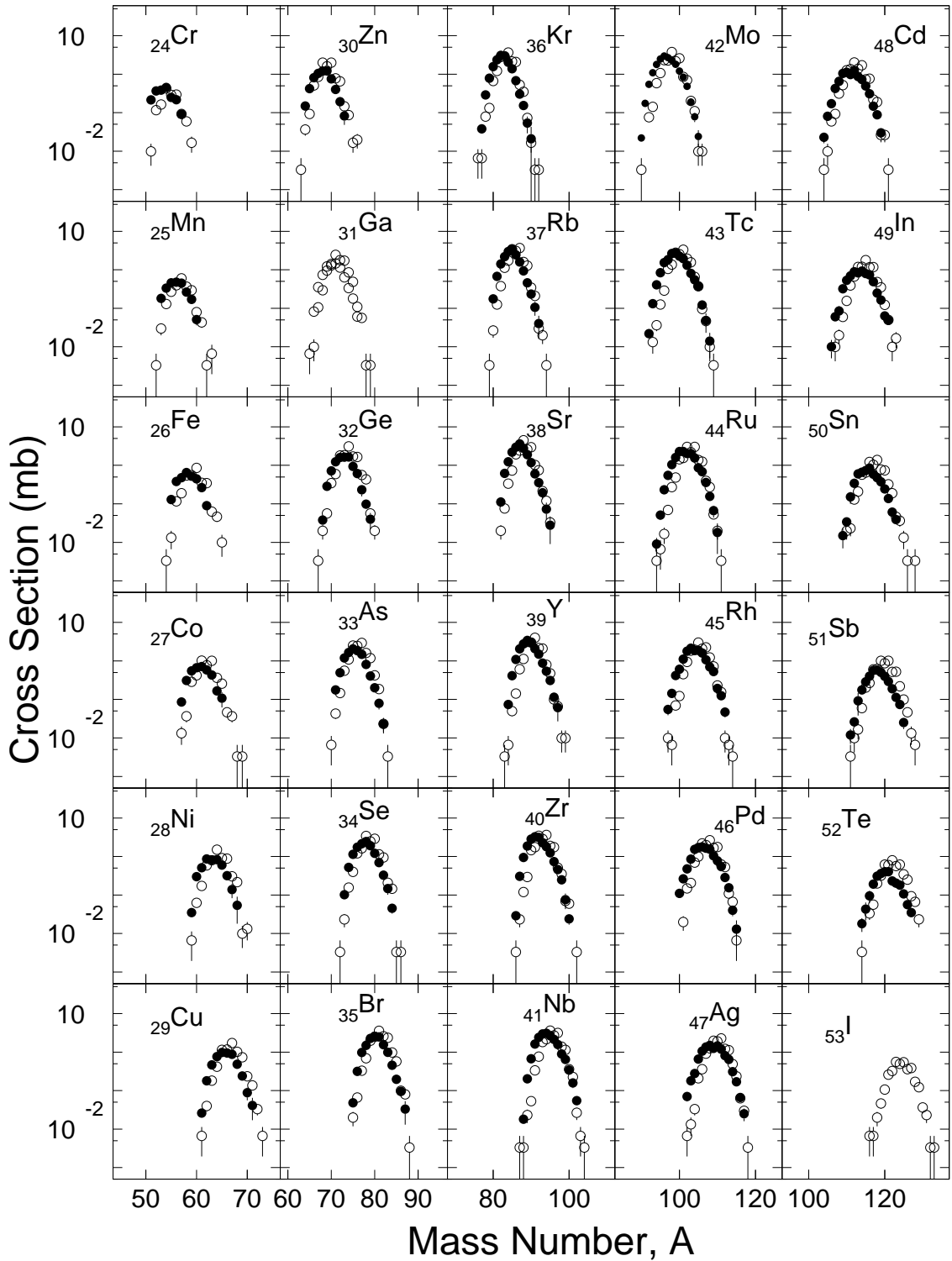}
\caption{
Comparison of all measured \cite{Enqvist01}
cross sections of fission  products from the reaction 1 GeV/nucleon
$^{208}$Pb on $p$ (filled circles) with LAQGSM+GEMINI (LAQGSM03.G1) 
results (open circles). Experimental data for isotopes
of Ga and I are not available, so we present only our predictions.
{\it t\_delay} = 75 and {\it sig\_delay} = 50 are used in GEMINI
to calculate this reaction.
}
\end{figure}
\clearpage

{\noindent
version of our codes allows us to
describe well many fission and fragmentation reactions,
especially on targets below the actinide region,
where the standard versions of our codes using GEM2 may not work 
well.}

We find that: 1) GEMINI merged with CEM or LAQGSM
provides reasonably good results for medium-heavy targets
without a fission delay time; 2) For preactinides, we have to use
{\it t\_delay} = 50--70 and {\it sig\_delay} = 1--50, otherwise
GEMINI provides too much fission---this 
may be related to the calculation of fission barriers of 
preactinides without strong ground-state shell corrections in the 2002
version of GEMINI we use;
3) The current version of GEMINI does not work well for low-energy
fission of actinides (see Fig.\ 51).

%\newpage
\begin{figure}[ht]                                                 %Fig. 51
\centering

%\vspace*{-30mm}
\includegraphics[width=100mm,angle=-0]{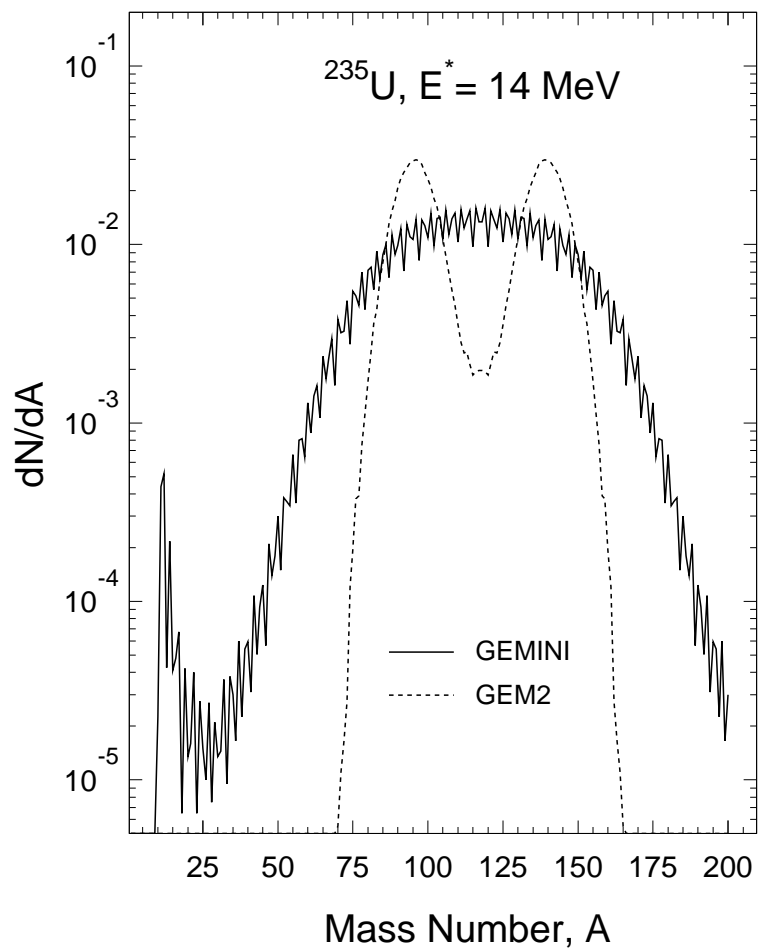}
\caption{
Mass distributions of fission fragments
from the compound nucleus $^{235}$U with an excitation of 14 MeV as
predicted by GEMINI (solid line) and GEM2 (dashed line),
respectively (adopted from \cite{Baznat04}).
No delay time in GEMINI for fission is used here. 
}
\end{figure}

The physical basis of these shortcomings of GEMINI are understood
and are discussed along with possible solutions to be employed
in future versions of GEMINI by Dr.\ Charity at this Workshop
\cite{GEMINI08}.\\

{\large\bf 11. Summary}\\

Improved versions of the cascade-exciton model of nuclear 
reactions and of the Los Alamos
quark-gluon string model have been developed recently at LANL 
and implemented in the codes CEM03.01 and LAQGSM03.01,
and in their slightly corrected 03.02 and 03.03 versions.
The recent versions of our codes describe quite well
and much better than their predecessors
a large variety of nuclear reactions of interest to ADS, 
as well as reactions at higher
energies of interest to NASA, accelerator shielding, astrophysics,
and other applications, up to $\sim 1$ TeV/nucleon.
(Energies above about 5 GeV are only accessible to LAQGSM.) 

What is more, our codes provide
reasonable results even for low-energy reactions, where they are
not easy to justify from a fundamental physics point of view
(see {\ e.g.}, \cite{ND2004}).

We observe that codes developed for applications must not only
describe reasonably well arbitrary reactions without any free
parameters, but also not require too much computing time.
Our own exercises on many different nuclear reactions
with MCNPX \cite{MCNPX} and LAHET \cite{LAHET3,LAHET}
transport codes using different event generators show that
the current version of CEM is a little faster than
the Bertini \cite{BertiniINC} and ISABEL
\cite{ISABEL} options for INC in these transport codes,
and is several times faster than the  
INCL/ABLA \cite{INCL,ABLA} option.
The computing time of LAQGSM 
for heavy-ion reactions was tested recently
by Dr.\ Gomez \cite{LAQGSM_vs_PHITS}  with
MCNPX in comparison with PHITS \cite{PHITS}, on simulations of  
a 400 MeV/nucleon uranium beam on a 0.2-cm thick lithium
target. He found that PHITS using the Quantum
Molecular Dynamics (QMD)
 model to simulate heavy-ion reactions
requires 210 times more computing time than LAQGSM does,
making it very difficult if not impossible to use in 
applications requiring fast results, for example in 
real-time beam monitoring
of positron emitting fragments used for PET tomography
in heavy-ion cancer therapy; LAQGSM does not have these
problems, providing quite fast and reliable results.

Both our CEM and LAQGSM event generators still have some
problems in a reliable description of some
fragments from intermediate-energy nuclear reactions on
medium-mass nuclei, just as other similar modern Monte-Carlo 
codes do. To address these problems, we developed 
``S'' and ``G'' modifications of our codes,
that consider multifragmentation of
nuclei formed after the preequilibrium stage of reactions when
their excitation energy is above $2$--$5 \times A$ MeV using 
the Statistical Multifragmentation Model (SMM)
code by Botvina {\it et al.} \cite{Botvina87},
\cite{SMM}--\cite{SMM08},
and the fission-like binary-decay model GEMINI by Charity
\cite{GEMINI}--\cite{GEMINI08}, respectively.
The ``S'' and ``G'' versions of our codes allow us
to describe many fragmentation reactions that we are not 
able to reproduce properly with the standard version of our codes.
However, there are still some
problems to be solved in understanding the ``true'' mechanisms
of fragment productions from many reactions, therefore we
consider the present ``S'' and ``G'' versions of our codes only
as working modifications and they are not yet implemented 
as event generators into our transport codes, in contrast
to our ``standard'' 03.01, 03.02, and 03.03 versions
used as event generators.

The latest versions of our
03.01, 03.02, and 03.03 CEM and LAQGSM codes
have been or are being incorporated 
as event generators into the transport codes MCNP6 \cite{MCNP6},
MCNPX \cite{MCNPX}, and MARS \cite{MARS}.
CEM03.01 was made available to the public via 
RSICC at Oak Ridge and NEA/OECD in Paris
as the Code Package PSR-0532.
We also plan to make LAQGSM03.01 and the latest 03.03 versions of our codes
available to the public via RSICC and NEA/OECD in the future.

\begin{center}
{\large\bf
Acknowledgments}
\end{center}

We thank the Organizers of this Workshop,
especially Dr.\ Sylvie Leray, for inviting us to present 
these lectures and for financial support. 
We are grateful to ICTP and IAEA for kind hospitality.
These lectures were written with support 
from the US Department of Energy.

%\vspace*{1.5cm}
%\newpage

\end{document}